\documentclass[aps,rmp,reprint, amsmath,amssymb,graphicx,longbibliography]{revtex4-1}

\usepackage{bm}
\usepackage{graphicx}		
\usepackage{color}
\usepackage{dsfont}

\begin{document}

\title{Nonequilibrium Physics in Biology}

\author{Xiaona Fang}
\affiliation{State Key Laboratory of Electroanalytical Chemistry, Changchun Institute of Applied Chemistry, Changchun, Jilin 130022, China}
\author{Karsten Kruse}
\affiliation{NCCR Chemical Biology,
Department of Biochemistry and Department of Theoretical Physics, University of Geneva, 30 quai Ernest-Ansermet, 1211 Geneva 4, Switzerland}
\author{Ting Lu}
\affiliation{Department of Bioengineering, University of Illinois at Urbana-Champaign, Urbana, IL 61801, USA}
\author{Jin Wang}
\affiliation{State Key Laboratory of Electroanalytical Chemistry, Changchun Institute of Applied Chemistry, Changchun, Jilin 130022, China}
\affiliation{Department of Chemistry and Physics, State University of New York, Stony Brook, NY 11794 USA.}
\thanks{To whom correspondence should be addressed. E-mail: jin.wang.1@stonybrook.edu}

\date{\today}

\begin{abstract}
Life is characterized by a myriad of complex dynamic processes allowing organisms to grow, reproduce, and evolve. Physical approaches for describing systems out of thermodynamic equilibrium have been  increasingly applied to living systems, which often exhibit phenomena unknown from those traditionally studied in physics. Spectacular advances in experimentation during the last decade or two, for example, in microscopy, single cell dynamics, in the reconstruction of sub- and multicellular systems outside of living organisms, or in high throughput data acquisition have yielded an unprecedented wealth of data about cell dynamics, genetic regulation, and organismal development. These data have motivated the development and refinement of concepts and tools to dissect the physical mechanisms underlying biological processes. Notably, the landscape and flux theory as well as active hydrodynamic gel theory have proven very useful in this endeavour. Together with concepts and tools developed in other areas of nonequilibrium physics, significant progresses have been made in unraveling the principles underlying efficient energy transport in photosynthesis, cellular regulatory networks, cellular movements and organization, embryonic development and cancer, neural network dynamics, population dynamics and ecology, as well as ageing, immune responses and evolution. Here, we review recent advances in nonequilibrium physics and survey their application to biological systems. We expect many of these results to be important cornerstones as the field continues to build our understanding of life.
\end{abstract}

\maketitle

\tableofcontents

\section{Introduction}

Life is never at equilibrium. Here, equilibrium refers to thermodynamic equilibrium such that detailed balance holds: transitions between two states occur on average at equal rates in either direction. Biological systems break this condition because they continuously exchange matter, infor- mation, or energy with the environment, for example, by taking up food or by absorbing energy in the form of light or heat. As a consequence, it is typically inappropriate to describe them as physical systems relaxing to equilibrium. But then, what concepts and frameworks are suitable to study living systems?

On molecular scales, life seems to be governed exclusively by physical laws \cite{SchrodingerWhatisLife}; no specific life force has been identified as being necessary to explain the dynamics of biomolecules. Energy acquired from the environment drives enzymatic reactions which favor specific molecular changes over others. However, it is currently unclear how this specificity is achieved.  Furthermore, large- scale features of living systems such as cell migration and division, consciousness, population organization, or evolution are poorly understood in terms of the molecular components that make up these systems. This stands in stark contrast to the behavior of equilibrium statistical systems, which can be understood in terms of their microscopic components. In this review, we aim to present recent developments in the field of nonequilibrium dynamics and thermodynamics that provide routes toward answering answering fundamental questions about living systems. They will likely prove instrumental to identify features common to a large variety of biological systems and elucidate how mechanics and chemistry act in concert to shape life.

Because living systems regularly showcase phenomena that defy purely molecular explanations, physical analysis is required. Simultaneously, living systems provide a frontier of physical research that is as rewarding as the study of the very small or the very large with philosophical implications of similar depth. Even if we set the discussion of topics like consciousness aside, living systems provide us with a plethora of phenomena that are alien to systems of inanimate matter, for example, spontaneous oscillations or flows. Beyond the detailed description of isolated phenomena, the physics of active matter~\cite{Marchetti2013RMP,Wang2015ADP} aims to develop a framework of concepts and tools that is universally applicable. Here, ``active'' refers to the property of living systems to be intrinsically out of thermodynamic equilibrium. That is, instead of being driven by an external field or gradient, the constituents of active matter themselves are driven out of equilibrium. For example, a molecule might undergo conformational changes as a bound molecule of Adenosine triphosphate (ATP) loses a phosphate group through hydrolysis and the replacement of Adenosine diphosphate (ADP) by ATP restores the original conformation. In presence of an excess of ATP, this will lead to a cycling of the molecule between the two conformations.

To reach the aim of a universal framework, results from various disciplines are exploited, notably, nonlinear dynamics, nonequilibrium thermodynamics, and nonequilibrium kinetics \cite{Jackson, VanKampen,Gardiner,Schnakenberg,Prigogine,Haken,Graham,QianBook,Hu, Sasai2003,Qian2009ME,Ao2008Comm,Wang2008PNAS, Marchetti2013RMP, Wang2015ADP}, as well as nonequilibrium thermodynamics \cite{VanKampen,Gardiner,Schnakenberg,Prigogine,Haken,Nicolas,Gaspard,QianBook,Seifert,Hatano,Qian2009ME,
Esposito,Wang2006PLOSCB,Ao2008Comm,Ge, Feng2011JCP, Zhang2012JCP, Zhang2014JCP, Wang2015ADP,Freidlin-Wenzel,SchussBook, Maier1993, Wolynes2005PNAS, Roma2005, Sneppen2002,Feng2010JPCL, Assaf2010, Lv2014PlosOne, Wang2010JCP,Feng2014CS}. In this endeavour it has proven particularly useful to start with concepts from equilibrium physics. A central concept is that of a potential or free energy landscape. It describes the evolution of a physical system towards equilibrium in terms of gradient descent towards the landscape's minima. This powerful picture has found its way into many other disciplines, notably, into biology \cite{Wolynes1994PT, Waddington, Wright, Fisher}. There, ``Waddington's landscape'' is used to illustrate the process of cell differentiation, when a stem cell specializes to become a liver, a brain, or some other kind of cell with a specific function \cite{Waddington}. Similarly, ``fitness landscapes'' are used to picture the course of evolution towards species that fit better and better to some environmental constraints \cite{Wright, Fisher}. 

However, the concept of energy landscapes needs to be generalized for nonequilibrium systems to explain phenomena such as limit cycles. One way to generalize the energy landscape of a system is to first consider its stochastic form. Under very general conditions, the probability distribution asymptotically reaches a steady state even for systems out of equilibrium. Gradients in the steady-state probability land- scape drive the deterministic ?mean field? dynamics onto one of the possible attractors, for example, a limit-cycle orbit. Typically, a second force results from a rotational flux~\cite{Xu2012JCP}, which drives the dynamics within an attractor, for example, the oscillation orbits within a limit cycle attractor~\cite{Wang2008PNAS}. These rotational fluxes are a consequence of energy or material being pumped into the system and thus tightly linked to entropy production and time- reversal symmetry breaking.

A different generalization of energy landscapes arises in the context of nonequilibrium thermodynamics, which is based on the rate of energy dissipation~\cite{deGroot:1985ue}. In this framework, one assumes spatially extended systems to be locally at thermodynamic equilibrium, but these equilibria may differ at different locations. In this way, a ?free energy landscape? of ?spatially heterogeneous equilibria? can be defined. Together with conservation laws and broken continuous symmetries, this assumption allows for a systematic framework to analyze deviations from thermodynamic equilibrium. For example, this approach yields the Navier-Stokes equation for an isotropic system of a single, conserved molecular species. Generalized hydro- dynamics have notably been developed for active matter and applied to various biological systems~\cite{Marchetti2013RMP}.
Both the stochastic system and the energy dissipation approaches allow for the distinction between driving forces resulting from thermodynamic equilibrium relaxation and from environmental coupling~\cite{Marchetti2013RMP, Wang2015ADP}.

Independently of their application to biological systems, the approaches raise immediate physics questions.  How do other equilibrium concepts as the thermodynamics and its fluctuations, optimal paths, kinetic rate, and fluctuation-dissipation theorem (FDT) generalize~\cite{Wang2010JCP,Feng2011JCP,Zhang2012JCP,Feng2014CS}? 
The work of Shannon ~\cite{Shannon} has clearly revealed a deep formal connection between concepts from equilibrium physics and information theory. How can information be incorporated in the nonequilibrium framework? Furthermore, biological systems are often characterized by their function. How can this be conceptualized and physically quantified? To give an example: biological organisms ``make decisions''. What is the physical basis of the process and to what extent do we need to incorporate it into the description of population dynamics?

Although physics has been applied to study biological processes for a long time, its popularity for this application is currently surging. Thanks to spectacular advances in genome sequencing, the molecular inventory of many biological organisms is now well known. Many proteins can be isolated and studied in reconstituted systems outside a cell, which allows tests of physical hypotheses in controlled environments. At the same time, proteins can be modified in a variety of ways in living cells to tune biological processes. The addition of fluorescent protein tags to functional proteins allows for researchers to probe their \textit{in vivo} behavior. Continuously progressing fluorescence imaging and electron microscope technology allows for the observation of cellular activities and organization in unprecedented detail. These developments together have led to enormous amounts of data that wait to be analyzed, and new data are continuously added. These experimental advances have a profound impact on expanding the concepts and tools used for describing active matter.

With this review, we intend to describe recent progress in the development of concepts in nonequilibrium physics and their application to biological systems. Since the topic is too vast to be covered fully in the present text, we selected topics that are in the focus of current research and have proven to be relevant for understanding important vital processes. The choice also reflects the topics the authors are familiar with. We will present a brief overview of the current status of concepts from nonequilibrium physics, where we highlight nonequilibrium potential and flux approach for dynamics/thermodynamics as well as the hydrodynamics approach for active matter. We then show how these concepts have been applied to a broad range of biological systems. Starting from molecular processes in the form of enzyme reactions and energy transport we will move up in scales by covering cellular processes in the form of cell fate decision making, cell cycle, differentiation and ageing, cell ensembles in the form of neural networks, tumors and immunity, and finally species and populations in the form of ecology, game theory and economy, and lastly evolution.  At the end, we give a brief summary and outlook of some current and future directions of the nonequilibrium physics in biology.

\section{Physical concepts for describing nonequilibrium systems}
\label{sec:physicalConcepts}

In this section, we will review some recently developed physical concepts for describing nonequilibrium dynamics. Special attention will be paid to the landscape and flux theory, which generalizes the notion of potentials to systems out of equilibrium. A more detailed review can be found in~\cite{Wang2015ADP}.

\subsection{Landscape and flux theory for nonequilibrium dynamics}

\subsubsection{Dynamical systems}

Consider a dynamical system and let ${\bf C}$ denote the vector of all system variables such as concentrations, momentum, polar order etc. In case of a
deterministic evolution and when the change with time of the system state ${\bf C}$ depends only on the current state, the 
dynamical equation reads
\begin{equation}
\label{eq:fundamentalDynamicEquation}
\dot{{\bf C}}={\bf F}({\bf C}),
\end{equation}
where ${\bf F}({\bf C})$ is the generalized driving force~\cite{Jackson}. This equation describes the evolution of a wide class of systems. 

An analysis of this equation usually starts by identifying the fixed points and then studying their local stability~\cite{Jackson}. Global stability needs to
be addressed separately since connections among the steady states are not always known from the local analysis, see Sect.~\ref{sec:globalStability}.
For potential systems, where the driving force can be expressed as the gradient of a potential (or energy) $U$, $\mathbf{F}=-\nabla U$, the global system behavior can be read off directly from the potential. Indeed, since $dU/dt=-\mathbf{F}\cdot\mathbf{F}$ along any trajectory, the minima of $U$ are the stable fixed points of a potential system. 

\subsubsection{Nonequilibrium potentials and rotational curl fluxes as the driving forces for dynamics}
\label{sec:noneqpotentials}

In general, the driving force cannot be expressed as the gradient of a potential. In this case, it is helpful to
explore a stochastic version of the deterministic
equation~(\ref{eq:fundamentalDynamicEquation}). The noise term can account for all factors not explicitly described by $\mathbf{F}$ and, if desired, one can eventually take the zero-noise limit~\cite{Swain02PNAS}. 
Stochastic trajectories are determined by the Langevin equation
\begin{equation}
\label{eq:langevin}
\dot{{\bf C}}={\bf F}({\bf C}) + {\bf \eta}({\bf C},t).
\end{equation}
Here, $\eta({\bf C},t)$ represents a stochastic force with the fluctuation strength measured by the auto-correlation function in time, $<{\bf \eta}({\bf
C},t) {\bf \eta}({\bf C},t')>=2 D \mathsf{D} \delta (t-t')$. In this expression, $D$ is an overall scale factor representing the magnitude of fluctuations
and $\mathsf{D}$ the diffusion matrix describing fluctuation anisotropies~\cite{VanKampen}. Instead of individual trajectories, one considers the probability distribution $P$ that 
evolves according to a linear deterministic equation, the Fokker-Planck
equation~\cite{VanKampen}:
\begin{equation}
\label{eq:FokkerPlanck}
\partial_t P({\bf C},t) + \nabla \cdot {\bf J}({\bf C},t) = 0
\end{equation}
with $\nabla$ denoting the gradient in state space. This equation describes probability conservation and 
the probability current ${\bf J}$ is given by
\begin{align}
{\bf J}({\bf C},t) &= {\bf F}({\bf C}) P ({\bf C},t)- \nabla \cdot [D \mathsf{D} P ({\bf C},t)] .
\end{align}
The first term describes an advective current that results from the driving force and the diffusion 
term captures the effects of fluctuations.

Eventually, the system typically reaches a steady state $P^{ss}$ with $\partial_tP^{ss}=0$. Consequently, the divergence of the steady state current vanishes. Potential systems in the steady state are in thermodynamic equilibrium, that is, the probability current vanishes itself such that detailed balance is obeyed and the probability is given by the Boltzmann distribution. In general, the steady state current does not vanish. However, it has to be purely rotational or curl.
The existence of a non-vanishing flux in steady state indicates that detailed balance is broken and the magnitude of the flux can be used to
measure the distance from thermodynamic equilibrium.

The driving force can 
be decomposed into a part that is the negative gradient of a nonequilibrium potential $U$ (or landscape) and a part involving a
rotational curl flux~\cite{Wang2008PNAS}. Explicitly,
\begin{align}
\label{eq:forceDecomposition}
\mathbf{F}= - D \mathsf{D} \cdot \nabla U + \nabla D \mathsf{D}  + \mathbf{J}^{ss}/P^{ss}
\end{align}
Here, $U$ is the negative logarithm of the steady state probability, $U=- \ln P^{ss}$. As its equilibrium analog, the nonequilibrium potential landscape is linked to
the steady state probability and provides a global quantification of the system behavior through its landscape \footnote{Typically, the diffusion coefficient tensor $D\mathsf{D}$ is constant and does not contribute to the driving force, although in general it contributes to the potential landscape.}. However, the nonequilibrium dynamics on the landscape also depends on the rotational curl flux. 
Whereas one can visualize equilibrium dynamics as a charged particle moving in an electric field,
nonequilibrium dynamics thus corresponds to a charged particle moving in an electric and a magnetic field.
Note that in contrast to the equilibrium
landscape, which is given a priori, the nonequilibrium potential landscape is associated to the steady state probability emerged from the stochastic dynamics. In turn, the rotational 
curl flux $\mathbf{J}^{ss}$ is associated to the steady state probability flux resulting from the coupling to the environment that allows for an exchange of matter, energy, or information~\cite{Xu2012JCP, Zhang2014JCP,Zeng2016ArXiv}. As it does not vanish, it breaks time reversal symmetry and thus creates a time arrow~\cite{Wang2010JCP, Feng2011JCP, Li2011BJ}. Furthermore, the rotational curl flux current extends through state space. Hence, unlike for equilibrium systems, the steady state can typically not be described fully by local properties.

There is some freedom in decomposing the driving force into a potential gradient and a rotational curl flux as one can always add the curl of a vector field
to the nonequilibrium potential $U$. The choice made above is somewhat natural as it leads to the equilibrium case, when the flux is zero, and is the
closest analog when one wants to describe the relaxation into steady state. It also allows for a generalization of
thermodynamics~\cite{Ge,Seifert1,Esposito,Feng2011JCP} and of the fluctuation-dissipation relation~\cite{Hatano, Seifert, Feng2011JCP}. 

There are several other ways of decomposing the stochastic dynamics. 
Some studies focus 
on finding a nonequilibrium potential~\cite{Graham,
Freidlin-Wenzel, Ao2004,Xing2010, Zhou2012} and associated analytical properties. Others emphasized 
the roles of the nonequilibrium curl flux in
addition to the nonequilibrium landscape in determining the dynamics \cite{Wang2008PNAS, Wang2010JCP, Feng2011JCP,Zhang2014JCP}. One approach aimed at finding a new type of
stochastic dynamics \cite{Ao2004}, although it may be challenging to obtain numerical solutions. Furthermore, the generality as well as uniqueness of the approach are still under discussion~\cite{Qian2014,
 LiTiejun2016JCP}.  Another approach, suggested by~\cite{Xing2010} used a projection operator to decompose the driving force. A recently proposed decomposition approach assumed orthogonality between the driving forces, which works only in the deterministic limit~\cite{Zhou2012}, reaching similar conclusions to those discussed earlier \cite{Wang2008PNAS, Wang2010JCP, Feng2011JCP} in the zero fluctuation limit. The force decomposition has been generalized from overdamped to underdamped dynamics~\cite{Risken, Qian2014, GeHao2014}. Another method decomposed discrete Markov chains into two parts: one that preserved and another that broke detailed balance~\cite{Schnakenberg, Qian04,Zia, Zhang2014JCP}. %
  
 \subsubsection{Thermodynamic origin of the rotational flux}

The landscape and flux can be obtained as mentioned once specific dynamics are given, although this statement is rather formal. To gain physical intuition, one can search for the origin of rotational curl flux. For an open system, the flux originates from environmental energy input, which breaks the detailed balance. In biology, the energy that drives the system away from equilibrium can be obtained from ATP hydrolysis. For example, the phosphorylation and dephosphorylation of ATP can provide an energy source for cellular functions, such as cell growth and division. One can couple ATP hydrolysis to specific protein reactions in molecular networks and other cell systems to explicitly quantify the energy input by the chemical potential (voltage) from the ATP/ADP concentration ratio for driving the associated nonequilibrium dynamics~\cite{Qian2007, Xu2012JCP}. This voltage gives rise to the rotational curl flux. Alternatively, one can phenomenologically couple the ATP chemical potential to thermodynamic forces as in the theory of active matter~\cite{Kruse:2004il,Marchetti2013RMP}. The quantitative connections from ATP that pump voltage to the flux driving nonequilibrium dynamics and entropy production or free energy cost have been studied in a few examples~\cite{Xu2012JCP}.

\subsubsection{Global stability and Lyapunov function for nonequilibrium systems}
\label{sec:globalStability}

The asymptotic dynamics and global stability of a system can be quantified, if it admits a Lyapunov function. A Lyapunov function $\phi$ is a real-valued function satisfying $\frac{d\phi}{dt}=\sum_iF_i\partial\phi/\partial C_i<0$ except in steady state~\cite{Jackson}.
For a deterministic system, a candidate Lyapunov function can be obtained by calculating the nonequilibrium potential and then taking the zero-noise limit.
To see this, consider the WKB ansatz up to leading order in $D$ for solving the Fokker-Planck equation~(\ref{eq:FokkerPlanck}), that is, $P \sim
\exp[-\phi_0/D]$, where $\phi_0$ is the leading-order term of the nonequilibrium potential. We thus arrive at the Hamilton-Jacobian
equation~\cite{Hu,Zhang2012JCP,Xu2013Non,Xu2014PlosOne}
\begin{align}\label{HJE}
\mathbf{F} \cdot \nabla \phi_0 + \nabla \phi_0 \cdot \mathsf{D} \cdot \nabla \phi_0 &= 0.
\end{align}
This equation implies
\begin{align}
\frac{d\phi_0}{dt} & = \mathbf{F} \cdot \nabla \phi_0
= - \nabla \phi_0 \cdot \mathsf{D} \cdot \nabla \phi_0 \leq 0
\label{eq:LyapunovZeroNoise}
\end{align}
showing that the nonequilibrium potential is a candidate Lyapunov function. 

\subsection{Discrete nonequilibrium dynamics}

In a biological context, many systems are characterized by discrete rather than continuous states. Think, for example, of the promotor site of a gene that
can be occupied or not by a transcription factor, molecular motors that have discrete binding sites on a cytoskeletal filament, or the number of
individuals in a population. We will now describe how the approaches described above for continuous systems can be adapted to the discrete case.

\subsubsection{The Master equation}

For a Markovian process in discrete state space, the analog of the Fokker-Planck equation determining the dynamics of the probability distribution $P$ is the Master
equation~\cite{VanKampen,Gardiner}:
\begin{equation}\label{eq:MasterEquation}
\frac{dP_i}{dt} = -\sum_jT_{ij}P_i + \sum_jT_{ji}P_j.
\end{equation}
Here $P_i$ is the probability of being in state $i$ and $T_{ij}$ denotes the transition rate from state $i$ to state $j$. The master equation reflects that
the probability to be in state $i$ decreases through transitions from state $i$ into any other state and increases through transitions from the other
states into state $i$. It implies conservation of probability, $d \sum_i P_i/dt=0$. Alternatively, we can write Eq.~(\ref{eq:MasterEquation}) as
\begin{align}
d\mathbf{P}/dt=\mathsf{M}^T\mathbf{P}
\end{align}
with the transition rate matrix  $\mathsf{M}$ given by $M_{ij}=T_{ij}$ for $i\neq j$ and $M_{ii}=(-1)\sum_jT_{ij}$. One can solve the master equation either
directly or by simulating the stochastic evolution of the system dynamics to gain information about the evolution of the probability with time and the steady state probability $P_i^{ss}$~\cite{Gillespie1976, Krauth:2006vv,Liang2010PNAS}.

In steady state, the flux between two states $i$ and $j$ is $F_{ij}^{ss}=T_{ji}P^{(ss)}_{j}-T_{ij} P^{(ss)}_{i}$. Detailed balance is satisfied if
$F_{ij}^{ss}=0$. In that case, the system is in equilibrium and
there is a potential $V_i$, such that
$P^{ss}_i\propto\exp\left\{-V_i\right\}$~\cite{Schnakenberg,Qian04,Zia}. However, $\frac{dP_i}{dt}=0$ only states that the sum of all fluxes
into and out of state $i$ is zero and thus $\sum_jF_{ij}^{ss}=0$, but not all terms individually.

\subsubsection{Decomposition of the transition matrix}

Similar to the flux in continuous systems, the transition rate matrix can be separated into a detailed-balance preserving part $\mathsf{D}$ and a
detailed-balance breaking part $\mathsf{C}$ by defining $C_{ij}=\max\{T_{ij}P^{ss}_i-T_{ji}P^{ss}_j, 0\}/P^{ss}_i$ for $i\neq j$ and
$C_{ii}=(-1)\sum_jC_{ij}$, and $D_{ij}=\min\{T_{ij}P^{ss}_i, T_{ji}P^{ss}_j\}/P^{ss}_i$, for $i\neq j$ and
$D_{ii}=(-1)\sum_jD_{ij}$~\cite{Schnakenberg,Qian04}. One can see that $\mathsf{M}=\mathsf{C}+\mathsf{D}$ and $\mathsf{D}^T\mathbf{P}^{ss}=0$. Since
$\mathsf{M}^T\mathbf{P}=(\mathsf{C}+\mathsf{D})^T\mathbf{P}^{ss}=0$, we also have $\mathsf{C}^T\mathbf{P}^{ss}=0$.
Whereas $\mathsf{D}$ preserves detailed balance, $D_{ij}{P_i}^{ss}=D_{ji}{P_j}^{ss}$, $\mathsf{C}$ captures the flux breaking detailed balance and
describes irreversible transitions, because $C_{ij}{P_i}^{ss}>0$ implies $C_{ji}{P_j}^{ss}=0$ and vice versa.

The steady state fluxes $F_{ij}^{ss}$ 
resulting from the detailed-balance breaking 
part of the driving force 
can be expressed in terms of fluxes along loops that
connect a state with itself, $i\to j\to k\cdots\to n\to i$~\cite{Schnakenberg, Qian04, Zia, Zhang2014JCP, Luo2016ArXiv}. 
In equilibrium, the flux along any loop is the same as for the corresponding reversed loop so that the net flux is zero. The flux loops provide additional information to the probability distribution for describing the nonequilibrium dynamics of a discrete system.


\subsection{Nonequilibrium paths}

An alternative formulation to the stochastic dynamic equation~(\ref{eq:langevin}) to describe the dynamical process is in terms of a path integral that determines the probability to go from
an initial state $\mathbf{C}_\text{initial}$ at $t=0$ to a final state $\mathbf{C}_\text{final}$ at time $t$~\cite{Onsager, Wiener, Feynman,
Ross, Wang2005BJ, Wang2006PRL, Wang2006BJ,Wang2010JCP,Wang2011PNAS,Feng2014CS,Feng2010JPCL, Zhang2013PNAS, Li2013PlosCB, Li2014JRSI, Roma2005, Sneppen2002,Maier1993, Wolynes2014PNAS}:
\begin{align}
P(\mathbf{C}_\text{final},t,\mathbf{C}_\text{initial},0)&=\int \mathcal{D} \mathbf{C}(t) \exp\left\{- S\left[\mathbf{C(t)}\right] \right\}
\label{eq:pathIntegral}
\end{align}
%
The integral over $\mathcal{D} \mathbf{C}(t)$ refers to the sum over all possible paths between the initial state $\mathbf{C}_\text{initial}$ at time $t=0$
and the final state $\mathbf{C}_\text{final}$ at time $t$. The action $S$ is the integral of the Langrangian along the path $\mathbf{C}(t)$, i.e.,
$S\left[\mathbf{C}(t)\right]=\int_0^tdt'\;L(\mathbf{C}\left(t')\right)$, which is given by
\begin{align}
L &= \frac{1}{2} \nabla \cdot \mathbf{F}
+ \frac{1}{4D} \left(\dot{\mathbf{C}} - \mathbf{F} \right) \cdot \mathsf{D}^{-1} \left(\dot{\mathbf{C}}- \mathbf{F} \right)
\end{align}
The first term results from the deterministic driving force, 
whereas the second
is a consequence of the Gaussian fluctuations $\eta$. 

For a potential system with $F=-D \mathsf{D}\cdot \nabla U$, 
the cross product terms in the action, $ -1/2 \int \frac{1}{\bf{D}} \cdot {\bf F} \cdot \dot{ \bf{C}} dt = -1/2 \int \frac{1}{\bf{D}} \cdot \bf{F} \cdot \bf{dC}$, are independent of the path and thus constant, such that they do not
 contribute to the optimal path equation. However, for non-potential systems they do contribute. In particular, the integral along a loop does not vanish in this case, which is akin to the 
Aharonov-Bohm effect in quantum mechanics and can be used to classify the
underlying topologies of nonequilibrium systems~\cite{Wang2010JCP, Feng2011JCP}. 

Often, the integral in Eq.~(\ref{eq:pathIntegral}) is approximated well by considering only the contribution of the so-called optimal path that is 
minimizing the action
\cite{Wang2010JCP,Wang2011PNAS,Feng2014CS,Feng2010JPCL, Zhang2013PNAS, Li2013PlosCB, Li2014JRSI}.
It is 
determined by the Euler-Lagrange equation
\begin{align}
\frac{d}{dt} \frac{\partial L}{\partial \dot{\mathbf{C}}}-\frac{\partial L }{\partial \mathbf{C}} = 0.
\label{eq:EulerLagrange}
\end{align}
Since the probabilities (\ref{eq:pathIntegral}) can be used to determine the nonequilibrium potential, optimal paths offer a possibility to reduce the computational effort for calculating the landscape from exponential to polynomial~\cite{Wang2010JCP}.

\begin{figure}[!ht]
\includegraphics[width=0.5\textwidth]{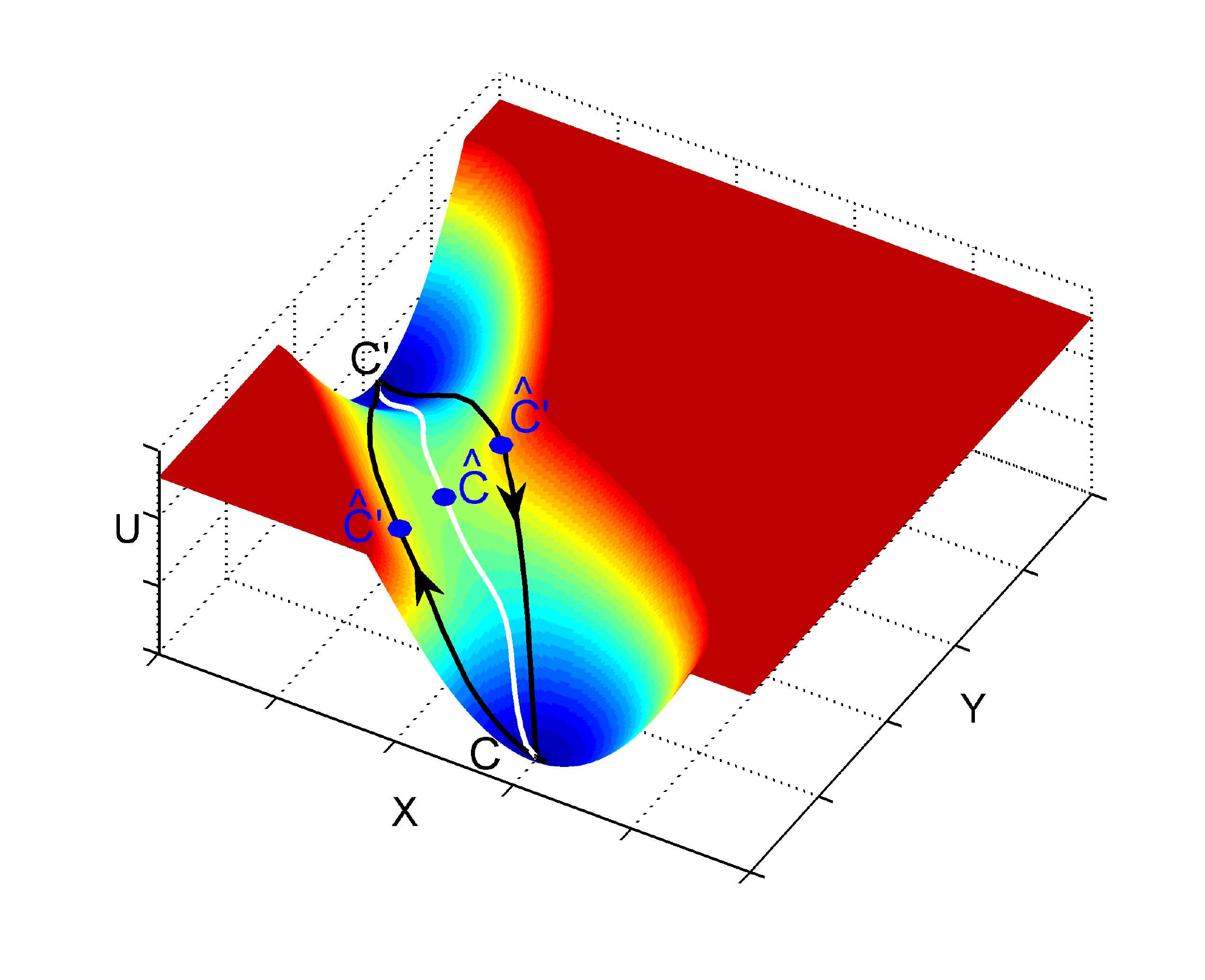}
\caption{ 3D illustration of a nonequilibrium landscape and nonequilibrium paths. The states $C$ and $C'$ are attractors of the system. The black lines indicate the optimal irreversible paths between them with $\hat{C'}$ denoting the maxima along them. The white line shows the steepest decent gradient path going through the saddle point $\hat C$. From~\cite{Feng2014CS}.}
\label{Feng2014CS_fig3}
\end{figure}

An example of optimal paths connecting two steady states 
is illustrated in Fig.~\ref{Feng2014CS_fig3}. Due to the 
curl flux force, optimal paths do in general not follow the landscape gradient and do not pass the saddle point $\hat{\mathbf{C}}$ between the two states' basins
of attraction. 
The example illustrates furthermore that optimal paths in non-potential systems are irreversible. 

It is often very challenging to solve solve the Euler-Lagrange equation (\ref{eq:EulerLagrange}). In practice, one can
directly evaluate the action of the paths by Monte Carlo methods and obtain the optimal path with minimum action through the Hamilton-Jacobi approach. This can
reduce the action calculations 
from a multi-dimensional integral into a one dimensional line integral \cite{Elber1, Elber2, Orland2006PRL, Wang2010JCP,Wang2011PNAS,Feng2014CS,Feng2010JPCL, Zhang2013PNAS, Li2013PlosCB, Li2014JRSI}.

\subsection{Nonequilibrium transition state theory}\label{sec:TransitionStateTheory}

In a stochastic potential system, transitions occur between states associated with potential minima. The transition path follows the optimal path and passes through a ``transition state'' that is associated with a saddle point of the potential. In the limit of small fluctuations, Kramers calculated the rate to be~\cite{Kramers}
\begin{align}
r&=\frac{D}{2\pi k_b T}\sqrt{U''_\textrm{min}|U''_\textrm{max}|}\exp\left[-\frac{\Delta U}{k_BT}\right].
\end{align}
$\Delta U$ is the potential difference between the initial and the transition state, $k_B T$ is the thermal energy, and $U''_\textrm{min}$ and $U''_\textrm{max}$ are the potential curvatures in the initial and the transition state, respectively. The latter are associated with the fluctuations around these states.

For nonequilibrium systems, the optimal paths do not necessarily pass a saddle point unless the
fluctuations approach zero.  Therefore, the transition state or Kramer's rate theory needs to be modified~\cite{Freidlin-Wenzel,SchussBook,Maier1993,Feng2014CS}. For small but finite fluctuations ~\cite{Feng2014CS}, one gets
\begin{align}
\label{rate2}
r &= \frac{\lambda_u (\hat{\mathbf{C}})}{2 \pi} \sqrt{\frac{\det \mathsf{M_{fluct}}(\mathbf{C})}{|\det \mathsf{M_{fluct}}(\hat{\mathbf{C'}})|}}
\mathrm{e}^{-S_\text{HJ}}.
\end{align}
Here, $r$ is the transition rate from state $\mathbf{C}$ to a state $\mathbf{C}' $. Furthermore, $\hat{\mathbf{C'}}$ denotes the point, where the optimal path
between $\mathbf{C}$ and $\mathbf{C}'$ crosses the line separating the two corresponding basins of attraction. 
It is determined by
$\mathsf{D}^{-1}\cdot\mathbf{F}=0$ along the optimal path. 
Furthermore, $S_\text{HJ}$ represents the
action along the optimal path. 
Finally, $\mathsf{M_{fluct}}$ denotes 
the Hessian of the action and $\lambda_u$ its unstable
eigenvalue. 
Similar to the Kramers' rate, these prefactors are associated with fluctuations around $\mathbf{C}$ and $\hat{\mathbf{C'}}$.
Note that $\hat{\mathbf{C'}}$ is different for transitions from $\mathbf{C}$ to $\mathbf{C}'$ and back, such that these rates do not
obey detailed balance.

\subsection{Nonequilibrium thermodynamics}
\label{sec:nonEqThermoDyn}


To develop nonequilibrium thermodynamics, one can start with the Shannon entropy of the system~\cite{Schnakenberg,Prigogine,Esposito,Ge, Ao2008Comm,
Wang2006PLOSCB, Feng2011JCP, Zhang2012JCP}
\begin{align}
\mathcal{S} &= - \int P(\mathbf{C},t) \ln P(\mathbf{C},t)\; d\mathbf{C}
\end{align}
The temporal evolution of the entropy can be decomposed into two parts:
\begin{align}
\dot{\mathcal{S}} &= \dot{\mathcal{S}}_t - \dot{\mathcal{S}}_e\\
\intertext{where}
\dot{\mathcal{S}}_t &= \int (\mathbf{J} \cdot (D\mathsf{D})^{-1} \cdot \mathbf{J}) \frac{1}{P} \;d\mathbf{C}\\
\intertext{represents the entropy production rate or the rate at which the total entropy of the system and the environment change, and} 
\dot{\mathcal{S}}_e &=  \int (\mathbf{J} \cdot (D\mathsf{D})^{-1} \cdot  \mathbf{F}')\;d\mathbf{C}
\end{align}
is the entropy flow 
into or out of the system from the environment. 
The entropy production rate $\dot{\mathcal{S}}_t$ is directly related to the nonequilibrium flux
$\mathbf{J}$ and is always larger than or equal to zero. In the expression of the entropy flux $\dot{\mathcal{S}}_e$, the effective force $\mathbf{F}'$ is
given by $\mathbf{F}'= \mathbf{F} - D \nabla \cdot \mathsf{D}$. The entropy flux can be positive (reduction of system entropy), which gives the opportunity of
creating order in the system as can be observed for living systems. At steady state, the entropy production rate is directly related to the rotational curl steady state probability flux and is equal to the entropy flux or heat dissipation.

Analogs of equilibrium thermodynamic quantities can be defined. If $U$ denotes again the nonequilibrium potential $U=-\ln P^{ss}$, then the nonequilibrium potential energy is given by
$\mathcal{U}=D\int U(\mathbf{C})P(\mathbf{C},t)\;d\mathbf{C}$ and the free energy $\mathcal{F}$ by $\mathcal{F}=D\int
P(\mathbf{C},t)\ln\left(P(\mathbf{C},t)/P^{ss}(\mathbf{C})\right)\;d\mathbf{C}$. These quantities are related through 
$\mathcal{F}=\mathcal{U}-D\mathcal{S}$.
The negative of the total entropy $-\mathcal{S}_t$ and the (nonequilibrium) free energy $\mathcal{F}$ are Lyapunov functions for the evolution of the probability $P$.
Note that the nonequilibrium potential energy and the nonequilibirum free energy (so sometimes called relative entropy) are defined in an analogous way to the energy and free energy in equilibrium statistical physics. 
The amplitude of the fluctuations represented by the scale factor $D$ effectively plays the role of the temperature. The actual thermal energy could be involved in $D$. 

Another form of nonequilibrium thermodynamics relates to the underlying dynamics (landscape
and flux) from the nonequilibrium fluctuation dissipation relation by evaluating the equal time correlation
functions of the flux velocity~\cite{Feng2011JCP}. The time derivative of the free energy can be written as
\begin{align}
\dot{\mathcal{F}} &= D\left\langle\mathbf{v}\cdot\nabla\ln\left[P/P^{ss}\right]\right\rangle,
\end{align}
where $\left\langle\cdots\right\rangle$ denotes the average with respect to $P$ and $\mathbf{v}=\mathbf{J}/P$ is the flux velocity. From
this
\begin{align}
\dot{\mathcal{F}} &= \left\langle\mathbf{v}^{ss}\cdot\mathsf{D}^{-1}\cdot\mathbf{v}^{ss}\right\rangle -
 \left\langle\mathbf{v}\cdot\mathsf{D}^{-1}\cdot\mathbf{v}\right\rangle\\
 &\equiv \dot Q-D\dot{\mathcal{S}_t},
\end{align}
where $\dot Q$ is the heat flowing out of the system. It follows that the rate of entropy production results from relaxation to the steady state along
gradients of the relative potential ($\dot{\mathcal{F}}$) and from a constant exchange of heat that keeps the system out of equilibrium in steady state
($\dot Q$)~\cite{Seifert1,Esposito, Ge, Feng2011JCP}. Thus the nonequilibrium thermodynamics in this form states that total entropy production is from both
non-stationary relaxation to the steady state and house keeping for maintaining the steady state~\cite{Ge, Feng2011JCP}. At steady state, the entropy production rate is equal to the house keeping heat for maintaining the steady state. Therefore, it is important to quantify entropy production as the nonequilibrium thermodynamic (or dissipation) cost for maintaining the steady state.

After defining the generalized thermodynamic force as
\begin{align}
\mathbf{X} &= (D\mathsf{D})^{-1}\cdot\mathbf{F}/P
\end{align}
one obtains
\begin{align}
\label{eq:Sdot}
\dot{\mathcal{S}_t} &= \int \mathbf{J}\cdot \mathbf{X}\;d\mathbf{C}.
\end{align}
Considering the individual components of the scalar product in the integral, on can thus identify pairs of conjugated thermodynamic forces and fluxes. They
form the basis of phenomenological (hydrodynamic) descriptions of nonequlibrium systems, where the thermodynamic fluxes can be expressed up to linear order in terms of the forces
\begin{align}\label{eq:onsager}
\mathbf{J} &= \mathsf{L}\mathbf{X}.
\end{align}
The matrix $\mathsf{L}$ respects the symmetries of the system (Curie principle) and contains the phenomenological coupling coefficients, which fulfill the Onsager reciprocal  relations~\cite{OnsagerRelation}.

The above considerations can be also applied to discrete systems. With the transition rate between state $i$ and $j$ again given by $T_{ij}$, the flux
between them is $F_{ij}=T_{ji}P_j-T_{ij}P_i$. We can also introduce the generalized thermodynamic potential
$A_{ij}\equiv\ln\left(\frac{T_{ij}P_i}{T_{ji}P_j}\right)$. For the entropy $S= -\sum_iP_i\ln P_i$ one then finds~\cite{Schnakenberg}
\begin{align}
\dot{S} &= \sum_{i,j} F_{ij}A_{ij}=\sum_{i,j}T_{ij}P_i\ln\left[\frac{T_{ij}P_i}{T_{ji}P_j}\right].
\end{align}
Similar to Eq.~(\ref{eq:Sdot}) the change in entropy is thus expressed as a sum of products of conjugated fluxes ($F_{ij}$) and forces ($A_{ij}$).

Since changes of entropy are intimately related to the existence of fluxes, the rate of entropy changes or entropy production can be used as a measure for how much a system is
out of equilibrium.

\subsubsection{Crooks' theorem and the Jarzynski relation}

When applying thermodynamics to ?small? systems where the number of molecules is on the order of $10^{10}$ or less, for example, on the scale of individual cells ($10^{4}$), fluctuations become significant. Consequently, it does not suffice to consider mean field thermodynamic quantities; distributions must be considered. A number of fluctuation theorems have been derived to relate these distributions to entropy production in case some time-dependent work process is applied to a system~\cite{Seifert2012RPP}. In particular, consider the change of the microstate $\mathbf{C}$ of a system coupled to a heat bath at temperature $T$. The change occurs along a path $\mathbf{C}(t)$ from $\mathbf{C}(0)=\mathbf{C}_\text{initial}$ to $\mathbf{C}(\tau)=\mathbf{C}_\text{final}$. Furthermore, let $\lambda$ denote a time-dependent system parameter. Then the distribution $P\left[\mathbf{C}(t)|\lambda(t)\right]$ of the path and the corresponding distribution for its reverse path $\bar{\mathbf{C}}$ fulfill~\cite{Crooks1,Crooks2}
\begin{align}
\frac{P\left[\mathbf{C}(t)|\lambda(t)\right]}{P\left[\bar{\mathbf{C}}(\tau-t)|\bar{\lambda}(\tau-t)\right]}&=\exp[\Delta S]= \exp[\beta (W-\Delta F)].
\end{align}
Here, $\Delta S$ and $\Delta F$ denote the differences in entropy and free energy between the initial and the final equilibrium state, whereas $W$ is the work applied to the system, and $\beta$ is inverse temperature. This was verified experimentally; see,  for example, \cite{Bustamenti2005Nature} and \cite{Seifert2005PRL}.

The fluctuation theorem implies the Jarzynski relation, namely, that the average exponential of the work performed equals the exponential of the free energy change~\cite{Jarzynski1, Jarzynski2}. Explicitly,
\begin{align}
\left\langle\exp[-\beta W]\right\rangle &= \exp[-\beta \Delta F].
\end{align}
This relation implies bounds on the possible paths along which work is extracted from the system by reducing its entropy. From a practical point of view it shows how to measure free energy differences by driving the system in arbitrary ways between two equilibrium states. This was applied experimentally to conformations of single RNA molecules \cite{Bustamenti}.

\subsubsection{Fluctuation-dissipation theorem for intrinsic nonequilibrium systems}
The conventional fluctuation dissipation theorem is important in linking the response of the system upon perturbations to the equilibrium fluctuations
\cite{Kubo}. This is useful for experimental efforts in extracting the system equilibrium fluctuations from its response, or vice versa. For nonequilibrium systems with broken detailed balance, a generalization of the FDT is
necessary ~\cite{Cugliandolo, Seifert1, Prost, Seifert2010EPL, Feng2011JCP}. In the landscape and flux representation, it takes the form~\cite{Feng2011JCP}
\begin{align}\label{R1}
 R^{\Omega}_i (t-t')& = - \langle \Omega(t) \partial_i \ln [P^{ss}({\bf x})] \rangle \\
 & = -\Big [ \langle  \Omega(t) \tilde{F}_k (t') D^{-1}_{ik} (t') \rangle
 + \langle \Omega(t)  v^{ss}_k (t') D^{-1}_{ik} (t') \rangle   \Big ],\label{eq:fdt}
 \end{align}
where $\Omega$ is an observable and $R$ the response to a perturbation. Furthermore $\tilde{F}_i  = F_i - \partial_j D_{ij}$ is again the generalized force and $v^{ss}= J^{ss}/P^{ss}$ the flux velocity. 

This expression reveals that there are two contributions to the system?s response. The first term on the right-hand side of
Eq.~(\ref{eq:fdt}) is analogous to the expression for equilibrium systems and results from correlations between observable ? and their driving force. The second term involves a correlation between the steady-state flux velocity and observ- able ?, which breaks the detailed balance. Thus, this term is absent in equilibrium systems. The general response therefore depends on both steady-state fluctuations and curl flux. Equation~(\ref{eq:fdt}) can be used to experimentally quantify the rotational curl flux by measuring the difference between the response function and the fluctuations around steady state.

\subsection{Nonequilibrium information dynamics}

\subsubsection{Nonequilibrium information landscape and flux, mutual information and entropy production}

The physical states of the systems and environments can be encoded into bits of information. Information flow is important for cellular signal transduction, development,
brain information processing \cite{Levenchenko, Bialek, Seifert2015PRE,Seifert_Bionetwork}. Speed and accuracy are crucial for the biological information transfer and processing. However, finding the keys for driving the fast efficient and accurate biological information transfer and processing is still challenging. Information dynamics is often stochastic, which can be characterized by probabilistic evolution \cite{Seifert2015PRE,Seifert_Bionetwork, Seifert2014JSM, Esposito2015PRE}.


%
The
information dynamics can be captured by the communications among different subsystems enabling information transfer \cite{Shannon}.  The stochastic information dynamics can be quantified by the probabilistic master equation including both the whole system Z and subsystems X and S, for example.  \cite{Seifert2015PRE,Seifert_Bionetwork, Seifert2014JSM, Esposito2015PRE, Zeng2016ArXiv}. It was found that the nonequilibrium information system can be globally quantified by the information landscape via the steady state distribution giving the weight of each information state while the information dynamics from state to state is determined by the both the information landscape and the information flux which measures the distance away from the equilibriumness \cite{Zeng2016ArXiv}.
The information theory \cite{Shannon} gives the mutual information measure for the capacity of the communications.
For a system $Z$, this is based on the \emph{Mutual Information Rate} (MIR) between its subsystems $X$ and $S$ in steady state. The MIR is defined on the probabilities of all possible time sequences, $P(Z^T)$, $P(X^T)$, and $P(S^T)$, ($P(Z^T)$: the probability of the whole system. $P(X^T)$ and $P(S^T)$: the marginal probability of the subsystems), and is given by
\begin{eqnarray}
  I(X,S)=\lim_{T\to \infty}\frac{1}{n}\sum_{Z^T}P(Z^T)\log\frac{P(Z^T)}{P(X^T)P(S^T)}.
\end{eqnarray}
It measures the efficient bits of information that $X$ and $S$ exchange with each other in unit time. A vanishing $I(X,S)$ indicates that $X$ and $S$ are independent of each other.
%
It was shown that the mutual information $I(X,S) $ can be decomposed to the time reversible equilibrium part $I_D(X,S)$ and time irreversible nonequilibrium part $I_B(X,S)$ directly related to the information flux so that the information communication capacity can be linked to the driving force of the information dynamics \cite{Zeng2016ArXiv}.
While the time reversible part of information exchange operates on both directions driven by the information landscape without extra energy input, the directional flow of information exchange does require energy input, and is related to the information flux $J_z$, one of the two driving forces of the information dynamics.
%
%
%
Furthermore, the information communication capacity is associated with the dissipation cost in maintaining it.
%
%
It turns out that the detailed balance breaking part of the mutual information can be decomposed as the difference between entropy production rate of the whole system and the individual subsystems \cite{Zeng2016ArXiv}.
\begin{equation}
 I_B(X,S)=\frac{1}{2}(EPR_z-EPR_x-EPR_s)
 \end{equation}
Here, $EPR_x$, $EPR_y$, and $EPR_z$ represent the entropy production rate of subsystem x, entropy production rate of subsystem y, and entropy production rate for the whole system z which includes the subsystems x and y.
The physical meaning of the above equation is that the residual dissipation cost characterized by the entropy production rates between the whole system and its subsystems gives a measure of the capacity of irreversible information exchange (mutual information) and quantifies the irreversible correlations between the subsystems. The eﬃcient irreversible informational transfer requires the energy input or dissipation cost. This can have direct impacts on the Jarzynski relationship and Crooks fluctuation theorem when including the information.

\subsubsection{Fluctuations in information thermodynamics}

For stochastic information systems, the thermodynamic fluctuations can be important beyond the average. The generalized fluctuation theorem involving information exchange characterized by mutual information I is given as~\cite{Sagawa}:
 \begin{align}
 \frac{P\left[\mathbf{C}(t)|\lambda(t)\right]}{P\left[\bar{\mathbf{C}}(\tau-t)|\bar{\lambda}(\tau-t)\right]}&= \exp[\beta (W-\Delta F)+I]
 \end{align}
 This states that the difference in statistical fluctuations forward and backward in time leads to the entropy production given by the difference between the work  and the free energy reduction from the mutual information. The generalized Jarzinsky relation~\cite{Sagawa} reads:
 \begin{align}
 \left\langle\exp\left[-\beta W-I\right]\right\rangle &= \exp[-\beta \Delta F].
 \end{align}
 These relations imply that the effect of information exchange acts as an extra work or an effective free energy reduction.

 Several studies were carried out
 on the information and nonequilibrium optimization on biological systems. \cite{Levenchenko, Bialek, Seifert_Bionetwork, Seifert2015PRE, Dill}. This includes maximizing information entropy or max caliber \cite{Dill}, maximizing mutual information \cite{Levenchenko, Bialek} for signal transduction and development, uncertainty, sensing and efficiency of information processing \cite{Seifert_Bionetwork, Seifert2015PRE, TuYuhai2012NaturePhys, Nemenman2007PlosOne, tenWolde2015PRL}. We expect more applications of the theoretical framework here to the biological information processing.
 
\subsection{ Gauge fields, time reversal symmetry breaking and underlying geometry for nonequilibrium systems}

Symmetry is at the heart of many physical laws. Continuous symmetries can be quantitatively described through their associated gauge fields~\cite{Peskin}. However, gauge theory can also be applied to nonequilibrium probabilistic dynamics \cite{Feng2011JCP,Polettini}. Indeed, the Fokker-Plank equation describing prob- abilistic evolution can be rewritten as
\begin{align}
\nabla_t P({\bf C},t) &= \nabla_i D_{ij} ({\bf C}) \nabla_j P ({\bf C}, t).
\end{align}
Here, the covariant derivatives with respect to the observable and time are defined as
\begin{align}
\nabla_i &= \partial_i + A_i \\ 
\nabla_t &= \partial_t + A_t , 
\end{align}
where $A_i=- \frac{1}{2} D^{-1}_{ij}\tilde{F}_j$ and $A_t=D_{ij} A_i A_j - \partial_i (D_{ij} A_j)$ form the components of an Abelian gauge field. 

The components $A_i$ introduce a curvature of the gauge field internal space
\begin{align}\label{curvature}
R_{ij}&=2 (\partial_i A_j -  \partial_j A_i)=\partial_i (D^{-1}_{jk} v_k) - \partial_j (D^{-1}_{ik} v_k),
\end{align}
where ${\bf v_{ss}} ({\bf C})={J_{ss}({\bf C})}/{P_{ss}({\bf C})}$ is the steady state flux velocity. In equilibrium, the steady-state flux is zero and the curvature vanishes, which corresponds to $R_{ij}=0$ and thus a flat internal space~\cite{Feng2011JCP}. Outside of thermodynamic equilibrium, the rotational flux typically does not vanish, therefore $R_{ij} \neq0$, yielding a curved internal space~\cite{Feng2011JCP}. Curvature of this internal space is a source of a global topological phase analogous to the quantum Berry phase~\cite{Wang2008PNAS,Wang2010JCP}. The heat dissipated along a closed loop is given by \cite{Feng2011JCP}
\begin{align}\label{sm_R}
T \Delta s_m^C = -  \oint_C A_i ({\bf x} ) d x_i 
&= -\frac{1}{2} \int_{\Sigma} d \sigma_{ij} R_{ij},
\end{align}
where $\Sigma$ is the surface spanned by the closed loop ${\it C}$ and $d
\sigma_{ij}$ an area element of this surface.

The dissipated heat $\Delta s_m$ in Eq.~(\ref{sm_R}) is equal to the entropy production at steady state. Through the
fluctuation theorem \cite{Crooks1, Crooks2}, time irreversibility emerges when entropy production is nonzero, caused by a flux that breaks the detailed balance \cite{Wang2010JCP,Feng2011JCP}. Thus, systems with nonzero curvature geometry break the detailed balance, showcase the emergence of the flux that explicitly breaks time-reversal symmetry, and generate dissipation.

\subsection{Multiple Landscapes, adiabaticity/non-adiabaticity and curl flux}

Up to now, we have been considering  nonequilibrium processes with one underlying landscape. Often, systems has multiple degrees of freedom
 coupled to each other. It is often useful to divide the systems into subsystems and study the intra-subsystem dynamics and inter-subsystem dynamics
in order to understand the details and obtain the global picture. In equilibrium systems, each subsystem can be described by a landscape and the whole
system can be described by the multiple coupled landscapes. For example, for electron transfer, both the intra-energy nuclear landscape and inter-energy
electronic landscape motion determine the dynamics \cite{MarcusET, WolynesET}. In equilibrium systems, the coupled landscape approach can be carried out
due to a priori known interaction potential landscape. For nonequilibrium systems, however, the coupled landscape approach requires the extensions of the
conventional equilibrium approach due to the absence of a priori known interaction potential landscape.

For nonequilibrium systems, both the adiabatic (strongly coupled) dynamics where inter-landscape motion is significantly faster than the intra-landscape motion
and non-adiabatic (weakly coupled) dynamics where inter-landscape motion is comparable or slower compared to the intra-landscape motion are important. The
multiple coupled landscapes are often technically challenging to study and visualize. This is due to continuous variable often used for intra-landscape
dynamics and discrete variable often used for inter-landscape dynamics as well as their couplings. One possible way of solving the problem is to convert
the discrete-continuous variables into the whole continuous representations. To realize this, one notices that although the discrete variables can only
take the discrete values, their associated occupations or probabilities are continuous variables. By introducing those additional variables, one can treat
the whole problem as the coupled continuous variable problem with extra degrees of freedom. Path integrals provide such a realization. Through mathematical
transformation, the coupled landscape dynamics can be treated as the dynamics under a single landscape with extended dimensions \cite{Zhang2013PNAS,
Chen2015PCCP}. In this way, one can have a unified global quantification of the nonequilibrium multi-landscape dynamics as determined by the landscape
gradient and curl flux on a single landscape with enlarged dimensions, making quantification and visualization easier for nonadiabatic dynamics. Details
are in \cite{Zhang2013PNAS, Chen2015PCCP}. Motors and gene regulations are possible good applications \cite{Zhang2013PNAS, Chen2015PCCP, Julicher1997RMP}.

\subsection{Organization principle of hierarchy and complexity of the dynamical systems at different scales}

Complex systems often involve different spatial and time scales. At different scales,
different phenomena and dynamics emerge. For example, when we heat up the water for coffee, in principle we can explore the microscopic molecular dynamics and see how the water changes to vapor. However, it is not practical to follow the dynamics of all the molecules involved. In addition, what we care about is whether or not the water is boiled. From an individual molecule trace, we will not be able to figure this out. In other words, watching the behavior of individual component does not give us information about the collective behavior automatically. The boiling behavior is a collective phenomena emerged from the interactions of many individual molecules at microscopic scales. Therefore large scale behavior emerged from microscopic scales can be very different from small scale behavior. A physical picture and unified quantitative theory is crucial for understanding this organization hierarchy and
emergent complexity of the systems \cite{Anderson, Hopfield, Wolynes1994PT, Wolynes1996PNAS, Laughlin2000PNAS, Haken2000Book, Prigogine1984Book}.
Four lines of thoughts are crucial: symmetry breaking, bifurcation,
phase transition, and the mechanism of emergent rare
events. By integrating these, a nonequilibrium landscape framework was suggested for mesocopic dynamics from the fast
dynamics at a small (microscopic) scale, the intra-basin motions
within each state  at an intermediate scale, and the
slow inter-basin switching with the kinetic rates
exponentially dependent on the system size at a larger scale \cite{Qian2016CPL}.
This nonequilibirum landscape framework represents the microscopic scale fast dynamics by a stochastic process
and the intermediate scale movements by a nonlinear dynamics.
Multiple attractors from larger scales representing the behavior and function emerge from the interactions of smaller scale systems.

\begin{figure}[t]
\includegraphics[width=0.5\textwidth]{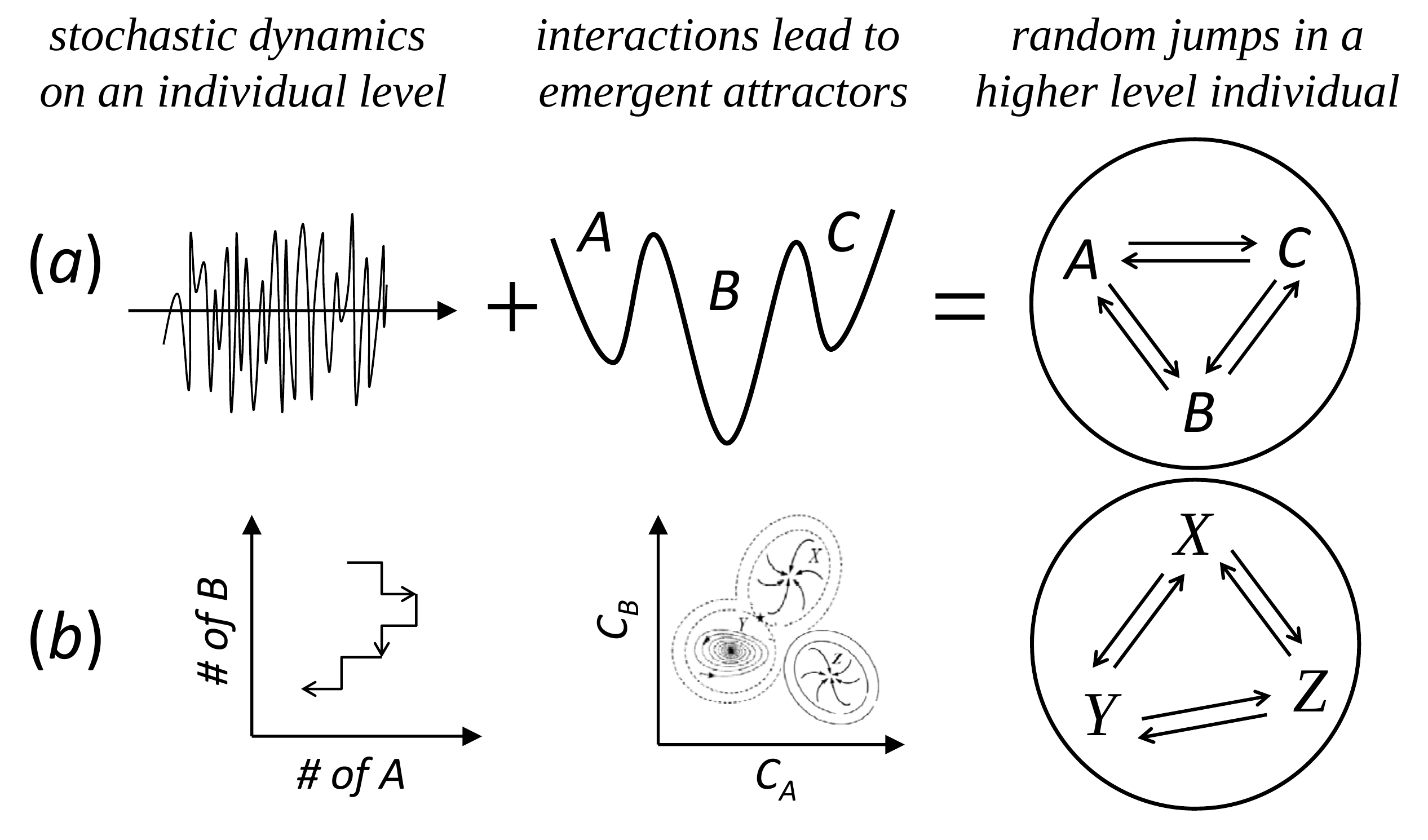}
\caption{Illustration of organizational hierarchy and complexity at different scales (a) A schematics showing rapid solvent-macromolecule collisions, as a source of stochasticity and together with a multi-energy-well landscape, gives rise to a kinetic jump process for an individual macromolecule with multiple states (shown within the circle). (b) A level higher, many interacting chemical individuals each with multiple discrete states form mesoscopic nonlinear reaction systems. (from Ref. \cite{Qian2016CPL}).} \label{Qian_CPL}
\end{figure}

Furthermore, each inter-basin transition/switching on the landscape can be described as a dynamic
symmetry breaking  exhibiting catastrophe
and phase transition breaking the ergodicity \cite{Qian2016CPL} shown in Fig.\ref{Qian_CPL}.  The dynamics of a nonlinear
mesoscopic system at the intermediate scale is stochastic.
Therefore, the basins of attractions and the switching among them on the nonequilibrium landscape
are both emergent phenomena. The resulting emergent inter-basin
stochastic dynamics provides the stochastic
element for the larger scale nonlinear dynamics (bigger spatial and longer time
scales). In fact, the mesoscopic landscape and flux emerges from the underlying micro dynamics of the system. At the mesoscopic scale, the emergent states in terms of the basin of attractions (e.g. water liquid and vapor gas phases) and their switchings become crucial in determining the behavior and function (e.g. water boiling) rather than the detailed microscopic individual dynamics (e.g. individual molecule motion).  The hierarchical structures of the protein dynamics at different scales in terms of its associated energy landscapes have been experimentally demonstrated \cite{Wolynes1994PT}. The nonequilibrium landscape framework discussed here
can help to capture the hierarchical structure and emergent complexity of the nonequilibrium biological system organization.


\subsection{Spatial nonequilibrium systems}
\label{sec:spatial}

So far we have discussed spatially homogeneous systems. However, spatially heterogeneous systems more accurately
represent biological phenomena such as organism development, motility, and cell structure. Another example lies in the
spatial organization of the neurons in the brain, where the relevant dynamic quantities depend on space~\cite{Bernard, Reinitz, Bray2000Book, Abbott2001Book,Marchetti2013RMP}. The dynamics and local stability of such systems have often been studied using deterministic or stochastic partial differential equations. The potential land- scape and rotational curl flux approach can be generalized to spatiotemporal nonequilibrium dynamics of such systems~\cite{Wu2013JPCB, Wu2013JCP, Wu2014JCP}.


\subsubsection{Landscape and flux decomposition}

To identify and quantify the driving force for spatial inhomogeneous nonequilibrium dynamical systems, we first define dynamical variables.
The dynamic variables are given by fields that depend on space and time. The set of dynamic variables is denoted by $\vec{\phi}(\vec{x},t)$ with components $\phi_a(\vec{x},t)$ representing the different state variables and $\vec{x}$ denoting a point in space. The Langevin equation for these fields then reads
\begin{align}
\partial_t{\phi_a(\vec{x},t)}&=F_a[\vec{x};\phi]+\xi_a[\vec{x},t;\vec{\phi}].
\end{align}
The deterministic driving force $\vec{F}$ now is a vector field and the noise term $\vec{\xi}$ is a stochastic force field. The latter can be expressed as
\begin{align}
\xi_a [\vec{x},t;\vec{\phi}]&=\sum_b\int d^3x'G_{ab}[\vec{x},\vec{x}';\vec{\phi}]\zeta_b(\vec{x}',t),\\
\intertext{where the random fields $\zeta_a$ obey}
\langle\zeta_a(\vec{x},t)\rangle&=0 \\
\langle\zeta_a(\vec{x},t)\zeta_b(\vec{x}',t')\rangle&=\delta_{ab}\delta^{(3)}(\vec{x}-\vec{x}')\delta(t-t').
\end{align}
In the previous equtaion, $\mathsf{G} [\vec{x},\vec{x}';\vec{\phi}]$ quantifies the dependence of the stochastic force field on space and on the dynamical variable. The stochastic fluctuation strength is characterized by the diffusion tensor $\mathsf{D}$ with
\begin{align}
D_{ab} (x,x'; [\phi] ) &= \frac{1}{2} \sum_{s} \int d^3y G_{as}[\vec{x},\vec{y};\vec{\phi}] G_{sb}[\vec{x}',\vec{y};\vec{\phi}].
\end{align}
Consequently, $\langle\xi [\vec{x},t;\vec{\phi}] \rangle = 0$ and $\langle\xi[\vec{x},t;\vec{\phi}] \xi [\vec{x'},t';\vec{\phi}]\rangle = 2 \mathsf{D} (x,x'; [\phi] ) \delta(t-t')$.

The state of the system is given by a probability functional $P[\vec{\phi}]$. It evolves in time according to a functional Fokker-Planck equation. It follows from the conservation equation for the probability,
\begin{align}
 \frac{\partial P[\vec{\phi}]}{\partial t}&=-\sum_a\int d^3x
\frac{\delta}{\delta\phi_a(\vec{x})} J_a[\vec{x};\vec{\phi}],
\end{align}
that is, by the functional divergence of a net probability flux. The flux field $J_a[\vec{x};\vec{\phi}]$ in turn can be split into two contributions \cite{Wu2013JCP,Wu2013JPCB,Wu2014JCP}
\begin{align}
J_a[\vec{x};\vec{\phi}]&=F_a[\vec{x};\vec{\phi}]P[\vec{\phi}]\nonumber\\
&-\sum_b\int d^3x'\frac{\delta}{\delta\phi_b(\vec{x}')}\left(D_{ab}[\vec{x},\vec{x}';\vec{\phi}]P[\vec{\phi}])\right).
\end{align}
In analogy with the spatially homogenous case, the driving force can be expressed in terms of the functional gradient of a nonequilibrium potential field landscape and a rotational curl flux field \cite{Wu2013JCP,Wu2013JPCB,Wu2014JCP}
\begin{align}
\tilde{F}_a[\vec{x};\vec{\phi}]&=-\sum_b\int d^3 x'
D_{ab}[\vec{x},\vec{x}';\vec{\phi}]\frac{\delta
U[\vec{\phi}]}{\delta \phi_b(\vec{x}')}
+\frac{J_a^{ss}[\vec{x};\vec{\phi}]}{P^{ss}[\vec{\phi}]}.
\end{align}
Here, $\tilde{F}_a[\vec{x};\vec{\phi}]=F_a[\vec{x};\vec{\phi}]-\sum_b\int d^3x'\frac{\delta}{\delta\phi_b(\vec{x}')}D_{ab}[\vec{x},\vec{x}';\vec{\phi}]$. The nonequilibrium potential landscape $U$ is linked to the steady state probability, $U[\vec{\phi}]=-\ln P^{ss}[\vec{\phi}]$, whereas the steady state probability flux field satisfies divergent free condition
\begin{align}
\sum_a\int d^3x \frac{\delta}{\delta\phi_a(\vec{x})}J_a^{ss}[\vec{x};\vec{\phi}]&=0.
\end{align}

To quantify the global stability of the spatial dependent nonequilibrium dynamical system, one can explore the Lyapunov functional, which is given by the intrinsic potential field landscape in absence of
fluctuations and by the free energy landscape functional for finite fluctuations \cite{Wu2013JCP,Wu2013JPCB,Wu2014JCP}. The monotonic decreasing feature of free energy functional is a reflection of the second law of thermodynamics of spatially dependent stochastic systems. The nonequilibrium thermodynamics can therefore be generalized from the well-mixed to the spatially dependent case
\cite{Wu2013JCP,Wu2013JPCB,Wu2014JCP}.


\subsubsection{Generalized hydrodynamics}
\label{sec:generalizedHydrodynamics}
Similarly to spatially homogenous systems, one can express the entropy production rate Eq.~(\ref{eq:Sdot}) also for spatially heterogeneous systems. Based on the entropy production rate, one can systematically derive a phenomenological description of the system's dynamics by assuming that the system is close to thermodynamic equilibrium~\cite{deGroot:1985ue}. Here, this condition states that locally the system is always in thermodynamic equilibrium. ``Locally'' means that one can virtually divide space into volume elements, such that the system is in thermodynamic equilibrium in each of them. The equilibria might differ between different elements, such that the exchange of energy and matter between adjacent volume elements leads to quasistatic changes of their states. This restricts this approach to so-called hydrodynamic modes, which are characterized by a relaxation time $\tau$ that increases with decreasing wave number $q$ as $\tau\propto q^{-2}$. They arise, notably, from conservation laws or broken continuous symmetries.

Due to the assumption of local thermodynamic equilibrium, one can define a free energy $\mathcal{F}$ for the system by summing the free energy of all volume elements. For constant temperature $T$, one can then express the entropy production rate as
\begin{align}
\dot{\mathcal{S}}_t &= -\frac{1}{T}\dot{\mathcal{F}} = \int d^3x\;\sum_i J_iX_i,
\end{align}
where the sum extends over all conjugated pairs of fluxes and forces. Expressions for the fluxes can be obtained by expanding them up to first order in the forces similar to Eq.~(\ref{eq:onsager}), where the now phenomenological coupling coefficients respect the Curie principle and the Onsager relations. For simple fluids, this approach leads to the Navier-Stokes equation, where the viscosities are the phenomenological coupling coefficients. It is therefore often called generalized hydrodynamics. It has been applied successfully to cellular processes, see Sect.~\ref{sec:cellularDynamics}. For strongly nonequilibrium systems, the local equilibrium assumption breaks down. A more general approach such as landscape and flux field is required to accurately describe the underlying nonequilibrium processes. 

\subsubsection{A strong nonequilibrium spatial dynamical system: turbulence}

As a specific example of a spatially heterogeneous system, consider the fluid dynamics of turbulent systems. Typically, the magnitude of the Reynolds number can inform if the inertial force dominates their viscous counterpart. For biological fluids, the Reynolds number is often low; however, they can still sometimes exhibit turbulent behavior, e.g., in bacterial suspensions~\cite{Dombrowski:2004eu,Wensink:2012hk}. The nonequilibrium behavior of turbulence~\cite{Goldenfeld} can be characterized through energy cascade \cite{TurbulenceIrreversibility1,TurbulenceIrreversibility2}. The notion of a cascade intuitively captures the energy flow from large to intermediate, then finally to small length scales where energy is dissipated~\cite{energycascade}, quantified by Kolmogorov's scaling laws~\cite{Kolmogorov1941a,Kolmogorov4/5,Landau,Kolmogorov1962}. A quantification of turbulence with explicit
detailed balance breaking description \cite{Wu2016ArXiv} can help to gain more insights into the nonequilibrium nature.

The potential landscape and flux field was quantified through the stochastically forced Navier-Stokes equations that governs fully developed turbulence \cite{Wu2016ArXiv}
\begin{align}\label{SNSESF}
\partial_t \mathbf{u}&=\mathbf{\Pi}^{s}(\nabla)\cdot\left(-\mathbf{u}\cdot\nabla\mathbf{u}\right)+\nu \Delta\mathbf{u}+\mathbf{f}^s.
\end{align}
Here, $\mathbf{u}$ denotes the flow velocity field and $\mathbf{\Pi}^{s}(\nabla)\cdot\left(-\mathbf{u}\cdot\nabla\mathbf{u}\right)=-\mathbf{u}\cdot\nabla\mathbf{u}-\nabla p$ represents hydrostatic pressure and convection. They constitute the deterministic driving force together with the viscous force $\nu \Delta\mathbf{u}$. Finally, $\mathbf{f}^s$ denotes the stochastic stirring force.

The probability functional evolves according to
\begin{align}
\partial_t P[\mathbf{u}]&=-\int d^3x\,\frac{\delta}{\mathbf{u}(\vec{x})}\cdot\mathbf{J}[\vec{x};\mathbf{u}].
\end{align}
At steady state, the flux field $\mathbf{J}^{ss}$ satisfies divergent free condition and is therefore a rotational curl flux. It is determined by both, a reversible pressure-convective force and an irreversible viscous force with diffusion in state space, as well as the stochastic stirring force \cite{Wu2016ArXiv}
\begin{align}
\mathbf{J}_{ss}(\mathbf{x})[\mathbf{u}]&=
\mathbf{J}^{ss}_\text{rev}[\vec{x};\mathbf{u}]+ \mathbf{J}^{ss}_\text{irr}[\vec{x});\mathbf{u}].
\end{align}
The viscous force decomposition then becomes  \cite{Wu2013JCP,Wu2014JCP, Wu2016ArXiv}
\begin{equation}\label{FDENEQNSE}
\nu \Delta\mathbf{u} =-\int
d\mathbf{x}'\,\mathsf{D}^{ss}(\mathbf{x}-\mathbf{x}')\cdot\frac{\delta}{\delta\mathbf{u}(\mathbf{x}')}\Phi[\mathbf{u}]+\frac{\mathbf{J}^{ss}_\text{irr}[\vec{x});\mathbf{u}]}{P^{ss}[\vec{x};\mathbf{u}]},
\end{equation}
where $\Phi[\mathbf{u}]=-\ln P^{ss}[\mathbf{u}]$ is the nonequilibrium potential landscape related to the steady-state probability
\cite{Wu2016ArXiv}. This landscape and flux field perspective of the nonequilibrium dynamics applies to both, stochastic fluid systems with low Reynolds
numbers as for biological fluids or high Reynolds numbers as in the case of turbulence.

It turns out that the energy transfer $\mathcal{T}$ associated with the energy cascade is tightly related to the breaking of detailed balance or irreversible flux
\begin{align}
\mathcal{T}(\mathbf{k})&=-\mathcal{R}\left\{\int \mathbf{u}^*(\mathbf{k})\cdot\mathbf{J}^{ss}_\text{irr}(\mathbf{k})[\mathbf{u}]\delta \mathbf{u}\right\}.
\end{align}
where $\mathcal{R}$ denotes the real part of the function. This relation leads to the $4/5$th scaling law for the third order structure function in turbulence \cite{Frisch1995Book, Wu2016ArXiv}. It also leads to Komogorov's $5/3$th scaling law for the second order structure function in turbulence under the hypothesis of self similarity  \cite{Wu2016ArXiv}. As seen, the driving force for the stochastic fluid (either physical/biological fluids or turbulence) systems is from the underlying potential landscape field
gradient and the curl probability flux field. The non-zero irreversible probability flux field indicates the detailed balance breaking which drives the flow for
the energy cascade.

\subsection{Nonequilibrium quantum landscape and flux}
\label{sec:quantum}

\subsubsection{Nonequilibrium Quantum Dynamics}

The nonequilibrium quantum phenomena is important in many branches of science. The quantum transport provides a good example.
\cite{Kim05,Laughlin83,Zhang06,Hinds05,Freed72,Kuznetsov99,Fleming07}. The electron and energy transports in biology and nanomaterials, within and between molecules
such as photosynthesis have been explored intensively in experiments \cite{Tour97,McEuen00,Umansky05,Mate88,Ho04,Ho02} and studied theoretically
\cite{Esposito06,Esposito09, Tanimura}. The unidirectional flow in nonequilibrium ultrafast electron transfer \cite{Zhong12} was seen in recent
experiments of photoreduction dynamics of oxidized photolyase \cite{Zhong13}. Furthermore, quantum effects in transports provide a test ground for the
nonequilibrium thermodynamics. This was made possible in single molecule junctions\cite{Natelson08}.   The coherence, representing pure quantum nature,
contributes to the transport \cite{Wolynes92} in addition to the populations. This was seen for quantum coherent excitation (charge) transport in light
harvesting complexes and photosynthetic reaction center \cite{Fleming07,Xu1992}. The nonequilibrium dynamics is also important for the quantum computations
and devices \cite{Nielson}.

However, the understanding of nonequilibrium quantum dynamics is often challenging due to lack of quantification of nonequilibriumness and the
relationship with quantum coherence.
There are several theoretical approaches for quantum transport, such as momentum balance and fluctuations in the mesoscopic systems \cite{Schoenmaker02},
the fluctuation-dissipation theorem \cite{Kubo57,Prelovsek05} and nonequilibrium Green's function method \cite{Saintjam71,Combesco71,Chang08}. However,
these formalisms can only be applied successfully in the near-equilibrium state. The quantum master equation (QME) provides a possible alternative for
studying the irreversible dynamics of quantum systems coupled to environments \cite{Breuer02,Spohn80,Haake73}, beyond near equilibrium regime. This
approach has been applied to decoherence dynamics in quantum optics \cite{Scully1997,Carmichael10}, chemical reactions  \cite{Mukamel99} and also condensed
matter systems \cite{Weiss12}.

In this section, we will illustrate an approach to nonequilibrium quantum dynamics in terms of population landscape, curl flux and coherence
\cite{Zhang2014JCP}. One can do so by exploring the energy \cite{Leitner10,Wolynes04} and charge transport \cite{Marcus08,Tao08,Tao09,Ohmine99} in
biomolecules. The energy transport is often coupled to two heat environments (bosonic baths ) with different temperatures (light harvesting complex under
light photonic and protein phononic baths) while the charge transport is often coupled to the two chemical environments (fermionic baths) with
different chemical potentials (electron transfer between two metals).

Starting from the original Hamiltonian coupled with two environments, one can derive the corresponding quantum master equation. From  there, one can
uncover and quantify the population landscape and curl flux for characterizing nonequilibrium quantum systems. One can see that the curl flux provides a measure of
detailed balance breaking and time-irreversibility, important in quantum transport. It turns out that nonequilibriumness from system environmental coupling
can significantly enhance the steady state coherence
 \cite{Zhang2014JCP, Sun2015AP}, in contrary to conventional wisdom that coherence always decays due to the system-environmental coupling
 \cite{OpenQuantumSystemBook, Manzano12}. One can explore the relationship among nonequilibriumness, curl flux, coherence, quantum transport and the
 energy/charge transfer efficiency.  Details are in later sections.


\subsubsection{Theory of nonequilibrium quantum dynamics in terms of flux, coherence and population landscape}

Here, one shows how a landscape and flux theory is developed \cite{Wang2008PNAS, Wang2011PNAS}
for the nonequilibrium quantum system\cite{Zhang2014JCP}. The general Hamiltonian of a quantum system interacting with $M$ environments can be written in
the form
\begin{equation}
\begin{split}
&
H_S=\sum_{n,m}H_{nm}|\psi_n\rangle\langle\psi_m|+\sum_{i=1}^M\sum_{\textbf{k},\sigma}\hbar\omega_{\textbf{k}\sigma}a_{\textbf{k}\sigma}^{(i)\dagger}a_{\textbf{k}\sigma}^{(i)}\\[0.2cm]
& H_{int}=\sum_{i,\langle
n,m\rangle}\sum_{\textbf{k},\sigma}g_{\textbf{k}\sigma}^{nm(i)}\left(|\psi_n\rangle\langle\psi_m|a_{\textbf{k}\sigma}^{(i)\dagger}+|\psi_m\rangle\langle\psi_n|a_{\textbf{k}\sigma}^{(i)}\right)
\end{split}
\label{01}
\end{equation}
where $\langle n,m\rangle$ denotes that only the pairs of states $n,m$ with energies $E_n<E_m$ are considered. The first term of $H_0$ is the Hamiltonian
of the system and the second term of $H_0$ is the Hamiltonian of the environments, while the term $H_{int}$  denotes the couplings between the system and
the environments.  Environments are often much larger in size than the system. One assumes no back reactions from system to environments. One can then
study system dynamics by averaging over environments , which leads to master equation for reduced density matrix\cite{Scully1997, Breuer02, Zhang2014JCP}:
\begin{equation}
\begin{split}
\frac{\partial\rho_S}{\partial t}=\frac{-1}{\hbar^2}\textup{Tr}_R\int_0^t ds &
\left[\tilde{H_{int}}(s),\left[\tilde{H}_{int}(t),\rho_S(t)\otimes\rho_R(0)\right]\right]\\
& +{\cal O}(g^2)
\end{split}
\label{02}
\end{equation}
The density matrix can be expanded in terms of coupling strength between system and environments: $\rho(t)=\rho_S(t)\otimes\rho_R(0)+\rho_c(t)$. Under weak
coupling, quantum master equation can be truncated to the second order. This gives the Redfield equation without secular approximation rather than the
Lindblad equation with, which can be written in Liouville space: $\dot{\rho_S} = \frac{i}{\bar{h}} [\rho_S, H_S] - \frac{1}{2
 \bar{h}^2} D(\rho_S) $ where $H_S$ is the system Hamiltonian and $D(\rho_S)$ is the dissipation operator from system-bath coupling. The density matrix then forms super vector
$|\dot{\rho}_S\rangle=\hat{\mathcal{M}}|\rho_S\rangle$. By separating the population (diagonal elements) $\rho_p$ and coherence (off-diagonal elements)
$\rho_c$ parts of the density matrix, one can write the matrix $\mathcal{M}$ in block form:
\begin{equation}
\begin{split}
 \begin{pmatrix}
   \dot{\rho}_{\textup{p}}\\[0.15cm]
   \dot{\rho}_{\textup{c}}\\
 \end{pmatrix}=\begin{pmatrix}
                \mathcal{M}_{\textup{p}} & \mathcal{M}_{\textup{pc}}\\[0.15cm]
                \mathcal{M}_{\textup{cp}} & \mathcal{M}_{\textup{c}}\\
               \end{pmatrix}
               \begin{pmatrix}
                \rho_{\textup{p}}\\[0.15cm]
                \rho_{\textup{c}}\\
              \end{pmatrix}
\end{split}
\label{03}
\end{equation}
Here, $\mathcal{M}_{\textup{p}}$ denotes transition matrix in population space. $\mathcal{M}_{\textup{c}}$ denotes transition matrix in coherence space
(non-off diagonal elements of density matrix)\cite{Zhang2014JCP}. $\mathcal{M}_{\textup{pc}}$ and  $\mathcal{M}_{\textup{cp}}$ denote coupling transition
matrix between population and coherence space.

At steady state, one can substitute coherence $\rho_c$ as a function of population $\rho_p$ to master equation \cite{Zhang2014JCP}.
 This leads to reduced quantum master equation at steady state for population $\rho_p$:
\begin{equation}
\begin{split}
\left(\mathcal{M}_{\textup{p}}-\mathcal{M}_{\textup{pc}}
\mathcal{M}_{\textup{c}}^{-1}\mathcal{M}_{\textup{cp}}\right)\rho_{\textup{p}}^{ss}=0
\end{split}
\label{08}
\end{equation}

Now one is able to define the transfer matrix as $T_{mn}=\mathcal{A}_{nn,mm}^{\textup{p}}\rho_{mm}^{\textup{p}}$ for $m\neq n$ where
$\mathcal{A}_{\textup{p}}\equiv\mathcal{M}_{\textup{p}}-\mathcal{M}_{\textup{pc}}\mathcal{M}_{\textup{c}}^{-1}\mathcal{M}_{\textup{cp}}$ . For $m=n$,
$T_{mn}=0$. Transfer matrix determines the temporal evolution of the density matrix. Therefore transfer matrix is the driving force for stochastic
probability evolution. One can further decompose the transfer matrix into symmetric and anti-symmetric part. $T_{mn} = \frac{T_{mn} + T_{nm}}{2} +
\frac{T_{mn} - T_{nm}}{2}$. One can see that symmetric part of the transfer matrix $\mathcal{A}_{nn,mm}^{\textup{p},S}=\frac{T_{mn} +
T_{nm}}{2}/\rho_{mm}^{\textup{p}}$ satisfies the detailed balance condition $(\frac{T_{mn} + T_{nm}}{2}/\rho_{mm}^{\textup{p}} ) \rho_{mm}^{\textup{p}} -
(\frac{T_{nm} + T_{mn}}{2}/\rho_{nn}^{\textup{p}}) \rho_{nn}^{\textup{p}} = 0 $. One can also check that the anti-symmetric part of the transfer matrix
$\mathcal{A}_{nn,mm}^{\textup{p},A}=\frac{T_{mn} - T_{nm}}{2}/\rho_{mm}^{\textup{p}}$ does not preserve the detailed balance $(\frac{T_{mn} -
T_{nm}}{2}/\rho_{mm}^{\textup{p}}) \rho_{mm}^{\textup{p}} - (\frac{T_{nm} - T_{mn}}{2}/\rho_{nn}^{\textup{p}} ) \rho_{nn}^{\textup{p}} \neq 0 $. This leads
to a non-zero steady state flux. One can see the quantum dynamics for the population is determined by the two driving forces. The one is detailed balance
preserving $\mathcal{A}_{nn,mm}^{\textup{p},S}=\frac{T_{mn} + T_{nm}}{2}/\rho_{mm}^{\textup{p}}$ force and determined by the steady state population
landscape, giving equilibrium part of the contribution to dynamics. The other from anti-symmetric driving force
$\mathcal{A}_{nn,mm}^{\textup{p},A}=\frac{T_{mn} - T_{nm}}{2}/\rho_{mm}^{\textup{p}}$ is the flux component of the driving force breaking detailed balance.
\cite{Zhang2014JCP}. The quantum flux can be further decomposed into sum of fluxes of various loops in state space\cite{Zhang2014JCP,Zhang2015NJP1}. The quantum flux is
thus a rotational curl. One can show how nonequilibriumness will influence the quantum coherence, transport, thermodynamics, fluctuation-dissipation relations and even underlying
geometry/topology in later sections \cite{Zhang2014JCP, Zhang2015NJP1, Zhang2015NJP2, Zhang2016SciRep, Zhang2016EPL}.



%

We will now describe how the concepts developed in Sect.~\ref{sec:physicalConcepts} have been applied to specific biological systems out of thermodynamic equilibrium on various length and time scales. 


\section{
Biomolecular systems and experimental quantification of flux}

As the fundamental building blocks of living organisms, biomolecules interact with each other to form complex molecular structures and dynamics.
Accompanied by energy consumption and exchanging matter, many biomolecular systems are far from equilibrium. This holds notably for elementary biochemical
reactions, e.g., non-Michaelis-Menten kinetics~\cite{min2006does,english2006ever}, molecular dynamics in space and time, e.g., Min-protein oscillations for
cell-division site selection~\cite{raskin1999rapid}, the organization of cytoskeletal 
structures, e.g., assembly 
of actin filaments 
and microtubule
~\cite{Kuhn2005,Fujiwara:2007ge, mitchison1984dynamic}, and complicated molecular machines, e.g., the bacterial flagellar rotation
motor~\cite{silverman1974flagellar}. We will use rhodamine oxidation
, cyanobacterial circadian rhythm, and energy transport in the light harvesting complex as three examples to illustrate nonequilibrium behavior occurring in molecular systems.



\subsection{Non-Michaelis-Menten enzyme kinetics }


In living cells, almost all biochemical reactions are catalyzed by enzymes which accelerate the conversion from substrates to products. Following Michaelis and Menten, 
the rate of the kinetics is obtained under the assumption that the substrate and the substrate-enzyme complex are in equilibrium~\cite{MMEnzyme, english2006ever, Xie2013Sci}. However, when there is energy input to or output from the system, the enzyme kinetics 
can deviate from the Michaelis-Menten 
behavior~\cite{Xie2006JPCB, Qian2002BPC, Cao2011JPCB} due to a rotational curl flux breaking the detailed balance \cite{Liu2016ArXiv}. 
The landscape and flux are determined theoretically as the driving forces for the nonequilibrium dynamics. Experimentally, the nonequilibrium landscape can be acquired by measuring the steady state distribution of the observables \cite{Fang2016ArXiv, Jiang2016ArXiv}.  Experimental quantification of the flux experimentally is more challenging but may be realized by measuring the deviation from the Michaelis-Menten kinetics ~\cite{Liu2016ArXiv}.

\begin{figure*}[t]
\includegraphics[width=1.0\textwidth]{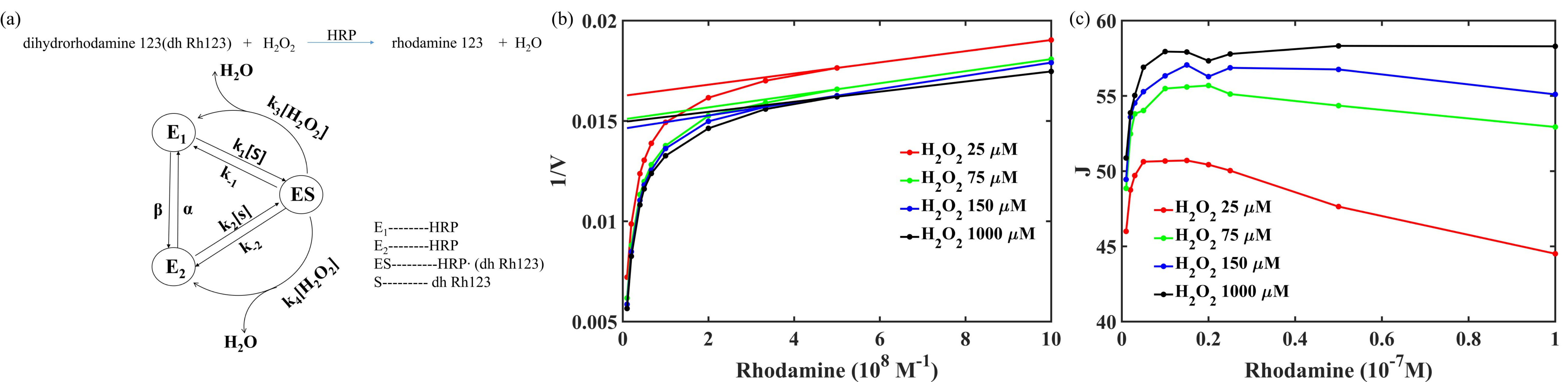}
\caption{ Scheme of enzyme reaction, non Michales-Menten kinetics and quantification of flux. (a): The simplest kinetic scheme with two unbound enzyme states. $E_1$ and $E_2$ denote the conformational states of enzymes, ES denotes
the intermediate state of the enzyme reaction and P denotes product. $\alpha$ and $\beta$ denote the conformation conversion rate. (b):Non-Michaelis
Menten (curved lines) versus Michaelis Menten enzyme kinetics (straight lines) with respect to rhodamine 123 concentrations at different substrate concentrations. (c):  Flux values with respect to rhodamine 123 concentrations at different substrate concentrations. (from Ref. \cite{Liu2016ArXiv}).}
\label{Liu2016ArXiv_fig1}
\end{figure*}

The deviation from Michaelis-Menten kinetics  
can be illustrated experimentally by
the catalysis of 
dihydrorhodamine 123
oxidation into fluorescent rhodamine 123
by the enzyme horseradish peroxidase 
in presence of hydrogen peroxide ( $H_2 O_2$). It is possible to study this reaction experimentally at the single molecule level, because the substrate as well as the enzyme do not fluoresce, whereas the product does~\cite{Rigler1, Rigler2, Rigler3}. The horseradish peroxidase has two different conformations, in both of which it can bind the substrate. The Master equation corresponding to the kinetic scheme, Fig.~\ref{Liu2016ArXiv_fig1}(a), is given by~\cite{Liu2016ArXiv} 
\begin{align}\label{A_Co}
\dot P&=
\left(  \begin{array}{ccc}
 -sk_1-\beta & \alpha & k_1 + k_3 \\
  \beta & -s k_2-\alpha  & k_{-2} +k_4 \\
  k_1 s & k_2 s  & -k_{-1} - k_{-2} - k_3 - k_4
\end{array}\right) P
\end{align}
Here $P$ is the vector $\left(P_1, P_2, P_{ES}\right)$ of the probabilities $P_1$, $P_2$, and $P_{ES}$ for the enzyme to be in conformational state 1 or 2
or binding the substrate, respectively.

The steady state probability flux is given by $J= \beta P^{ss} _1 -\alpha P^{ss} _2$, where $P^{ss}_i$ is the steady-state probability for being in state $i$. The flux is zero and detailed balance
holds if
\begin{align}
\frac{\alpha}{\beta}&=\frac{k_{-1}k_2 + k_2 k_3}{k_{-2}k_1 + k_1 k_4}.
\end{align}
In this case Michaelis-Menten kinetics emerges:
\begin{align}
\frac{1}{v} &= C_0 + \frac{C_1}{[S]},
\end{align}
where $v$ is the reaction rate and $[S]$ the substrate concentration. $C_0$ and $C_1$ are constants that depend on the molecular rates. The inverse of the Michaelis-Menten rate is a linear function of the inverse of the substrate concentration.
If detailed balance is
broken, the flux is nonzero and the enzyme reaction rate is
\begin{align}
\frac{1}{v} &= C_0 + \frac{C_1}{[S]} +\frac{C_2}{[S]+\lambda}
\end{align}
with an additional dependence on the substrate concentration containing the constants $\lambda$ and $C_2$. In this case, the inverse reaction rate is no longer a linear function of the inverse substrate concentration as is observed experimentally for rhodamine oxidation by the horseradish peroxidase and thus deviates from Mechalias-Menten kinetics
, Fig.~\ref{Liu2016ArXiv_fig1}(b). Note that 
the extra term $\frac{C_2}{[S]+\lambda}$ results 
from the presence of a 
flux loop, but not directly from having more than one conformational state of the enzyme. The deviation from Michaelis-Menten kinetics is thus a consequence of breaking detailed balance, which in this case originates from heat absorption by the reaction~\cite{Liu2016ArXiv}. 
In 
more complex systems each additional flux loop $i$ contributes an additional term $\frac{C_i}{[S]+\lambda_i}$. 

Exploiting the correlation function of the
experimental fluorescence signals
, one can obtain 
the kinetic rate parameters and finally quantify the enzymatic rate and the probability flux 
as a function of the substrate
concentration~\cite{Liu2016ArXiv} shown in
Fig.~\ref{Liu2016ArXiv_fig1} (b,c). Clearly, the enzyme rate (curved line for inverse of enzyme rate versus inverse substrate) deviates significantly from the conventional Michaelis-Menten rate (straight line for inverse of enzyme rate versus inverse substrate). The nonzero fluxes are quantified for different substrate concentrations.  \cite{Liu2016ArXiv} As discussed, non-MM rate can be used to quantify the degree of detailed balance breaking through the corresponding rotational curl 
flux.

Therefore, the flux breaking the detailed balance which is now quantified experimentally as a major driving force for nonequilibrium dynamics  \cite{Liu2016ArXiv} in addition to the landscape \cite{Fang2016ArXiv, Jiang2016ArXiv}  can lead to non-Michaelis-Menten enzyme kinetics. It is important
to note that the breaking down of the detailed balance is originated from the energy imbalance of the enzyme reaction due to energy flow (heat absorption) into the reaction.
\cite{Liu2016ArXiv}


\subsection{Bacterial circadian rhythm}
\label{sec:circadianRhythm}

\begin{figure*}[t]
\includegraphics[width=0.8\textwidth]{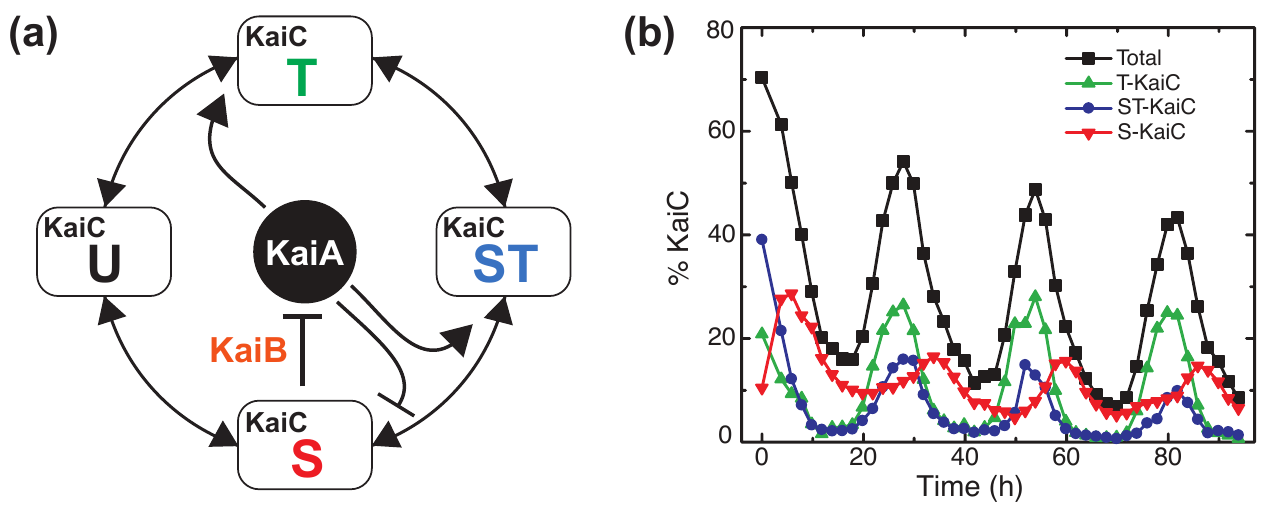}
\caption{The KaiABC circadian system of the cyanobacterium~\emph{Synechococcus elongatus}. (a) Reaction network. Two-headed arrows indicate transitions between phosphorylation states of KaiC, arrows emanating from KaiA indicate promotion and inhibition of transitions, the arrow from S-KaiC to KaiA, suppression of KaiA activity by KaiB in presence of S-KaiC. (b) Concentrations of phosphorylated KaiC as a function of time. From~\cite{rust2007ordered}. }
\label{fig-exp-KaiABC}
\end{figure*}

Circadian rhythms are biological processes that display sustainable oscillation of about 24 hours and allow organisms to anticipate environmental changes that occur regularly during a day. Typically, they are generated through negative feedback regulation of so-called clock genes at the level of transcription or translation ~\cite{Takahashi, Goldbeter1, tenWolde1, tenWolde2, tenWolde3, tenWolde4, Novak}. In contrast
, the circadian rhythm of the cyanobacterium \textit{Synechococcus elongatus}~\cite{ishiura1998expression,wang2009robustness} results from mere interaction of the 
proteins KaiA, KaiB, and KaiC. Remarkably, after addition of 
ATP they generate circadian oscillations \textit{in vitro} with only a weak dependence on temperature~\cite{tomita2005no,rust2007ordered,nakajima2005reconstitution}. 
Of the three proteins, KaiC is a hexameric enzyme that can be 
phosphorylated 
at two of its amino acids: serine 431 (S431) and threonine 432 (T432). The enzyme 
can be in four different states: fully unphosphorylated (U-KaiC), partially 
phosphorylated either at S431 (S-KaiC) or 
at T432 (T-KaiC), as well as fully phosphorylated at both, S431 and T432 (ST-KaiC). Furthermore, KaiA 
promotes KaiC phosphorylation, whereas KaiB antagonizes 
the activity of KaiA, Fig.~\ref{fig-exp-KaiABC}(a).

The phosphorylation state of KaiC changes in a cyclic way: KaiA promotes the transition from U-KaiC to T-KaiC and then into ST-KaiC. Afterwards it transforms into S-KaiC and finally into U-KaiC. The state ST-KaiC is effectively long-lived, because KaiA promotes changes from S-KaiC back to ST-KaiC. Only when S-KaiC has reached a threshold value, the rate transitions from ST-KaiC to S-KaiC increases rapidly, because S-KaiC inhibits KaiA through KaiB. After S-KaiC has turned into U-KaiC, KaiA is reactivated and a new cycle begins, Fig.~\ref{fig-exp-KaiABC}(b). Note, however, that the specific activation of KaiB by S-KaiC and the role of KaiA in rephosphorylating S-KaiC and thus generating ST-KaiC remain to be confirmed.

The nonequilibrium landscape of the KaiABC system has the form of a Mexican hat, whereas the nonequilibrium flux drives the system along the hat's valley. Together this explains the stability of the oscillations. The entropy flow associated with the nonequilibrium flux and force, which is ultimately caused by ATP hydrolysis involved in the phosphorylation kinetics,
provides the thermodynamic cost for maintaining the robust and coherent circadian oscillation.

\subsection{Nonequilibrium quantum transports in biomolecules}

Representative examples of quantum mechanical biological processes include photosynthetic energy absorption, olfaction, bird magnetoreception,  and electron/proton transports in enzyms
~\cite{brookes2017quantum}. As these processes involve the conversion of energy into usable forms for chemical transformations, they are out of equilibrium by nature.

To illustrate the nonequilibrium quantum dynamics of these processes,  we explore the landscape and flux in a two site and two level model system \cite{Zhang2014JCP} coupled with two temperature or two chemical potential environments where analytical solutions can be obtained.
The two site and two level systems have been widely investigated in condensed matter physics, chemistry, quantum
optics and information \cite{Leggett,Wolynes1988,WeissBook,BenJacob_1985,QuantumOpticsBook,Nielson,Preskill,Landauer1,Deutsch,
Bennett,Unruh,Palma} for exploring quantum dissipations and coherence.  The system dynamics coupled with a single environment is often assumed. The steady
state often becomes an equilibrium state with detailed balance and significantly reduced or zero coherence. There are examples of systems coupled to multiple environments, such
as  energy and charge transfer in photosynthesis \cite{PhotoSynthesisBook, Fleming07},  nano-quantum transport \cite{NanoTransportBook}. For these, the
final steady state is often not an equilibrium state and quantum coherence is not necessarily zero at steady state \cite{Zhang2014JCP, Sun2015AP,
Zhang2015NJP1, Zhang2015NJP2, Zhang2016SciRep}. How would the nonequilibriumness influence the quantum coherence and transport efficiency? These issues are major challenges for photosynthesis \cite{PhotoSynthesisBook, Fleming07}.

\subsubsection{An analytical model for nonequilibrium quantum energy/charge transfers in biomolecules}

To address these issues, let us consider a two site system coupled with two environments. Each environment is in equilibrium with a different temperature or chemical potential environment, either obeying Bose or Fermi statistics. The two site system connected by the tunneling can be used to describe the quantum transport while the temperature or chemical potential difference measures the degree of the nonequilibriumness as they set up a nonequilibrium thermal or chemical battery or pump for the system. We can quantify the asymmetry of the transition matrix and the flux as well as the coherence and quantum transport efficiency with respect to the nonequilibriumness measured by the temperature or chemical potential difference. The origin of the nonzero flux can then be identified as the temperature or chemical potential difference.

\begin{figure*}[t]
\includegraphics[width=0.8\textwidth]{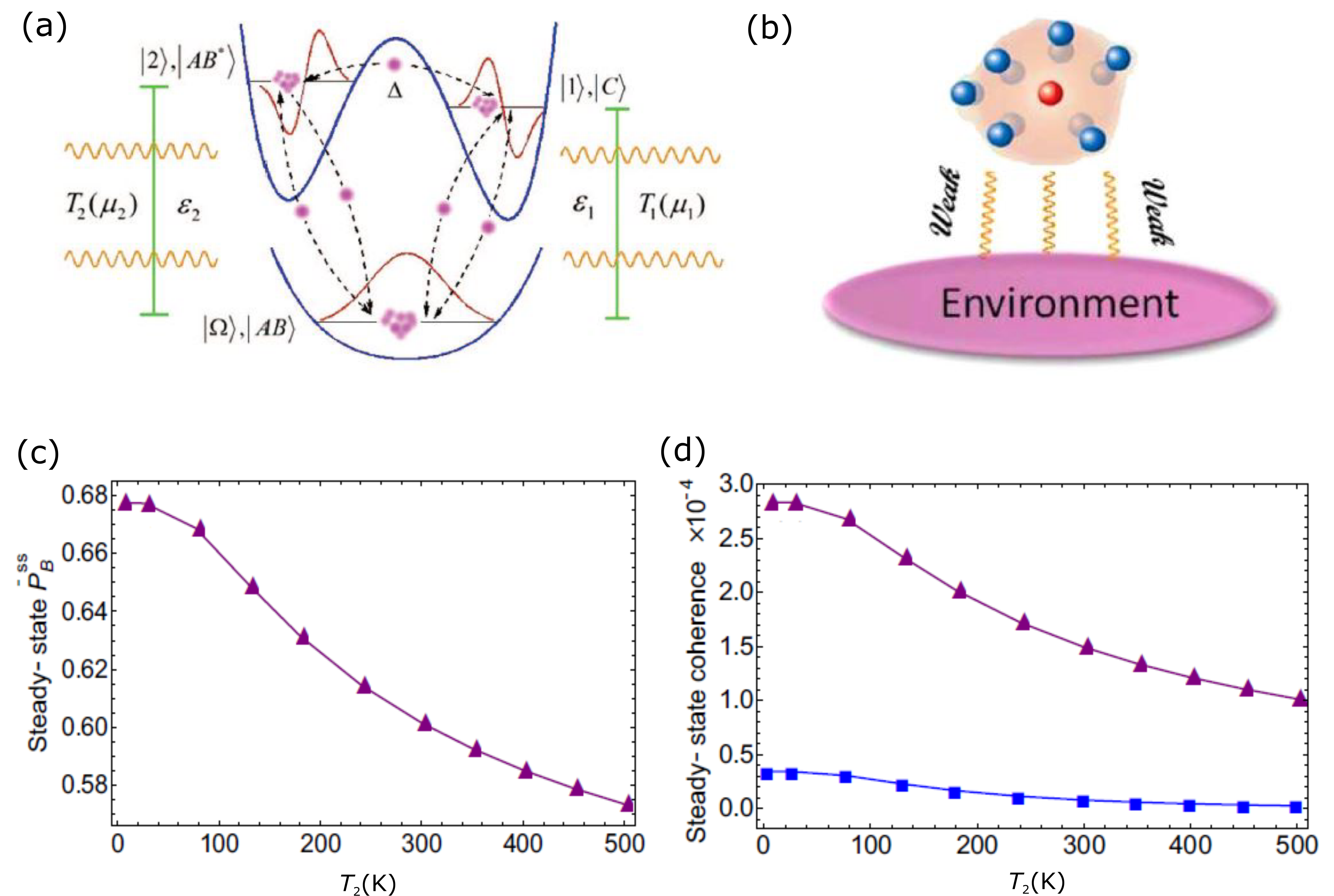}
\caption{Model for nonequilibrium transport, polarons and quantum transport efficiency. (a) Two site two level quantum transport coupled with the two environmental baths (b). Polarons formed from strong interactions of excitons and vibrons lead to weak interactions with the environments and long time coherence. (c) Quantum transport efficiency represented by  the steady-state population on pigment B (energy transfer is from pigment A to pigment B) with respect to the temperature of low-frequency fluctuations (low temperature here corresponding to high temperature difference between radiation bath and protein bath); (d) Steady-state quantum coherence varies as a function of the temperature of low frequency fluctuations. In (d) the purple and the blue lines denote the electronic (localized) coherence and excitonic (delocalized) coherence respectively. (from Ref. \cite{Zhang2014JCP} and Ref. \cite{Zhang2016SciRep}}\label{fig:quantum}
\end{figure*}

The two sites describe the transfer while the two levels describe the ground state and the excitation shown in Fig. \ref{fig:quantum} (a) (same ground state sharing between two sites). After being excited from ground state, the energy
transfer in molecules is often from donor to acceptor sites. For simplicity, one can assume the difference in excitation energies of the two sites are
small in the near resonance regime, $\varepsilon_2-\varepsilon_1\ll \textup{min}(\varepsilon_1,\varepsilon_2)$.

The energy transport in biomolecules can be described by the system interacting with two thermal environments in different temperatures. The free and
interaction parts of the Hamiltonian are
\begin{equation}
\begin{split}
&
H_S=E_g|\Omega\rangle\langle\Omega|+\varepsilon_1\eta_1^{\dagger}\eta_1+\varepsilon_2\eta_2^{\dagger}\eta_2+\Delta(\eta_1^{\dagger}\eta_2+\eta_2^{\dagger}\eta_1)\\[0.2cm]
& \quad\quad\quad\quad
H_R=\sum_{\textbf{k},p}\hbar\omega_{\textbf{k}p}a_{\textbf{k}p}^{\dagger}a_{\textbf{k}p}+\sum_{\textbf{q},s}\hbar\omega_{\textbf{q}s}b_{\textbf{q}s}^{\dagger}b_{\textbf{q}s}
\end{split}
\label{2}
\end{equation}
\begin{equation}
H_{int}=\sum_{\textbf{k},p}\lambda_{\textbf{k}p}\left(\eta_2^{\dagger}a_{\textbf{k}p}+\eta_2a_{\textbf{k}p}^{\dagger}\right)+\sum_{\textbf{q},s}\lambda_{\textbf{q}s}\left(\eta_1^{\dagger}b_{\textbf{q}s}+\eta_1
b_{\textbf{q}s}^{\dagger}\right)
\label{4}
\end{equation}
where $|\Omega>$ represents the ground state. $\eta$ and $\eta^{\dagger}$ represent the exciton annihilation and creation operators occupying the specific
excited site (two sites) for the system. Excitons are bosons but obey anti-commuting relationship within the site when constrained to only two energy levels, while they obey commuting relationship between sites \cite{Mukamel2009ChemREv}.
The annihilation and creation operators $a(b)$ and $a^{\dagger}(b^{\dagger})$ of the environments (reservoirs) satisfy Bose-Einstein commutation relations:
They ($a$ and $b$)
represent two environmental baths in equilibrium at different temperatures, obeying Bose statistics. Through the coupling with the system, the environments with different temperatures create the nonequilibriumness for the system. $\Delta$ represents the electronic coupling (tunnelling strength) between the two sites. Once the Hamiltonian is specified, using rotating wave approximation, one can follow the procedures outlined in the previous section to trace out the environments and derive the reduced system master equation (Redfield equation without secular approximation) under Markovian approximation \cite{Zhang2014JCP}. The analytical expressions for the elements in $\mathcal{M}$ are given in reference \cite{Zhang2014JCP}. For the system  coupled to  Fermionic environments, one can follow similar procedure and obtain reduced master equation \cite{Zhang2014JCP}.

\paragraph{Curl quantum flux versus nonequilibriumness and tunnelling at steady state}

Following the procedure outline in the previous section,
one can decompose the driving force for the population evolution quantified by the kinetic transfer $T$-matrix into the following form \cite{Zhang2014JCP,Zhang2015NJP1}
\begin{equation}
T=\begin{pmatrix}
   0 & \mathcal{A}_{gg}^{11}\rho_{gg} & \mathcal{A}_{22}^{gg}\rho_{22}\\
   \mathcal{A}_{gg}^{11}\rho_{gg} & 0 & \mathcal{A}_{11}^{22}\rho_{11}\\
   \mathcal{A}_{22}^{gg}\rho_{22} & \mathcal{A}_{11}^{22}\rho_{11} & 0\\
  \end{pmatrix}
 +\begin{pmatrix}
   0 & 0 & \mathcal{J}_q\\
   \mathcal{J}_q & 0 & 0\\
   0 & \mathcal{J}_q & 0\\
 \end{pmatrix}
 \label{18}
\end{equation}
Here the flux is
$\mathcal{J}_q=\mathcal{A}_{22}^{11}\rho_{22}-\mathcal{A}_{11}^{22}\rho_{11}$. In Eq.(\ref{18}) the $1^{\textup{st}}$ part of the transfer matrix describes
the equilibrium under detailed balance; The $2^{\textup{nd}}$ part represents the circular flow called the 'curl nonequilibrium quantum flux',
crucial for quantum transport.
The analytical expressions for quantum flux can be derived for energy transport in bosonic (radiation or phonon) and
fermionic environments (leads)($\hbar\omega\equiv\varepsilon_2-\varepsilon_1$) \cite{Zhang2014JCP}
\begin{equation}
\begin{split}
\mathcal{J}_q^b=\frac{2\Gamma}{\hbar^2}\frac{v^b\frac{\Delta^2}{\hbar^2\omega^2}}{1+4u^b\frac{\Delta^2}{\hbar^2\omega^2}},\quad
\mathcal{J}_q^f=\frac{2\Gamma}{\hbar^2}\frac{v^f\frac{\Delta^2}{\hbar^2\omega^2}}{1+4u^f\frac{\Delta^2}{\hbar^2\omega^2}}
\end{split}
\label{20}
\end{equation}
where the forms of functions of $u$ and $v$ are given as
\begin{equation}
\begin{split}
& v^b=\frac{\left(n_{\varepsilon}^{T_2}-n_{\varepsilon}^{T_1}\right)\left(\bar{n}_{\varepsilon}+2\right)}{\left(1+2\bar{n}_{\varepsilon}+3n_{\varepsilon}^{T_1}n_{\varepsilon}^{T_2}\right)\left[1+\frac{\Gamma^2}{\hbar^4\omega^2}\left(\bar{n}_{\varepsilon}+2\right)^2\right]}\\[0.2cm] & u^b=\frac{\left(\bar{n}_{\varepsilon}+2\right)\left(3\bar{n}_{\varepsilon}+2\right)}{4\left(1+2\bar{n}_{\varepsilon}+3n_{\varepsilon}^{T_1}n_{\varepsilon}^{T_2}\right)\left[1+\frac{\Gamma^2}{\hbar^4\omega^2}\left(\bar{n}_{\varepsilon}+2\right)^2\right]}\\[0.2cm]
& v^f=\frac{\left(n_{\varepsilon}^{\mu_2}-n_{\varepsilon}^{\mu_1}\right)\left(2-\bar{n}_{\varepsilon}\right)}{\left[1+\frac{\Gamma^2}{\hbar^4\omega^2}\left(2-\bar{n}_{\varepsilon}\right)^2\right]\left(1-n_{\varepsilon}^{\mu_1}n_{\varepsilon}^{\mu_2}\right)}\\[0.2cm] & u^f=\frac{\left(1-\frac{\bar{n}_{\varepsilon}^2}{4}\right)}{\left[1+\frac{\Gamma^2}{\hbar^4\omega^2}\left(2-\bar{n}_{\varepsilon}\right)^2\right]\left(1-n_{\varepsilon}^{\mu_1}n_{\varepsilon}^{\mu_2}\right)}
\end{split}
\end{equation}
where $\bar{n}_{\varepsilon}\equiv n_{\varepsilon}^{T_1}+n_{\varepsilon}^{T_2}$ or $n_{\varepsilon}^{\mu_1}+n_{\varepsilon}^{\mu_2}$. $n_{\varepsilon}^{T}=\frac{1}{e^{\frac{\varepsilon }{kT} } -1 } $ is the particle occupation for bosons with energy $\varepsilon$ at temperature $T$, while $n_{\varepsilon}^{\mu}=\frac{1}{e^{\frac{\varepsilon-\mu }{kT} } -1 } $ is the particle occupation for fermions with energy $\varepsilon$ at chemical potential $\mu$ and temperature $T$.

The  function $v $ (the superscript f for fermions and b for bosons)  as the occupation difference provides a measure for the effective voltage and detailed balance breaking induced from environments. Therefore, the effective potential $v$ is directly related to the temperature difference of the bosonic baths or the chemical potential difference of the fermionic baths. The function $u$ (the superscript f for fermions and b for bosons)  is a modulation factor for the flux and transport efficiency. As seen when the temperature difference or chemical potential difference is zero for the two baths, the effective voltage is zero, the flux is zero and the detailed balance is preserved. Therefore, $v$ quantifies the degree of nonequilibriumness away from the equilibrium. The definition of others are given in \cite{Zhang2014JCP}.  One sees that the
nonequilibrium quantum flux is governed by two ingredients:  the nonequilibriumness and tunnelling for driving the transports.


\paragraph{Enhancement of steady state coherence and entanglement from nonequilibriumness}

From the reduced quantum master equation, one can quantify the steady state quantum flux and the coherence to uncover their
relationship \cite{Zhang2014JCP}:
$\mathcal{J}_q^{b(f)}=\frac{2\Delta}{\hbar}\times|\textup{Im}\rho_{12}|$.  From this, one can conclude that at fixed tunneling strength, the increase of
the nonequilibrium flux will lead to the increase of steady state coherence in a linear fashion.  One can also find
\begin{equation}
|\textup{Im}\rho_{12}|=\frac{\Gamma v}{\hbar^2\omega}\frac{\frac{\Delta}{\hbar\omega}}{1+4u\frac{\Delta^2}{\hbar^2\omega^2}}
\label{21}
\end{equation}
The steady state quantum coherence is promoted by the nonequilibrium effective voltage from the difference in two temperatures or chemical potentials of the environments at fixed tunneling.  This is in
contrast to the system coupled to a single  environment with often decoherence at equilibrium state.  The environments can create the nonequilibriumness for maintaining
non-zero quantum coherence, suggesting a possible application to quantum information devices for keeping coherence through nonequilibrium
driving. \cite{Zhang2014JCP}.
On the other hand, the quantum entanglement is a quantum nature where a quantum state must be described for the system as a whole.  In other words, the quantum state of each component of the system such as particle or quibit cannot be described independently of the state of the other(s). The quantum entanglement can be quantified by the concurrence for low dimensional system and negativity for high dimensional systems \cite{Concurrence, Negativity}. It has recently been shown that steady state entanglement can be enhanced by the nonequilibriumness characterized by the temperature difference or chemical potential difference of the environments \cite{Wang2018ArXiv,Zimboras2005PRA, Segal2011PRA, Brandes2007PRB, Paris2007PRA, Burgarth2008PRA}.
It is worth mentioning that  the off-diagonal elements of the steady state density matrix in the localized basis in this case is zero when nonequilibrium voltage and therefore the flux is zero. This is even true when we change to the eigen state delocalized basis. However, this is not necessarily general for other systems. It is because the coherence is basis dependent. What is important is the fact that the steady state coherence and entanglement can be enhanced when increasing the nonequilibriumness.
In the example of spin chains, the nonequilibriumness also enhances the dynamical coherence, the entanglement and fidelity \cite{Zhang2017PRB}. A way of understanding this is from the global nature of the nonequilibrium flux spanning the state space leading to the enhancement of the quantum global nature characterized by the coherence.

\paragraph{Quantum energy transfer efficiency at steady state}

 The energy transfer efficiency can be  introduced in terms of the steady state quantum flux \cite{Zhang2014JCP}, so
 that $\eta=\mathcal{J}_q/(\mathcal{J}_q+\mathcal{A}_{22}^{gg}\rho_{22})$. One sees that
\begin{equation}
\begin{split}
&
\eta^b=\frac{\left(n_{\varepsilon}^{T_2}-n_{\varepsilon}^{T_1}\right)\frac{\Delta^2}{\hbar^2\omega^2}}{n_{\varepsilon}^{T_2}\left[B\left(T_1,T_2,\omega\right)+\left(\bar{n}_{\varepsilon}+2\right)\frac{\Delta^2}{\hbar^2\omega^2}\right]}\\[0.2cm]
&
\eta^f=\frac{\left(n_{\varepsilon}^{\mu_2}-n_{\varepsilon}^{\mu_1}\right)\frac{\Delta^2}{\hbar^2\omega^2}}{n_{\varepsilon}^{\mu_2}\left[F\left(\mu_1,\mu_2,T,\omega\right)+\left(2-\bar{n}_{\varepsilon}\right)\frac{\Delta^2}{\hbar^2\omega^2}\right]}
\end{split}
\label{25}
\end{equation}
where the definition of two functions $B$ and $F$ are given in reference \cite{Zhang2014JCP}.
At fixed tunneling, the environments characterized by temperature or chemical potential difference enhance the
transfer efficiency. Tunneling also increases the efficiency. The transfer efficiency is significantly higher for fermionic environments due
to Pauli exclusion principle.
\cite{Zhang2014JCP}


\paragraph{Dissipation and quantum thermodynamics at steady state}

The heat
dissipation through heat current  and entropy production rate (EPR) measures the thermodynamic cost of transport.  From first and second laws
of thermodynamics with the energy
conservation and the positivity of total entropy production in the nonequilibrium process, one sees: $\dot{Q_1} - \dot{Q_2} = \dot{E}$ and
$\dot{S_{env}}+\dot{S} = \dot{S_{tot}}$. $\dot{Q_1}=Tr [H_S D^1(\rho_s)]$ and $\dot{Q_2}=Tr [H_S D^2(\rho_s)]$ ($D^1$ and $D^2$ are the dissipation operators
from the system-environmental (1 and 2) coupling. )
are the energy
flowing into the system from high-temperature and low-temperature environments, respectively \cite{Valente2014PRE, Zhang2015NJP1}.
 $\dot{S}$ and $\dot{S_{tot}}$
are the rate of system entropy and total
entropy production, respectively.
Increasing nonequilibrium voltage enhances heat current $\dot{Q}$ and thermodynamic cost via entropy production $\dot{S}_{tot}=-\frac{\dot{Q_1}}{T_1} +
\frac{\dot{Q_2}}{T_2}$, \cite{Zhang2014JCP, Zhang2015NJP1}, closely related to the presence of quantum curl flux for driving nonequilibrium quantum dynamics.


\subsubsection{Long time quantum coherence and efficient energy transport of the light-harvesting complex}

The light-harvesting complex is a protein complex increasing the number of absorbed photons by the photo-system of photosynthetic organisms. It does so by transferring energy and electric charges efficiently to the photosynthetic reaction center. Experiments suggest that this process involves long-time quantum coherence at ambient temperatures~\cite{PhotoSynthesisBook, Fleming07}. Great efforts have been taken towards understanding the mechanism underlying efficient energy transfer by the light-harvesting complex~\cite{Silby, Scully, Jang, Plenio, Novelli}.


In energy transfer, the electronic excitons are coupled with the molecular vibrational phonon environments, the two site model mentioned earlier can be generalized to the $N$-site excitonic system (with $H_{sys_{ext}}$) connected by the tunneling coupled (with $H_{int}$) to the phonon (with  $H_{env_{phonon}}$) environment at room temperature and radiation environment at a higher temperature ( with $H_{env_{rad}}$ ) by the energy function as $H=H_{sys_{ext}}+H_{env_{rad}} + H_{env_{phonon}} + H_{int}$ \cite{Zhang2016SciRep}.
Previous investigations often
assume that the phonon environments fluctuate much faster than the excitation system where effect of phonons can be averaged. 
However, recent studies show that some
discrete intramolecular vibrations have a similar lifetime as excitons~ 
\cite{Mancal, Jonas, Moran, Olaya-Castro2011}. Therefore, the phonon dynamics can have a crucial effect on energy transport when energy quanta of
vibrational modes are in resonance with energy splitting of excitons
\cite{Olaya-Castro2014, Plenio, Plenio2, Romero}. 
The persistence of quantum coherence originating from exciton-phonon coupling 
has been observed in experiments \cite{Novelli}.

The effects of quasi-resonant coupling between excitons and phonons for the lifetime of quantum coherence has been studied in an effective analytical theory~\cite{Zhang2016SciRep}. There, a general scenario was investigated in which bare electrons/excitons are surrounded by discrete and continuous vibrational phonon modes ($H_{env_{phonon}} = H_{env_{{phonon}_{discrete}}}+ H_{env_{{phonon}_{continuous}}} $) and radiation environments. The near resonant coupling between the electron/exciton system and discrete phonon modes can lead to their strong interactions and the formation of polarons. In this case, the discrete vibrational modes originally from the phonon environments no longer weakly interact with the exciton system anymore. Instead, due to the strong interactions with the excitons they become part of the system in the form of composite as polarons while the remaining phonon modes become effectively the new environments along with the radiation baths.

In more details, this leads to an effective Hamiltonian  from the original one $\tilde{H_{eff}}=H_S+H_{ph}+H_{env_{rad}}+ H_{int}$ \cite{Zhang2016SciRep}.
 $H_S$ involves the renormalized exciton on site energy, phonon mediated exciton-exciton interactions, the electronic coupling renormalized by the exciton-vibrational (discrete phonon mode) interactions. $H_{ph}$ denotes the energy of the remaining phonon environments. The discrete phonon modes strongly interact with excitons and form the polarons characterized in the last term in $H_S$. $H_{int}$ describes the coupling of the new composite polarons to the remaining phonon modes.

Due to the off resonances between the residual vibrational modes and the energy splitting of excitons, the resulting polarons are only weakly coupled to the remaining phonon environments. \cite{Silby2, Silby, Scully, Jang, Olaya-Castro2014, Plenio, Plenio2, Romero, Zhang2016SciRep}. The weak coupling of polarons to the remaining phonon environment leads to less dissipation and can thus sustain quantum coherence significantly longer than the bare excitons alone, shown in Fig.~\ref{fig:quantum}(b). However, coherence alone can not guarantee the efficient energy transfer. For this detailed balance must be broken from the coupling of the light-harvesting complex to the nonequilibrium environments, for example, a coupling of high-temperature photons and low temperature vibrational modes to the protein, which can funnel the path and subsequently facilitate the coherent and unidirectional energy flow of excitations to the photosynthetic reaction center, Fig.~\ref{fig:quantum} (c) (d) \cite{Zhang2016SciRep}. While the long time survival of the dynamical coherence is dominated by the suppression of exciton-environment interaction, the non-equilibriumness is crucial for efficient energy transfer (Fig.~\ref{fig:quantum} (c)) and the steady state coherence (Fig.~\ref{fig:quantum} (d)) \cite{Zhang2014JCP,Esposito06,Zhang2015PCCP,Zhang2016SciRep}.


\section{Gene regulatory circuit motifs and experimental quantification of landscapes}

Genes encode for proteins, which provide the fundamental infrastructure for a functional cell. A certain class of proteins called transcription factors feed back on the expression of genes by binding to specific sites on the DNA called promoters, thus altering the rate of transcription. Specifically, ``activators'' increase the rate, whereas ``repressors'' inhibit expression. Together, genes and transcription factors form complex regulatory networks that have profound functional roles, for example, in decision making, differentiation, and development. In addition to naturally existing networks, synthetic counterparts are now routinely implanted into living cells.

\subsection{Naturally existing circuit motifs: lambda phage and bacteria competence }

We use bacterial phage infection and natural competence as representative examples to highlight nonequilibrium behaviors of endogenous gene networks.

\subsubsection{Landscape quantification of cell fates and their decision making of lambda phage}

Phage lambda is a bacterial virus that infects the bacterium \textit{Escherichia coli}~\cite{ptashne2004genetic,balazsi2011cellular}. The infection
process involves three steps, including phage attachment to the bacterial cell wall, injection of its DNA into the host, and execution of its transcriptional
circuitry, which determines different next-step strategies. Specifically, the phage can either integrate its own DNA into the host chromosome, resulting in the state
of lysogeny, where the virus remains `silent'. Alternatively, it triggers the lytic cycle of 
self-replication and assembly, eventually causing lysis of the host.

\begin{figure*}[t]
\includegraphics[width=0.8\textwidth]{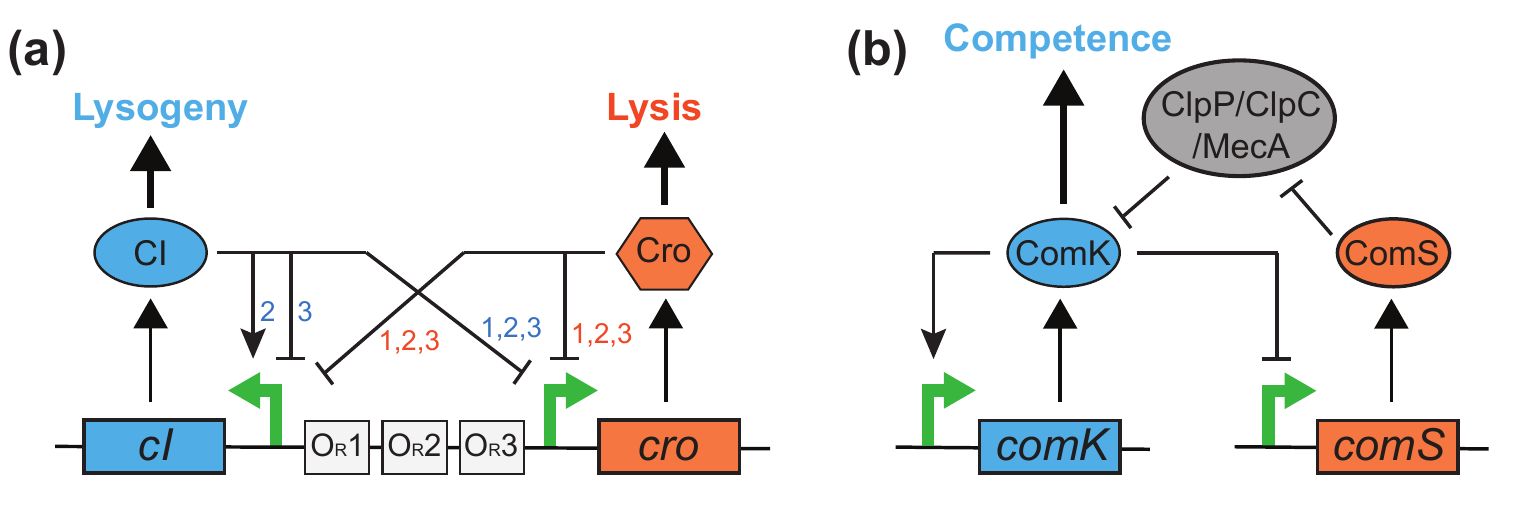}
\caption{Examples of natural gene regulatory networks.  (a) The genetic switch  between the lytic and the lysogenic life cycles of the bacterial phage $\lambda$~\cite{ptashne2004genetic}. CI is expressed in the lysogenic cycle and Cro is expressed in the lytic cycle. (b) The gene network underlying competence in \textit{Bacillus subtilis}~\cite{suel2006excitable}.  ComK is a master regulator of competence; ComS inhibits ComK degradation by the ClpP-ClpC-MecA complex. }
\label{fig-exp-natural-cellnetworks}
\end{figure*}

The decision of the lambda phage to enter the lysogenic or lytic cycle is enabled through an underlying switch controlled by the two genes cI and cro. The two genes encode the transcriptional repressor proteins CI and Cro (Fig.~\ref{fig-exp-natural-cellnetworks}a)~\cite{ptashne2004genetic}. In lysogeny, \textit{cI} is expressed while, in lysis, \textit{cro} is expressed.

Together mutual repression of the two genes through transcriptional factor binding at a shared promoter constitutes the core of this two-state switch, although there are additional processes involved in viral decision making \cite{oppenheim2005switches,ptashne2004genetic}.

The nonequilibrium landscape provides a profound physical understanding of the lambda phage switch. The landscape can be inferred from the steady-state protein distribution, which can be obtained by simultaneous fluorescent labeling of several gene products and tracing the expression of individual genes over time~\cite{Xie2006Sci, Elowitz2002Sci}. Note that due to distinct maturation times of different fluorescent labels care has to be taken. A co-localization method was suggested to resolve the issue~\cite{Pogliano,Fang2016ArXiv}. It is based on using the same fluorescent labels for different genes, but at different locations of the cell. From the real time traces of the two gene expression levels, one can obtain joint histograms and therefore quantify the landscape directly from the experiments. The possible cell fates can then identified with the attractors of the landscape and the cell-fate decision-making process can be quantified by investigating the transitions between the corresponding basins of attraction~\cite{Fang2016ArXiv, Wang2011PNAS, Balazsi2013Cell, Wang2010BJ, Xu2014PlosOne2, Li2013PlosCB, Li2015CR}.

\begin{figure*}[t]
\includegraphics[width=0.8\textwidth]{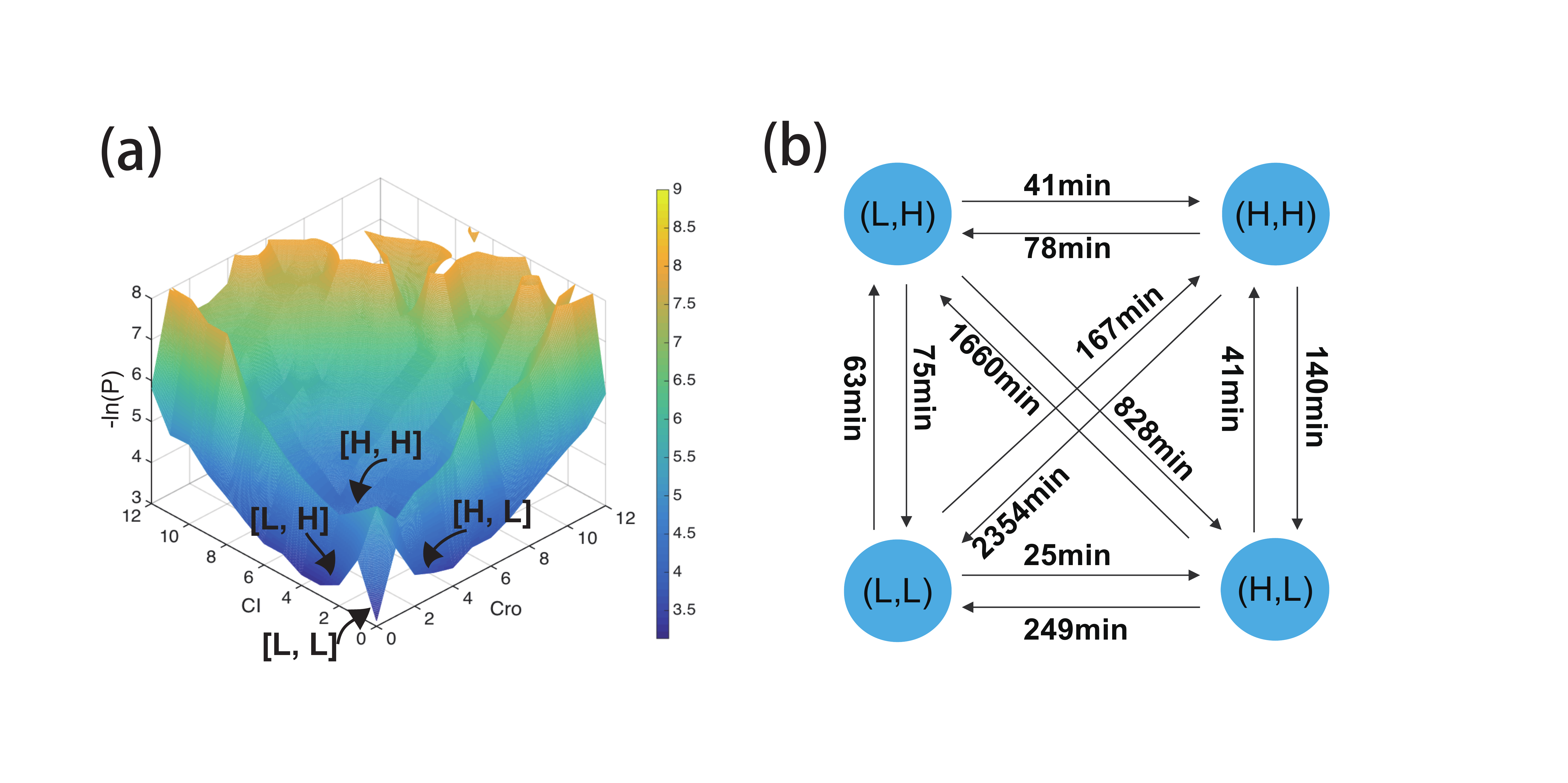}
\caption{ Nonequilibrium landscape and switching times of Lambda Phage: (a) 2D histogram as the landscape of CI and Cro production per 5~min. (b) Switching times between states of (Cro, CI)as (L,H),(L,L),(H,L),(H,H). (from Ref. \cite{Fang2016ArXiv}).} \label{Fang2016ArXiv_fig}
\end{figure*}

For the lambda phage switch, 
the underlying nonequilibrium 
landscape of CI and Cro shows four distinct states of (CI, Cro) with (high, low), (low, high), (high, high) and (low, low) expression levels~\cite{Fang2016ArXiv}, Fig.~\ref{Fang2016ArXiv_fig}a. The lysogenic and the lytic cycle are associated with the (high, low) and the (low, high) states, respectively. In a system when the effective binding and unbinding rates of the transcription factors to the genes are large compared to their synthesis and degradation rates, in the so called adiabatic limit, one expects either high levels of CI with low levels of Cro or \textit{vice versa}. The existence of 
two additional states is thus surprising which contrasts 
the conventional wisdom of only two existing states \cite{PtashneBook, Little}. In a system when the effective binding and unbinding rates of the transcription factors to the genes are slower than or comparable to their synthesis and degradation rates, in the so called non-adiabatic limit, all four states of (high, low), (low, high), (high, high) and (low, low)  are expected~\cite{Hornos2005PRE, Wolynes2007JCP, Feng2011JPCB, Feng2012SR, Zhang2013PNAS, Li2014JRSI, Chen2016SR}. 

The landscape obtained from the joint histogram of the real time trace of CI and Cro contains additional information that can be extracted by considering the basins of attraction of the four states. They have 
different depths and widths and the barriers between the basins also differ from each other implying distinct transition rates between the various states, Fig.~\ref{Fang2016ArXiv_fig}b. The residence times of individual states and switching times betweens states can be obtained from the experimental real-time traces of the CI and Cro expression levels by means of a hidden Markov model~\cite{HMMModel}. 
Perhaps less obviously, one can also infer the processes underlying switching. For example, switching from the (low, high) to the (high, low) state occurs preferentially via the (high, high) state rather than directly~\cite{Wolynes2007JCP, Fang2016ArXiv}. 

The new method of co-localization enables experimental monitoring the real time traces of the CI and Cro genes simultaneously. This leads to the quantification for cell fate decision making process in terms of the underlying nonequilibrium landscape and nontrivial cell fate states as well as their associated switchings upon environmental and genetic influences on regulations among genes.

\subsubsection{Bacterial competence}

The transition of \textit{Bacillus subtilis} from a vegetative state, in which it produces asexually, to a `competent' state, in which it can take up DNA from
the extracellular milieu, is another example of a bacterium switching between two states~\cite{grossman1995genetic,schultz2007molecular}. When facing nutrient
limitation, \textit{B. subtilis} cells often develop into spores, whereas a small fraction of the population is competent to use exogenous DNA as a food source or as a genetic material for an enhanced mutation rate and evolvability. In this case, however, making a decision between different fates does not rely on a genetic switch, but rather on an excitable network. The underlying competence regulatory circuit is centered around ComK, a master regulator that activates the expression of a set of competence genes (Fig.~\ref{fig-exp-natural-cellnetworks}b)~\cite{suel2006excitable}. ComK activates its own production, whereas its degradation is subject to the
multi-component molecular complex MecA. At the same time, ComK degradation is suppressed by ComS, a peptide that competes with ComK for the MecA complex.
Additionally, there is an indirect negative feedback between ComK and ComS. 

Together, ComK, ComS, and the MecA complex form an entangled regulatory network that involves both negative and positive feedback, which can generate excitable
dynamics involving pulses of ComK production and hence bacterial competence~\cite{suel2006excitable,suel2007tunability}. Starting from a stochastic 
increase,
the ComK level is amplified by autoregulation and then quickly increases to a maximal ComK expression, which leads to the transition to competence. At the
same time, a high level of ComK causes the suppression of ComS production, which, in turn, causes rapid ComK degradation by the MecA complex and eventually
termination of the ComK pulse. Due to the molecular noise in 
the system, ComK excitation 
occurs continuously. 
A theoretical
model was suggested to account for the noise controlled nonequilibrium transitions into and from competence
with non-adiabaticity of comparable time scale of binding/unbinding relative to the synthesis/degradation. Taking non-adiabaticity into account, the model prediction is in 
 better agreement with the experiments ~\cite{grossman1995genetic,schultz2007molecular}. It suggests again the non-adiabatic fluctuations
can be crucial for biological
functions.


\subsection{Synthetic regulatory circuit motifs: genetic switch and oscillation, self regulator}

With the advent of synthetic biology, a vast array of engineered gene networks have been successfully created since the year 2000. Examples include switches, oscillators, communication modules, patterning devices, and others which are often out of equilibrium~\cite{cameron2014brief}.

\subsubsection{Genetic switches}

\begin{figure*}[!t]
\includegraphics[width=0.8\textwidth]{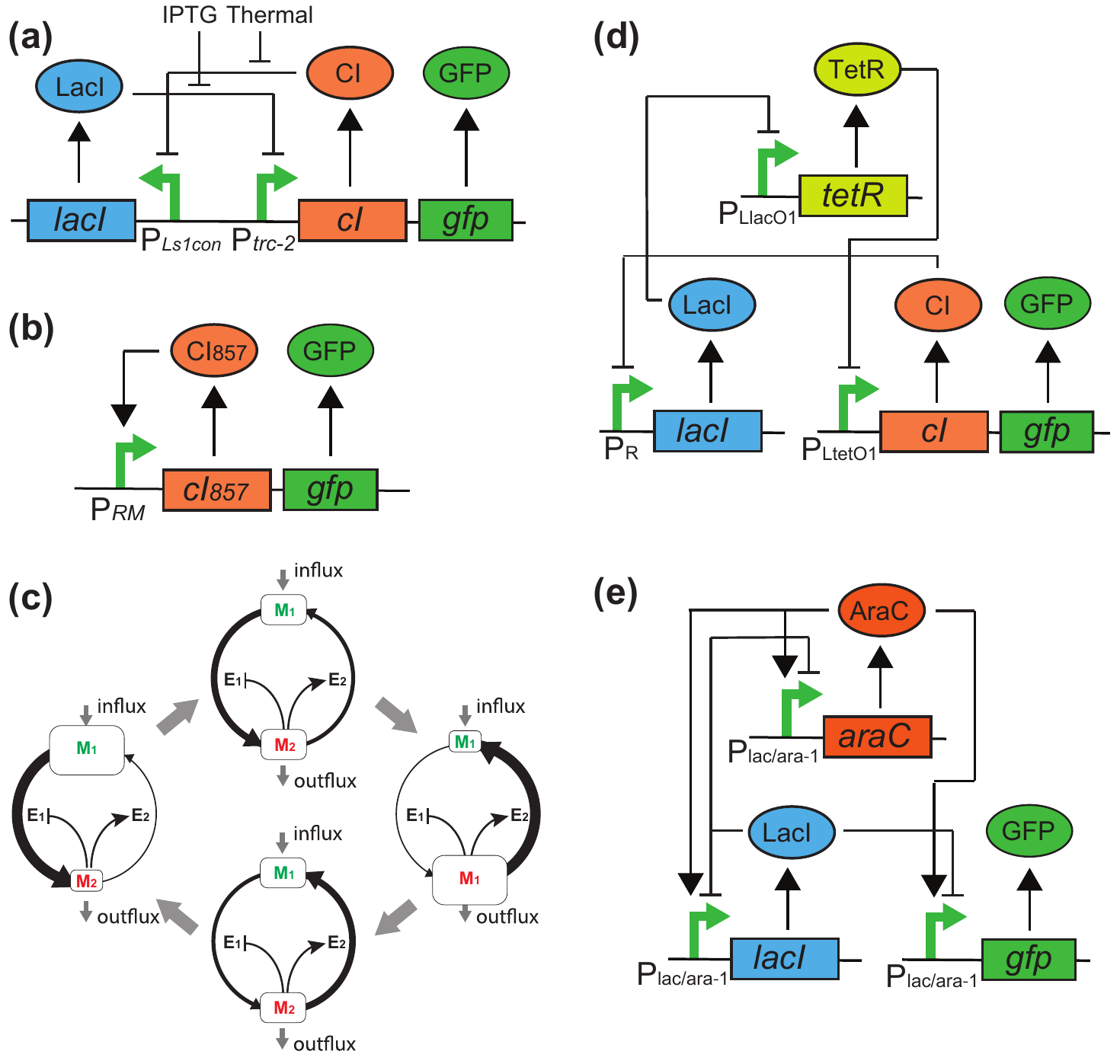}
\caption{Examples of synthetic nonequilibrium gene networks. (a) Genetic toggle switch~\cite{gardner2000construction}. (b) Auto-activation gene switch~\cite{isaacs2003prediction}. (c) Repressive oscillator~\cite{elowitz2000synthetic}. (d) A metabolic oscillator~\cite{fung2005synthetic}. (e) A fast and robust gene oscillator~\cite{stricker2008fast}.}
\label{fig-exp-synthetic-networks}
\end{figure*}

The toggle switch constructed by Collins and colleagues \cite{gardner2000construction} is a simplified version of the lambda phage switch previously discussed. It consists of two genes that encode transcriptional repressors and two corresponding (or `cognate') promoters. The genes and promoters are arranged to allow the repressor encoded from one gene to inhibit the expression of the other and vice versa, thus creating a circuit of mutual inhibition (Fig.~\ref{fig-exp-synthetic-networks}a).

The mutual suppression topology of the network can, in principle, generate bistability, a dynamic property that enables the existence of two stable states of a system. Indeed, the circuit remains stable in both a state of high expression of one gene and low expression of the other and a state with the inverse expression profile, Fig.~\ref{fig-exp-synthetic-networks}(a) At the population level, the cells exhibit a bimodal distribution characteristic of a bistable system. 

Thus, the bistability demonstrated by the toggle switch should be possible in a network involving a single, self-activating gene. This idea was tested by creating an autoregulatory circuit using the right operator site of lambda phage and the \textit{cI} gene (Fig.~\ref{fig-exp-synthetic-networks}b)~\cite{isaacs2003prediction}.

\subsubsection{Self repressor and experimental quantification of landscape}

In the extant gene regulatory networks, self-repression is much more common than self-activation~\cite{AlonSystemsBiologyBook}. This motif can accelerate responses and increase the robustness of the steady state expression level. In
the adiabatic case, when
binding and unbinding of the repressor are fast compared to its synthesis and degradation~\cite{SheaAckers}, the dynamic effect of regulatory binding to and
unbinding from the promoter averages out and a landscape with a single basin of attraction emerges. In contrast, 
if binding and
unbinding are of the same order or slower than synthesis and degradation, the gene has some chance of being expressed in spite of the presence of
repressor proteins
and states of high expression state in addition to the repressed low expression state
can appear~\cite{Hornos2005PRE, Feng2011JPCB}. For a bimodal distribution of the expression levels, one obtains a
nonequilibrium landscape with two basins of attraction~\cite{Jiang2016ArXiv}.

\begin{figure}[t]
\includegraphics[width=0.49\textwidth]{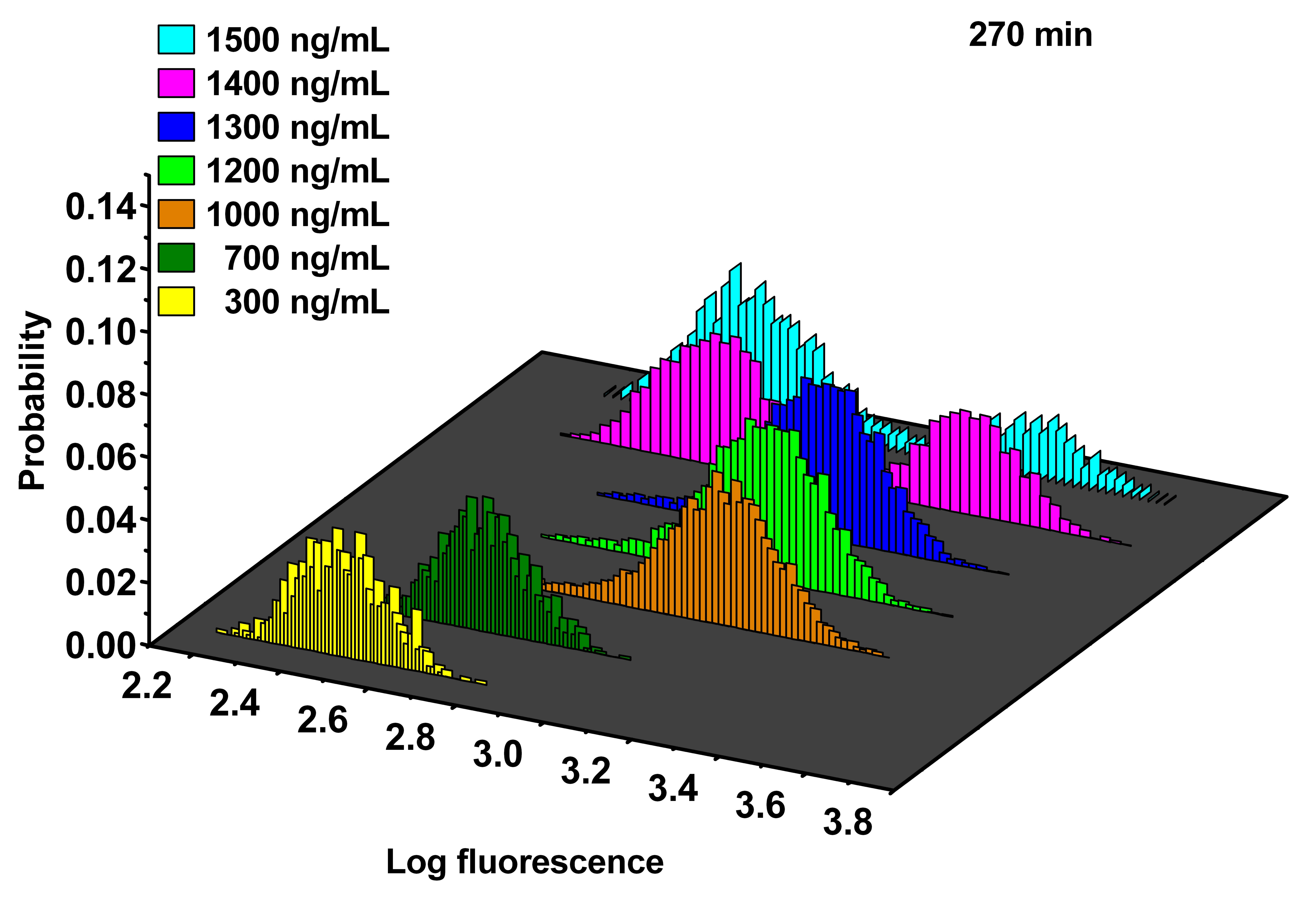}
\caption{  Experimental expression distributions of the self-repressing gene circuit (MG::PR-8T) at different aTc concentrations observed under a
microscope. (from Ref. \cite{Jiang2016ArXiv}). The negative logarithm of the distribution quantifies the landscape} \label{Jiang2016ArXiv_fig}
\end{figure}

Experimentally, a self-repressing gene circuit based on the P$_{tet}$ promoter and its repressor TetR was designed and implemented in \textit{E.~coli}~\cite{TetR, Balazsi2011PNAS}. Repressor binding can be controlled by a so-called inducer: binding of the inducer leads to a conformational change and subsequent dissociation of the repressor from the promoter.
As a result, the distributions of TetR expression change with the inducer concentration, Fig.~\ref{Jiang2016ArXiv_fig}. With increasing inducer concentrations, the mean
repressor concentration increases as binding of the inducer effectively reduces the affinity of the repressor for the promoter and the system eventually becomes nonadiabatic. 
For inducer concentrations above a
critical level, the experimental expression distribution becomes bimodal. The landscape can thus be quantified as described above. The residence times in each state and associated switching rates can be obtained by the real time trace analysis through the hidden Markov chain~\cite{Jiang2016ArXiv}.

The two typical TetR concentrations in the bimodal case correspond to different cell fates. Through the real-time traces of TetR concentrations, one can
quantify the landscape of cell fates and the associated decision making (switching) speed between the fates \cite{Jiang2016ArXiv} related to the life time of each state.



\subsubsection{Genetic oscillators}

Adding one more repressor to the toggle switch previously discussed, such that cyclic chain of repressors R$_i$ represses R$_{i+1}$, alters the behavior significantly. Explicitly, when R$_1$ represses R$_2$ , R$_2$ represses R$_3$, and R$_3$ represses R1 , a network called the repressilator is formed that can generate spontaneous oscillations~\cite{elowitz2000synthetic} (Fig.~\ref{fig-exp-synthetic-networks}c). Quantitative analysis of the system suggests that in addition to sustained limit-cycle oscillation shown in the previously described experiment, the circuit exhibits different dynamic modes, including damped oscillations and bistability depend- ing on parameter values. Indeed, the latter two types of dynamics were observed in another synthetic circuit implemented in \emph{E. coli}~\cite{atkinson2003development}. Oscillations can be generated not only by synthetic gene networks but also by metabolic networks. In the study by Fung ~\emph{et al}~\cite{fung2005synthetic}, an oscillatory circuit called the metabolator was created by integrating cellular metabolism with transcriptional regulation (Fig.~\ref{fig-exp-synthetic-networks}d). 

These examples demonstrated that oscillation can be generated using rationally designed circuits. However, they all are subject to a common challenge -- each lacks circuit performance robustness. This difficulty was addressed by~\cite{stricker2008fast} by creating a persistent genetic oscillator (Fig.~\ref{fig-exp-synthetic-networks}e) that involves an activator gene araC and a repressor gene~\emph{lacI} that are co-regulated by a hybrid promoter~P$_{lac/ara-1}$. The resulting intertwined positive and negative feedback loops confer the circuit robust oscillations~\cite{Feng2012BJ}. In experiment, the circuit was found to oscillate over a wide range of experimental conditions~\cite{stricker2008fast}. Thus, it will be interesting to explore the relationship between the circuit structure and the landscape or flux topography underlying robust oscillations.


\section{Gene regulatory network: cell cycle}

The cell cycle encompasses key processes of life, from growth and DNA replication to division \cite{CellCycleBook}. Many diseases involve cell cycle dysfunction; for example, cancer cells grow faster and divide more frequently than healthy cells \cite{CancerBiologyBook}. The eukaryotic cell cycle consists of two coordinated phases of growth, interphase, and division \cite{CellCycleBook}. The interphase is distinguished further into a first gap phase G$_1$, during which the cell accumulates mass; a synthesis phase S, during which the DNA is replicated; and a second gap phase G$_2$, during which the cell continues to grow. During the subsequent division or mitosis phase M, the cell typically divides into two daughter cells. Progression through the cell cycle is tightly controlled by genetic networks~\cite{Tyson2004MBC, TysonFissionYeast, Wang2010PNAS, Li2014PNAS}. From a physics perspective, genetic control of the cell cycle is naturally considered as a limit cycle. This is indeed the case for the cell cycle of embryonic frog cells~\cite{FerrellJr:2011je}. For yeast and mammalian cells, several scenarios have been proposed, including treating the cell cycle as a discrete attractor~\cite{Tang04}, using bifurcations~\cite{Tyson2004MBC, TysonFissionYeast}, and using the limit cycle approach~\cite{Wang2010PNAS, Li2014PNAS,Lv}.


\subsection{Embryonic cell cycle in frogs}

As mentioned in Sec.~\ref{sec:circadianRhythm}, limit cycle oscillations typically rely on a negative feedback loop. For the cell cycle in embryonic cells of the African clawed frog \textit{Xenopus laevis}, the core
negative feedback involves two genes. One of them encodes for a `Cyclin', 
the other for the Cyclin-dependent kinase Cdk1. Although the complete network is
rather involved, the essence of the network can be captured by a two-component model~\cite{Ferrell2013, Ferrell2014}, 
which can be cast into two 
ordinary differential
equations for the Cyclin concentration $Cyc$ and the Cdk1 concentration $Cdk1$:
\begin{align}
\label{eq:dcycdt}
\frac{d}{d t}Cyc&=k_\text{s}-k_\text{d}Cyc \\
\label{eq:dcdk1dt}
\frac{d}{d t}Cdk1&=k_\text{s}+k_\text{cdc}(Cyc-Cdk1)
-\left(k_\text{Wee1}+k_\text{d}\right)Cdk1.
\end{align}
The first equation describes the synthesis and degradation of cyclin with respective rates $k_\text{s}$ and $k_\text{d}$. Whereas $k_\text{s}$ is constant, $k_\text{d}$ increases with increasing Cdk1 concentration. The dependence is captured by a Hill function 
\begin{align}
k_\text{d}=a_\text{d}+b_\text{d} \frac{Cdk1^n}{K^n+Cdk1^n},
\end{align}
where $K$ is the value at which the Cdk1-dependent part has reached the half of its maximal value. The Hill exponent $n$ has a high value, which makes
the dependence on Cdk1 `ultrasensitive'. Equation~(\ref{eq:dcdk1dt}) describes activation and inactivation of Cdk1. The rates $k_\text{cdc}$ and
$k_\text{Wee1}$ depend on $Cdk1$ in a sigmoidal fashion described by a Hill function 
with proper values of $K$ and $n$. These dependencies
effectively 
account for the influence of other components in the cell cycle network. 

The ultrasensitive dependence of the rates on Cdk1 leads to a time delay in the effect of Cdk1 on degradation. In this way, cyclin first accumulates
accompanied by a moderate increase of Cdk1. After Cdk1 has passed a threshold its activation is dramatically increased, which leads to a dramatic
decrease in the amount of cyclin and simultaneously a dramatic deactivation of Cdk1, upon which cyclin accumulates again.

\begin{figure*}[t]
\includegraphics[width=0.7\textwidth]{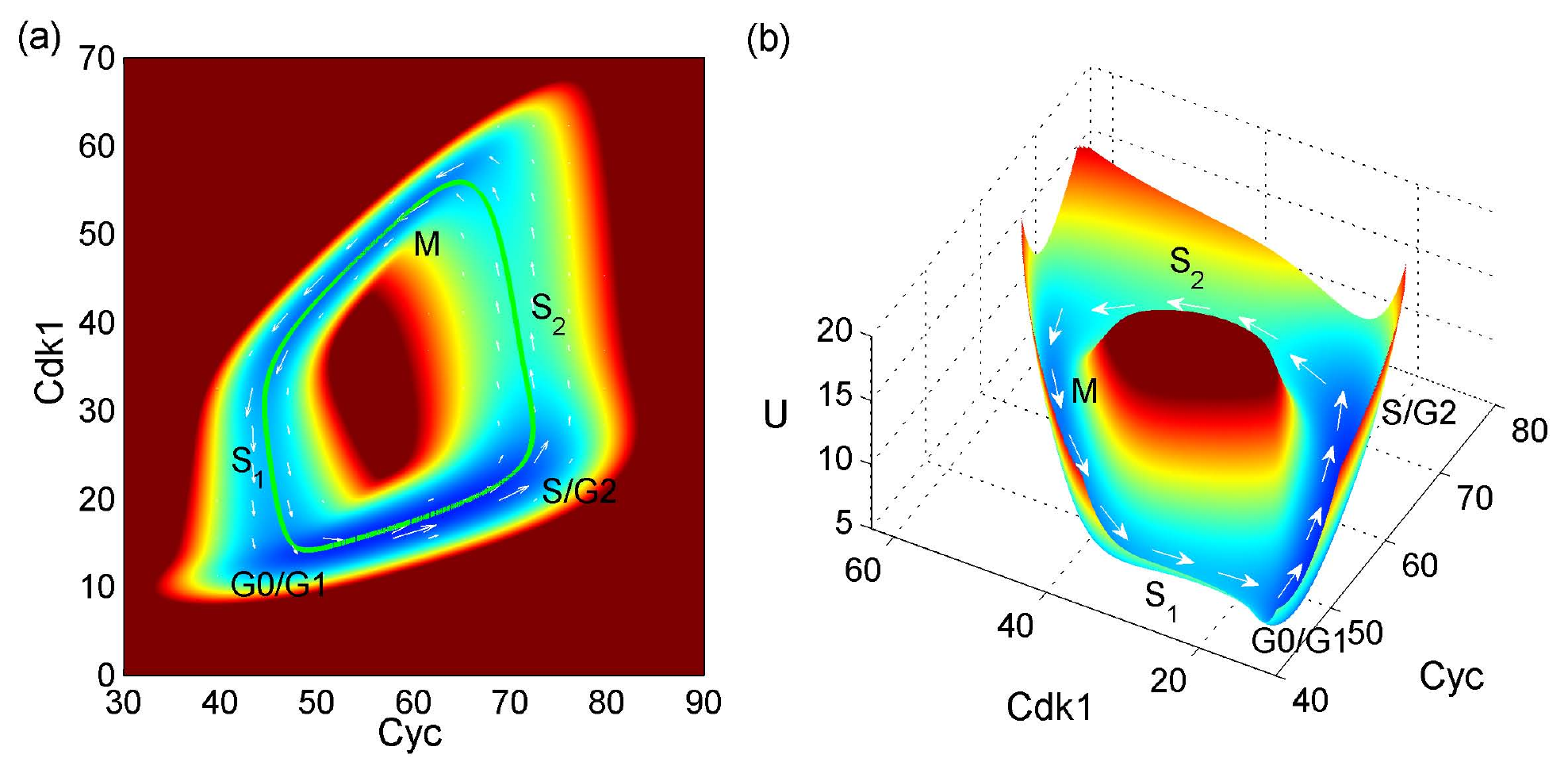}
\caption{Nonequilibrium landscape $U$ and flux (arrows) for the cell cycle dynamics (\ref{eq:dcycdt}) and (\ref{eq:dcdk1dt}) in 2D (a) and 3D (b). The different phases of the cell cycle are indicated. From~\cite{Zhang2016ArXiv3}.}
\label{Zhang2016ArXiv3_fig3}
\end{figure*}

As explained in Sec.~\ref{sec:noneqpotentials}, one can characterize the dynamics of Eqs.~(\ref{eq:dcycdt}) and (\ref{eq:dcdk1dt}) in terms of a
nonequilibrium potential landscape and a rotational curl flux \cite{Zhang2016ArXiv3}, Fig.~\ref{Zhang2016ArXiv3_fig3}. The landscape presents two attractor basin valleys and two saddle points.
The bottom basin valley is narrow and stretched. It
quantifies respectively the G0/G1 phase and S/G2 phase on each side of
the basin valley. The
top basin valley quantifies the M phase.
The transition states s$_1$ and s$_2$ correspond to transitions from M to G0/
G1 when a
cell matures and the division occurs
and from S/G2 to M which can guarantee that DNA
replication is achieved before reaching the next phase M. They are associated with so-called `check points' 
assuring that the cell is ready to enter the next phase of the cycle. The system is
periodically driven by the rotational curl flux from one basin of attraction to the other subsequently via s$_1$ and s$_2$. While the landscape guarantees the
stability of the cell cycle path, the rotational curl flux guarantees the stable flow. This gives a global and physical
picture for cell cycle seen in several species~\cite{Wang2010PNAS, Li2014PNAS, Zhang2016ArXiv3, Luo2016ArXiv}.

Note, that the speed at which the cycle is traversed depends on both, the rotational curl flux and the transition states. The cell cycle is greatly accelerated
for cancerous cells. The energy pump is the origin of the flux and energy dissipation in terms of the nutrition supply. To slow the cell cycle speed down for treatment, one can thus either decrease the flux through limiting the supply of nutrients or increase the barrier
of check points by adjusting the associated key regulators.

%
\subsection{Origins of single cell life through replication by energy pump}

The cell cycle speed correlates quantitatively with energy dissipation~\cite{Li2014PNAS,Zhang2016ArXiv3}which, as mentioned earlier, is directly related to the degree that the detailed balance is broken. A faster progression through the cell cycle requires more energy consumption, plus there is an energy threshold to overcome in forming a stable cell cycle. Replication is a signature feature of living systems. As seen here replication cannot proceed without an energy pump. Therefore, a neces- sary condition for life to begin is that an energy pump into the system must exist. The degree of nonequilibrium in thermo- dynamics and associated dynamics in terms of flux are thus required and can be quantified for the origin of single cell life \cite{Englander2013JCP, Li2014PNAS,Zhang2016ArXiv3}. Life may begin from cycles. The complexity of life may be built from multiplicative cycles and their associations.


\subsection{Cell cycle in fission yeast}

\begin{figure*}[!ht]
\includegraphics[width=1.0\textwidth]{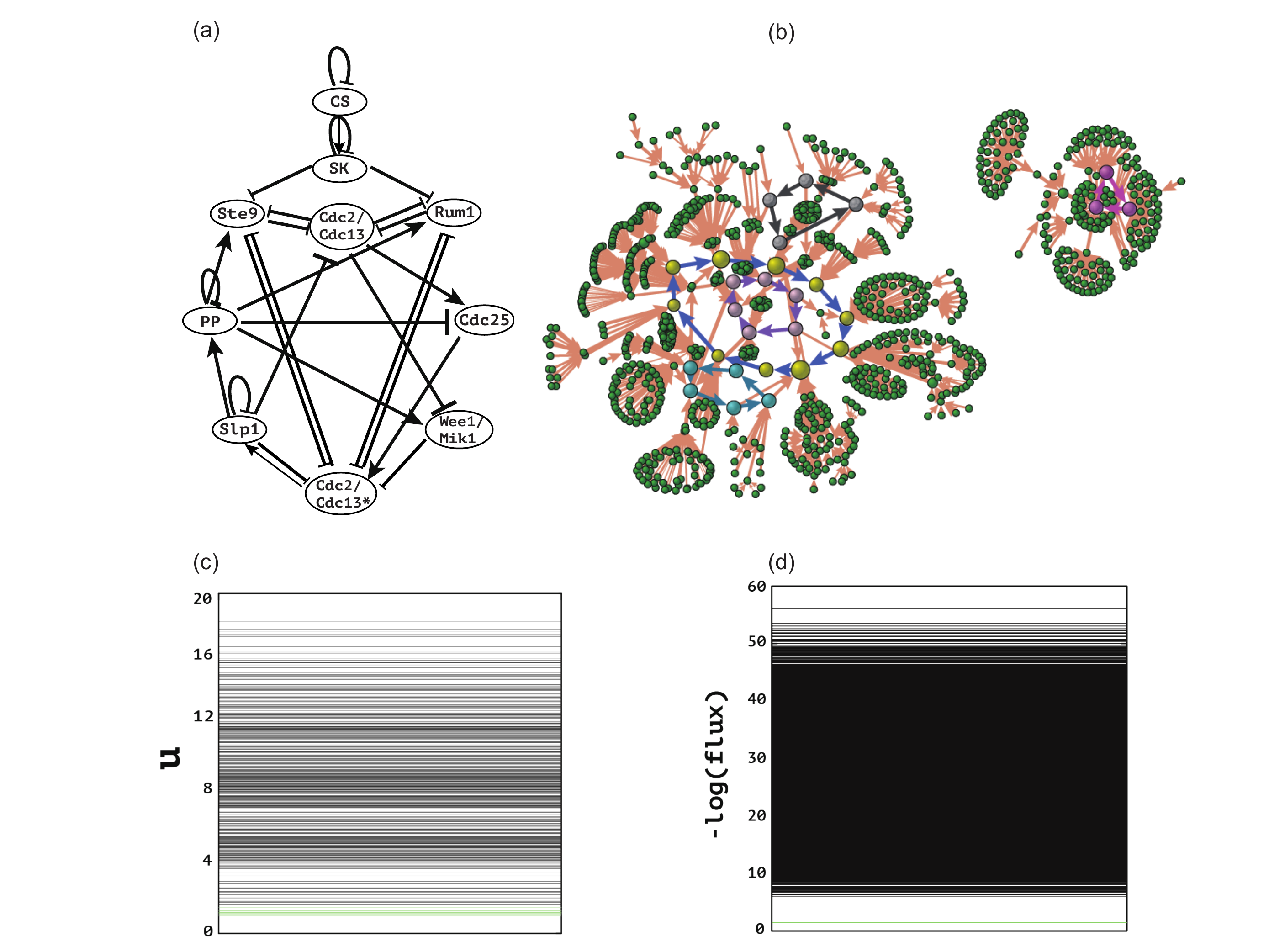}
\caption{(a) Simplified gene regulatory network of the fission yeast cell cycle. Arrows indicate activation,
repressive interactions are indicated by $-|$. (b) Potential  and flux landscapes. Each state is represented by a
dot of a size representing their steady state probability, together forming the potential landscape. The flux landscape is represented by the different flux loops. The blue
flux loop is
the dominant one representing the biological oscillation path. (c) Potential landscape spectrum. The value of the potential landscape of each state is represented by a
horizontal line. (d) Flux spectrum for all
flux loops. The
flux along each loop is represented by a horizontal line. In (c) and
(d) the green lines correspond to the states and the
flux loop of the "native" biological oscillation cycle path.}\label{fig:fspect}
\end{figure*}

The gene regulatory network controlling the fission yeast cell cycle is complex and involves a few hundred genes~\cite{TysonFissionYeast}. Even
a simplified network based on experimental studies still involves 10 key genes 
\cite{Bornholdt2008PLoSOne}, Fig.~\ref{fig:fspect}(a). This network can be further simplified by
reducing the states of the individual genes to be on and off only. The corresponding ``boolean'' network is a discrete dynamic system with $2^{10}$
states. Boolean networks are particularly suited to explore the global dynamics and wiring topology of networks~\cite{Kauffman69,Tang04,Wang2007BJ, Wang2008PRE}.

In the presence of fluctuations, one can follow the master equation for the stochastic evolutionary dynamics of the fission yeast cell cycle. One can map out the landscape through the steady-state solution of the corresponding master equation (\ref{eq:MasterEquation}), where the transition rates $T_{ij}$ are eventually determined by the original gene regulatory network. The resulting landscape has the form of a Mexican hat and the cell cycle path corresponds to the valley of the hat, Fig.~\ref{fig:fspect}(c). The cell cycle path is stable when states on the path have much lower potentials than those outside the path relative to the spectrum?s standard deviation. This gap between the potential minimum and the average of other states not on the cell cycle, relative to variance, leads to a funneled potential landscape toward the cell cycle path and guarantees the stability, since states on the oscillation path have much higher weights than other states. However, this analysis cannot guarantee directional flow for oscillations, Fig. \ref{fig:fspect}(c).

The steady-state nonequilibrium probability flux provides a driving force in addition to the landscape gradient for nonequilibrium networks. Flux can be obtained from the steady state solution of the master equation based on the underlying gene regulatory networks \cite{Wang2007BJ, Wang2008PRE, Luo2016ArXiv}. As mentioned previously, the flux originates from the energy pump through the nutrition supply. The nonequilibrium flux can be further decomposed into flux loops  \cite{Luo2016ArXiv}; doing so forms a nonequilibrium flux landscape, Fig. \ref{fig:fspect}(b). When there is a distinct separation between the nonequilibrium flux from the native biological cycle, Fig. \ref{fig:fspect}(b) and the rest relative to variance, the cell cycle becomes the dominant loop compared to the other possible loops, as in Fig. \ref{fig:fspect}(d). Therefore, a funneled nonequilibrium flux landscape provides a physical mechanism to guarantee stable cell cycle flow \cite{Luo2016ArXiv}.

By performing a global sensitivity analysis on the topog- raphy of the potential landscape and the flux loop landscape upon changes of genes and their mutual regulations, the identities of key genes and regulatory motifs for the network are revealed. This provides a possible way to control the cell cycle speed in the prevention or treatment of cancer.

It is worthwhile to note that the described gene network regulating the cell cycle of \textit{Xenopus} consists of two genes, although the described fission yeast network is much more complex and contains many more genes. This difference in the literature can be attributed to the fact that yeast has been studied as a model organism for decades, compared to \textit{Xenopus}. Remarkably, similar cell cycle mechanisms are found in both organisms.

\section{Cellular structure and dynamics}\label{sec:cellularDynamics}

In this section, we turn to processes of intracellular organization. After discussing nonequilibrium aggregation and phase-separation phenomena, we focus on the
cytoskeleton, the paradigmatic active gel that determines cell mechanical properties and drives vital cellular processes. Finally, we will briefly address
nonequilibrium aspects of cell signalling. Although landscape and flux theory has not been applied to the cellular structure and dynamics due to the technical  challenges, other approaches such as
local thermal equilibrium approach, hydrodynamics approach, and active particle dynamics~\cite{Marchetti2013RMP, WolynesCytoskeleton} have been successfully applied to this exciting research area which we will briefly review some interesting perspectives.

\subsection{Nonequilibrium phase separation}

In animal and plant cells, important functional subunits are segregated into compartments that are surrounded by lipid bilayers. A plethora of proteins
attach to and are embedded in these fluid membranes and can arrange into functional assemblies. In addition, there are important functional
three-dimensional cellular subunits that lack a delimiting membrane. These structures can result from phase separation with continuous exchange of matter
and energy determining their size distributions and dynamics.

\paragraph{Lipid rafts}

The equilibrium state of multicomponent lipid bilayers is either well mixed or the different kinds of lipids phase segregate into macroscopic domains of an
extension that scales with the system size. For membranes of living cells, however, there is ample evidence for lipid microdomains of either different
composition or a different liquid phase compared to the environment  -- often called ``rafts''. They have a comparatively small size of 10-200~nm and a
lifetime of several milliseconds~\cite{Pike:2006cu}. At equilibrium, domains of this size should only exist in presence of long-distance
interactions~\cite{Seul:1995bz}. A possible origin of such long-range interactions relevant for cells is a coupling of membrane curvature and lipid
composition~\cite{Baumgart:2003kk}. Indeed, mixtures of lipids with different intrinsic curvatures have been observed to segregate into small budding
domains~\cite{Baumgart:2003kk}. In mixtures of cholesterol and lipids with high and low melting temperatures, long-lived microdomains form, but are
probably due kinetic arrest in the coarsening process and are absent in the equilibrium state~\cite{Veatch:2003wj}.

Lipid nanodomains can robustly form by nonequilibrium processes, for example, in presence of lipid exchange between the cytosol and the
membrane~\cite{Gheber:1999uh,Turner:2005dt,Foret:2005dm,Fan:2008bj} or of chemical reactions~\cite{Huberman:1976hi,Glotzer:1995ce}. Let $\phi$ be the
volume fraction of a lipid in a binary mixture with constant total density $n_0$. Then~\cite{Huberman:1976hi,Glotzer:1995ce,Foret:2005dm}
\begin{align}
\label{eq:lipidexchange}
\partial_t\phi &=\mu\nabla^2\frac{\delta F}{\delta\phi}-\frac{1}{\tau}\left(\phi-\bar\phi\right).
\end{align}
Here, $\mu$ is an effective mobility and $F$ the corresponding free energy
\begin{align}
F[\phi]&=n_0 k_BT\int d^2\mathbf{r} \left[\frac{\xi_0^2}{2}\left(\nabla\phi\right)^2+f\left(\phi\right)\right].\\
\intertext{For}
f\left(\phi\right)&=\phi\ln\phi+\left(1-\phi\right)\ln\left(1-\phi\right)+\chi\phi\left(1-\phi\right)
\end{align}
the first term in Eq.~(\ref{eq:lipidexchange}) is the familiar Cahn-Hilliard current. The parameter $\chi$ determines the strength of lipid-lipid
interactions and demixing occurs for $\chi>2$. The second term describes relaxation to the stationary density $\bar\phi$ with a characteristic time $\tau$
that is determined by the rates of lipid integration into the membrane and dissolution into the cytosol. For biologically relevant parameters, raft-like
domains of sizes between 20~nm and 200~nm are formed for exchange times $\tau$ between $10^{-4}$~s and 1~s. Importantly, the mobility of cytosolic lipids
is many orders of magnitude larger than for lipids in membranes, such that the spatial distribution of cytosolic lipids is essentially homogenous. In this
way, dissociation leads to efficient mixing, which prevents the formation of macroscopic domains.

\paragraph{Clusters of membrane-associated proteins}

\begin{figure}
\includegraphics[width=0.5\textwidth]{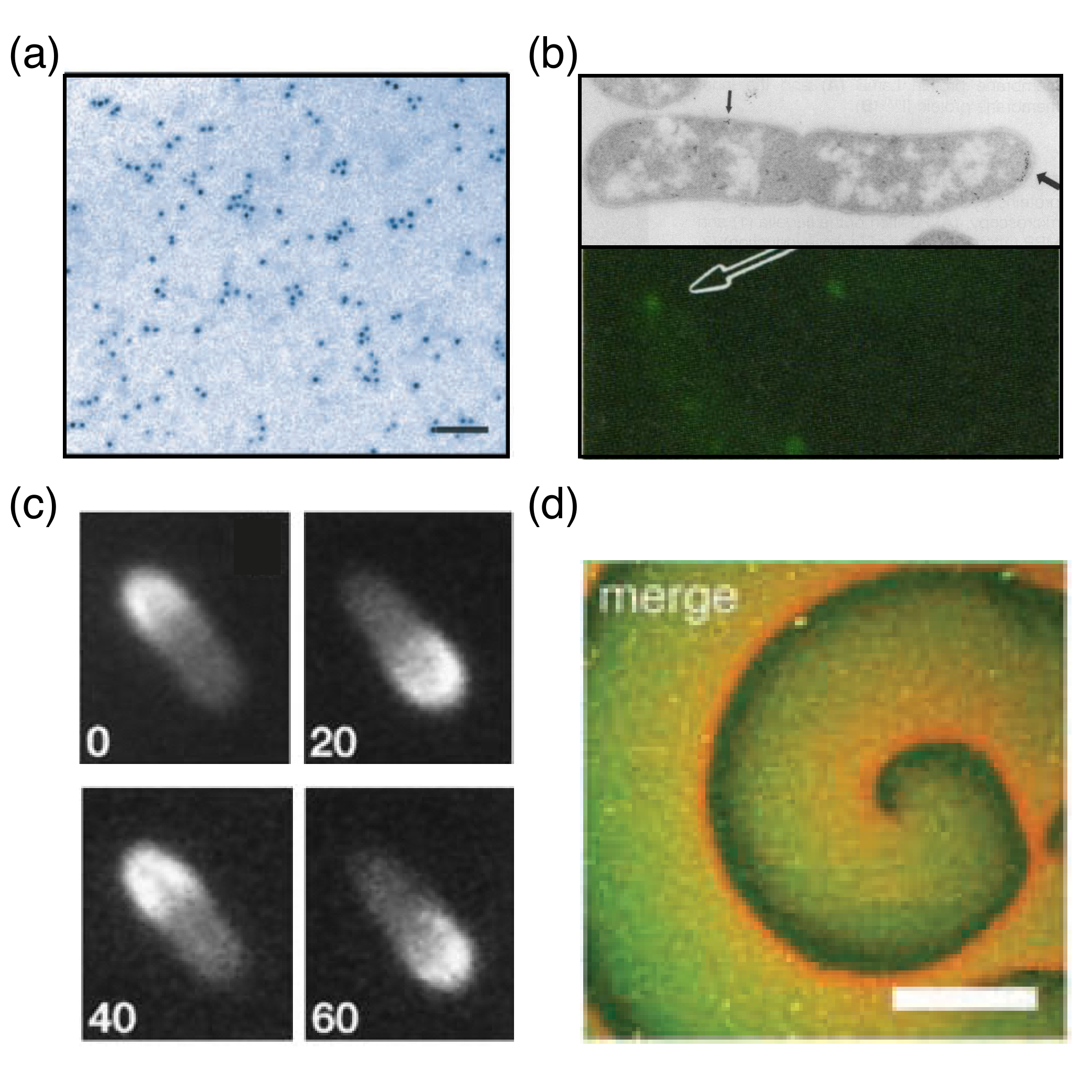}
\caption{\label{fig:clusters}Examples of protein clusters on cell membranes. (a) Electron microscopy image illustrating K-ras nanoclusters. Scale bar 50~nm.
From~\cite{Prior:2003jh}. (b) Electron micrograph (top) and fluorescence image (bottom) of \textit{E.~coli} illustrating clustering of chemoreceptor Tsr.
From~\cite{Maddock:1993tu}. (c) Min-protein oscillations in \textit{E.~coli}. Numbers indicate time in seconds. From~\cite{Raskin:1999vr}. (d) Spiral of MinD
(green) and MinE (red) on a supported lipid bilayer. Scale bar 50~$\mu$m. From~\cite{Loose:2008ca}.}
\end{figure}

Through the preferential localization of proteins to specific lipids the existence of lipid rafts implies the existence of protein clusters. Examples of
protein clusters are protein coats during early phases of endocytosis~\cite{Faini:2013bv}, receptors~\cite{Maddock:1993tu,Uhles:2003bi}, SNARE
proteins~\cite{Low:2006ip}, and proteins involved in signaling~\cite{Douglass:2005fy,Tian:2007tn,Goswami:2008cj,Fairn:2011kp,Bonny:2016em}. Similar
mechanisms as evoked for the formation of lipid rafts have been considered in this context~\cite{Destainville:2008kp,Sieber:2007er},
Fig.~\ref{fig:clusters}(a,b).

Some of these clusters form at specific positions. For example, receptor clusters in the rod-shaped bacterium \textit{Escherichia coli} form at the cell
ends~\cite{Maddock:1993tu}. Similarly, Spo0J/Soj in the like-wise rod-shaped bacterium \textit{Bacillus subtillus} localizes to the cell ends, although
they bind to DNA and not to membrane~\cite{Quisel:1999uh,Marston:1999uj}. Cell polarity in budding yeast \textit{Saccharomyces cerevisiae} is established
by localization of Cdc42~\cite{Fairn:2011kp}. Cues like membrane-curvature or -composition or specific DNA sequences might be involved in positioning these
aggregates. However, in presence of cooperative effects during attachment, proteins can self-organize in clusters at specific locations by a Turing-like
mechanism~\cite{Turing:1952ja}. A simple example is, when binding to the membrane is facilitated by molecules already bound to the membrane, whereas
detachment from the membrane is spontaneous. The processes can be captured by the following toy model~\cite{Wettmann:2014kg}, where the distributions of
cytosolic and membrane-bound molecules, $c$ and $m$, along an interval of length $L$ evolve according to
\begin{align}
\label{eq:dcdt}
\partial_t c &= \phantom{D}\partial_x^2 c - \left(1+m^2\right)c + km\\
\label{eq:dmdt}
\partial_t m &=D\partial_x^2m + \left(1+m^2\right)c - km
\end{align}
with $k$ being the detachment rate. For sufficiently small values of the ratio $D$ of membrane-bound and cytosolic diffusion, the homogenous state can
become unstable leading to a maximum of $m$ at one end of the interval. In a rod-shaped bacterium this would correspond to accumulation of proteins at one
cell end. An analysis of the corresponding stochastic dynamics that accounts for the randomness inherent in molecular reactions shows that the life-time of
such clusters increases with molecule number and thus cell size~\cite{Wettmann:2014kg}. Similar observations had been made in detailed descriptions of the
Spo0J/Soj proteins~\cite{Doubrovinski:2005cm}.

\paragraph{Surface waves}

The total number of particles $\int dx\left(c+m\right)$ is conserved by the above dynamics, which presents an interesting twist to the original Turing
mechanism. On general gorunds, one can expect the spontaneous emergence of traveling waves in such systems~\cite{Kessler:2016gh}. This phenomenon is
exemplified by the proteins MinD and MinE of the rod-shaped bacterium \textit{Escherichia coli}~\cite{Loose:2011dd}. MinD and MinE direct the protein MinC,
which inhibits assembly of the cell division machinery, to the vicinity of the cell poles and thus localize division in the cell center. However, the Min
proteins are not statically distributed at to the two cell ends but rather shuttle periodically between them with a period of about a
minute~\cite{Raskin:1999vr}, Fig.~\ref{fig:clusters}(c).

MinD is an ATPase that clusters on the membrane after binding MinD by a mechanism that is still poorly understood~\cite{Hu:2002hs}. At the membrane, it
recruits MinE that catalyzes ATP hydrolysis and drives MinD off the membrane. In presence of suitable cooperative effects during the formation of MinD
clusters on the membrane, the oscillatory behavior observed \textit{in vivo} can emerge spontaneously~\cite{Huang:2003bc}. Although MinD binding to the
membrane was experimentally found to be cooperative, the molecular details are unknown. Currently, in most theoretical descriptions of the Min-protein
dynamics a process similar to the cooperative binding present in Eqs.~(\ref{eq:dcdt})-(\ref{eq:dmdt}) is assumed~\cite{Huang:2003bc,Bonny:2013gt}, but
alternative mechanisms have been proposed, where MinD forms complexes only after binding to the membrane~\cite{Kruse:2002ip,Walsh:2015jy,Petrasek:2015es}.

Physical studies of the mechanism generating the oscillatory patterns \textit{in vivo} notably involved observing the Min-protein dynamics in different
geometries. Varying the cell length revealed transitions from a bistable regime in short bacteria to standing and then to traveling
waves~\cite{Bonny:2013gt}. The wave dynamics could be reconstituted in an open geometry \textit{in vitro} on supported lipid bilayers~\cite{Loose:2008ca},
Fig.~\ref{fig:clusters}(d). This allowed for further molecular characterization of the Min-protein dynamics~\cite{Loose:2011il,Schweizer:2012ba}. The
\textit{in vitro} approach has been extended to study waves in confined geometries~\cite{Zieske:2013fi,Caspi:2016ia}. There the patterns found in living
cells could be reproduced further supporting a common mechanism underlying the patterns in bacteria and in the reconstituted systems.

\subsection{Adhesion domains}

Cells adhere to other cells, a substrate, or the extracellular matrix via transmembrane proteins like integrins and cadherins~\cite{Schwarz:2013em}. These
form initially submicrometer sized circular domains, which then mature into larger complexes of up to 10~$\mu$m that also involve cytoplasmic proteins,
notably components of the cytoskeleton, see below. Through the coupling to the cytoskeleton, adhesion domains are subject to mechanical forces that are
necessary for maturation~\cite{Balaban:2001tr}. The molecular mechanism involved in force sensing could notably depend on a force-dependent lifetime of
individual adhesion bonds. In addition, there is a feedback form the adhesion domains to the organization of the actin cytoskeleton~\cite{Drees:2005kt}. A
general theory of force-dependent formation of adhesion domains seems to be currently missing.

\paragraph{Membrane-less organelles}

In addition to membrane domains, cells contain three-dimensional membrane-less functional units. Originally suggested for so-called P granules in the
developing nematode \textit{Caenorhabditis elegans}, such functional units can be described as liquid droplets that form through a nucleation
process~\cite{Brangwynne:2009bm}. P granules are asymmetrically distributed in \textit{C.~elegans} embryos at the one-cell stage. After the first cell
division, the daughter cell rich in P granules will eventually form a line of germ cells, whereas the other daughter cell will give rise to a somatic cell
line. The asymmetric distribution is induced by a gradient in the distribution of the protein Mex-5 that promotes P-granule disassembly and thus
effectively increases the concentration at which P-granule components saturate. The asymmetric state is out of equilbrium as droplets continuously form in
regions of low Mex-5 concentrations and dissolve in regions of high Mex-5 concentrations~\cite{Lee:2013hk}. Diffusion then leads to a permanent net droplet
flux between the two regions and an oppositely directed flux of P-granule components.

P granules are liquid-like droplets with a viscosity of 1 Pa$\cdot$s, that is, 3 orders of magnitude higher than the viscosity of water, and a surface
tension of the order of 1 $\mu$N/m, that is, 5 orders of magnitude smaller than the air-water surface tension and allows rapid droplet formation. Other
compartments have been found to similarly form liquid-like droplets~\cite{Brangwynne:2011bp}. Prominent examples are centrosomes~\cite{Zwicker:2014gh} that
serve as microtubule organizing centers and metaphase spindles that arrange the chromosomes during division, which can be described as liquid crystalline
droplets~\cite{Reber:2013fr}. In the nucleolus, a large subcompartment of the nucleus in which ribosomes are created, subdomain structure has been ascribed
to droplets of immiscible liquid-like phases~\cite{Feric:2016if}.

Above the saturation threshold, P granules continue to grow and fuse, which leads to Oswald ripening~\cite{Brangwynne:2009bm}. In contrast, the two
centrosomes present in cell division do not fuse, which indeed would be detrimental to the segregation of the chromosomes. The mechanism suppressing Oswald
ripening in that case is similar to that proposed for limiting the size of lipid rafts: Constituents of the droplet continuously change between two
different states with different physical properties, such that, in one state, the protein prefers the condensed phase, whereas the other form is soluble in
the cytoplasm. In the case of centrosomes, this could be the case for spindle defect protein-5 that assumes different conformations depending on its
phosphorylation states~\cite{Zwicker:2014gh}. Several of such ``active droplets'' can stably co-exist in which case they assume equal
sizes~\cite{Zwicker:2014gh}. This is important as centrosome size seems to be directly controlling spindle length in
\textit{C.~elegans}~\cite{Greenan:2010gu}.

\subsection{The cytoskeleton - an active material}

The cytoskeleton is a network of filamentous protein assemblies that interact with a plethora of proteins, which regulate filament length, cross-link
filaments and generate active stresses. It is involved in vital processes like cell division and migration and also determines cellular mechanical
properties. From a physical point of view, the cytoskeleton is an active material as its constituents are kept out of thermodynamic equilibrium by the
hydrolysis of nucleoside triphosphates (NTP). We will present first generic physical properties of cytoskeletal systems and then turn to two biological
applications that are of current interest, cell migration and the actin cortex.

\subsubsection{Filaments}

Cytoskeletal filaments fall into two classes. Actin and tubulin, which, respectively, form actin filaments and microtubules, can bind nucleotides and
assume different states with different binding affinities depending on the nucleotide bound. In addition, actin filaments and microtubules are structurally
polar assemblies. The structural polarity is also expressed in different exchange kinetics at the two ends that are commonly referred to the plus- and
minus-ends, respectively, with exchange being more rapid at the plus-end~\cite{Kuhn2005,Fujiwara:2007ge}. In contrast, proteins like vimentin or keratin
that form intermediate filaments, which seem to play a mostly structural role, exist in only one state and their assemblies are non-polar.

\paragraph{Filament length dynamics and treadmilling} The coupling to NTP-hydrolysis in combination with their structural polarity leads to assembly
kinetics of actin filaments and microtubules that are alien to commonly studied polymers. In particular, during assembly there can be an overshoot in the
average filament lengths~\cite{Brooks:2008hc} or even oscillations~\cite{Carlier:1987un}. Most spectacularly, it can also lead to treadmilling, when
filaments show net growth at the plus- and net shortening at the minus-end~\cite{Wegner:1976ks,Margolis:1978vz}. Observations of filament treadmilling have
been reported \textit{in vivo}~\cite{WatermanStorer:1997ug,Rzadzinska2004} and \textit{in vitro}~\cite{Carlier:1997tp,Panda:1999tw}. It relies on the
establishment of an NTP-gradient along the assembling filament~\cite{Erlenkamper:2013cy}: Filament subunits with NTP bound have a higher affinity to bind
to other subunits than NDP-bound subunits. In cells, hydrolysis of NTP bound to a filament subunit is essentially irreversible, such that the fraction of
NDP-bound subunits increases towards the minus-end. In this way, assembly occurs preferentially at the plus end in form of NTP-subunit attachment, whereas
detachment occurs preferentially at the minus-end, where NDP-subunits leave the filament. This gradient also implies an effective length dependence of the
depolymerization rate, which can lead to a finite typical filament length~\cite{Erlenkamper:2013cy,Mohapatra:2016kq}. Other mechanisms of length-dependent
assembly and disassembly rates exist that involve proteins influencing the growth or shrinkage of cytoskeletal filaments~\cite{Mohapatra:2016kq}, see
below.

\paragraph{Nucleation promoting factors}  Filament assembly is directly utilized by cells to form protrusions and for migration. The polymerization of
filaments anchored in a network or to a substrate can generate forces onto an object either directly~\cite{Dogterom1997,Footer:2007dw} or by generating
stresses in non-flat networks~\cite{Prost:2002vu}. To see how stresses can arise from polymerization, note first that filaments do not appear spontaneously
for the conditions present in cells. Instead they require factors that promote the formation of filament nuclei, which can then grow either spontaneously
or with the help of elongation factors. Important factors promoting the nucleation of actin filaments are complexes of actin related proteins 2 and 3
(Arp2/3) and members of the formin family. These are typically active only in the vicinity of a membrane. As a consequence, actin gels grow by adding
material at the interface with a surface which leads to mechanical stresses if the surface is curved. These stresses are exploited, for example, by the
bacterium \textit{Listeria monocytogenes} for propulsion in the cytoplasm of her host cell~\cite{Prost:2002vu}.

The mechanisms underlying nucleation by the Arp2/3 complex and formins are different. The Arp2/3 complex branches new filaments from existing filaments.
This process is used in particular to extend the leading edge of cells crawling on a substrate as it provides new free plus ends that can grow, whereas
elongation of older plus ends is dampened by capping proteins. The interplay of Arp2/3 and capping proteins can lead to a variety of force-velocity curves
for the advancing leading edge~\cite{Carlsson:2003jo,Weichsel:2010dl,Schreiber:2010fy}.

Animal cells typically assemble a thin actin sheet below their plasma membrane. Filaments in this actin cortex nucleated by the Arp2/3 complex and formins
form two subpopulations that can be distinguished through their different turnover rates~\cite{Fritzsche:2013bf}. Furthermore, formin nucleated filaments
are typically 10 times longer than Arp2/3-nucleated filaments~\cite{Fritzsche:2016fx}, which in turn effects the gel's mechanical
properties~\cite{Bai:2011if,Fritzsche:2016fx}. The mechanical properties of actin gels in presence of filament turnover are only beginning to be
explored~\cite{Hiraiwa:2016ku}.

\subsubsection{Motors}

\paragraph{Single molecular motors} Cytoskeletal motor proteins assure directed long-range transport in cells and generate mechanical stresses in the cytoskeletal
network. Members of the myosin super family interact with actin filaments, whereas kinesins and dyneins interact with microtubules. They are ATPases that
have various conformational states depending on the nucleotide bound. Thermodynamics shows that for motor proteins to move directionally, isotropy and
detailed balance must be broken~\cite{Julicher1997RMP}. Molecular motors are characterized by their force-velocity relation and their persistence. In many
cases, the force-velocity relation is well approximated by
\begin{align}
v = v_0\left(1 - \frac{f}{f_s}\right).
\end{align}
Here, $v_0$ is the motor velocity in absence of a load, $f_s$ the stall force at which the motor stops advancing, and $f$ the magnitude of the force
opposing motor movement. Although, forces larger than the stall force should lead to backward motion of the motor, in practice, this is hardly observed.
Instead the motor readily detaches under such conditions.

The persistence is given by the average length a motor walks along the filament before detachment. It equals the ratio of the stepping and the detachment
rates times the size of a single step. The detachment rate depends on the applied force and Kramers rate theory suggests
\begin{align}
k_\mathrm{off}\left(f\right)&= k_\mathrm{off,0}\exp\left\{-\frac{|f|a}{k_BT}\right\},
\end{align}
where $k_BT$ is thermal energy and $a$ a molecular length scale. However, myosin motors show catch-bond behavior, such that the detachment rate initially
decreases with increasing applied force~\cite{Guo:2006ff}.

\begin{figure*}
\includegraphics[width=0.8\textwidth]{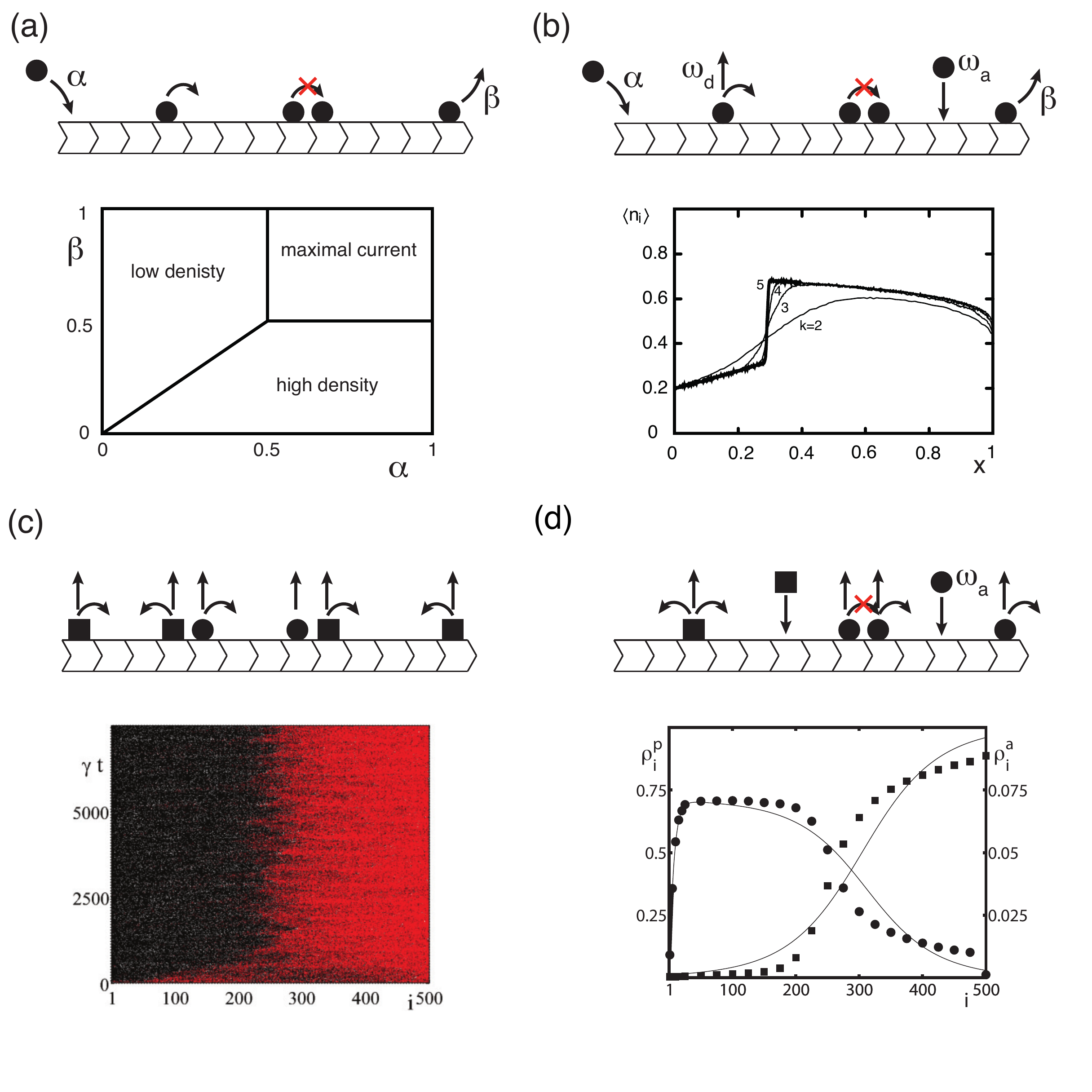}
\caption{\label{fig:latticeGases}Collective behavior of motors on a single filament. (a) Totally asymmetric exclusion process. Top: illustration of
processes. Bottom: phase diagram. (b) Totally assymmetric exclusion process with Langmuir kinetics. Top: illustration of processes. In addition to (a),
empty sites anywhere along the lattice are occupied at rate $\omega_a$ and particles can detach at rate $\omega_D$ anywhere from the lattice. Bottom:
steady state  density profiles revealing a domain boundary between low and high denisty phases. From~\cite{Parmeggiani:2003ht}. (c) Exclusion process of two
particles species. Top: illustration of processes. The dynamics of circular particles as in (b), but without additional entry rate $\alpha$ and exit rate
$\beta$ at the ends. In addition, square particles hop at the same rate to free neighboring sites, but cannot hop off at the lattice ends. For clarity,
processes have been distributed on two lattices. Bottom: Left: Space-time plot of the densities of square (red) and circle (white) particles. Right: Steady
state density profiles of square and circle particles. Form~\cite{Johann:2014ca}.}
\end{figure*}

\paragraph{Many motors on a single filament} Collective transport phenomena are commonly studied using models in which the filament is represented by a
one-dimensional lattice with $N$ sites and motors by particles that occupy the sites. Particles hop at specific rates to neighboring sites. Often steric
interactions between the motors are accounted for by an exclusion principle such that each lattice site can be occupied by one particle at most. If
particles enter the lattice at one end and leave it at the opposite end and if furthermore hops occur only into one direction, one obtains the totally
asymmetric exclusion process (TASEP)~\cite{Schutz:1993cd,Derrida:1993vu}. Depending on the rates of particle entry and exit at the boundaries, the process
displays a low density and a high density phase as well as a maximal current phase~\cite{Krug:1991zz}, Fig.~\ref{fig:latticeGases}(a). Adding attachment and
detachment of particles in the bulk (Langmuir kinetics), stable walls between high and low density domains can be formed~\cite{Parmeggiani:2003ht},
Fig.~\ref{fig:latticeGases}(b). If diffusing particles that cannot hop off the lattice et the ends are added, the two species can
segregate~\cite{Johann:2014ca,Pinkoviezky:2017gy}, Fig.~\ref{fig:latticeGases}(c). Similar segregation phenomena have been observed in mixtures of active and
passive swimmers~\cite{McCandlish:2012ix} and might also be relevant for the segregation of transcribed and not transcribed DNA in the cell
nucleus~\cite{Grosberg:2015vs}.

\paragraph{Length regulation involving molecular motors} Some molecular motors are capable of removing subunits at filament
ends~\cite{Desai:1999uv,Hunter:2003uy,Varga:2006era}. For example, Kinesin 8 attaches anywhere along a microtubule and then moves towards its plus end. In
this way, a concentration gradient of motors along the microtubule is established that leads to an effective length-dependent depolymerization rate at the
plus end~\cite{Varga:2006era,Varga:2009jy,Reese:2011bz}, which can be exploited to regulate the length of
microtubules~\cite{Johann:2012kk,Melbinger:2012hp}. For actin cables a similar mechanism was proposed that, in contrast, depended on a length-dependent
assembly rate induced by molecular motors~\cite{Mohapatra:2015cc}. In this way, molecular motors provide specific realizations of a general strategy to
regulate the length of cytoskeletal filaments~\cite{Mohapatra:2016kq}. 

Molecular motors are also involved in axonal length sensing and regulation~\cite{Rishal:2012jg}. However, the mechanism does not rely on gradients.
Instead, it has been proposed that motors moving towards the axon tip transport a promoter of axon extension that at the same time elicits the transport
into the opposite direction of a factor that inhibits further transport of the extension promoter. In a simple way, this process can be captured by a
system of coupled delayed differential equations for the concentration $u_E$ of the extension promoter at the axon tip and $u_I$ of the
inhibitor~\cite{Karamched:2015ha}:
\begin{align}
\frac{d}{dt}u_E(t) &= I_0 - \gamma_Eu_E(t) - W_If(u_I(t-\tau))\\
\frac{d}{dt}u_I(t) &=  -\gamma_Iu_I(t) + W_Ef(u_E(t-\tau)),
\end{align}
where $f$ is a sigmoidal function and $\tau$ is the time motors of velocity $v$ need to traverse the axon of length $L$, $\tau=L/v$. This system can
generate an oscillation in $u_E$ with a length-dependent period. This oscillating concentration of $u_E$ can be transformed into a signal with a period-
and thus length-dependent mean concentration allowing a nerve cell to sense and regulate its axon's length via a threshold
mechanism~\cite{Bressloff:2015gg}.

\paragraph{Bidirectional motion} For directional transport in cells, vesicles are typically bound to several motors. In this case, the motors are
mechanically coupled to each other, because pulling of one motor exerts a force on the other motors and thereby changes their velocities as well as their
persistences. If a vesicle is bound to motors of opposite directionality, bidirectional transport can occur~\cite{Grill:2005di,Mueller:2008bt}. Indeed, in
a situation, when a vesicle is stalled due to motors of acting antagonistically, a fluctuation by which one motor detaches leaves the other motors of the
same directionality carrying a higher load. This increases their detachment rates and thus starts an avalanche of detachment events of motors of one
species. The vesicle will thus move into one direction until again a stalled situation occurs. At the end of the stall period, the vesicle can move in
either direction. Whether this "tug-of-war" mechanism underlies cellular bidirectional transport is still debated~\cite{Klein:2014ej}.

Bidirectional transport can also be observed for a single motor type and in absence of load-dependent detachment
rates~\cite{julicher:1995je,Badoual:2002ky}. Experimentally, such bidirectional motion has been observed in gliding assays, where filaments move on a
substrate covered with motors~\cite{Riveline:1998uh}, for non-directional motors~\cite{Endow:2000bk}, and for non-polar bundles of actin filaments moving
on a carpet of myosin motors~\cite{Gilboa:2009vt}. The rate of switching between the two directions of motion was calculated along the lines presented in Sect.~\ref{sec:TransitionStateTheory}~\cite{Guerin:2011ed}.

\paragraph{Spontaneous motor oscillations}

In addition to the chemical and genetic oscillators discussed above, cells can also present mechanical oscillations that are not the result of an underlying chemical pulse generator. Many cells possess filamentous protrusions called cilia or flagella that periodically change their shape or produce traveling waves. These include spermatozoa and algae that have one or two flagella as well as \textit{paramecium} that is covered by a dense carpet of cilia~\cite{Bray2000Book}. Their periodic deformations propel these cells in fluid environments. Strikingly, the main constituents of the appendages, microtubules and associated molecular motors, have found in reconstitution experiments to spontaneously produce very similar patterns~\cite{Sanchez:2011jn}.

J\"ulicher and Prost found early that ensembles of molecular motors coupled to an elastic element can
spontaneously develop oscillations~\cite{Julicher:1997ua}. They studied the case, where motors are rigidly bound to a common backbone and switch between
two internal states. An alternative possibility is that the motors detach from the filament in a force-dependent manner~\cite{Grill:2005di}. The
oscillatory regimes are distinct as was shown in a model for "soft" motors that comprises both cases~\cite{Guerin:2010hr}. Spontaneous motor oscillations
have been found in muscle sarcomeres~\cite{Yasuda:1996ta,Guenther:2007cy,Sato:2013hf} and could be essential for the beating of eukaryotic
flagella~\cite{Camalet:2000uh}.

%

\subsubsection{Filament networks}

\paragraph{Reconstituted filament networks} Contraction of muscle sarcomeres relies on a crystal arrangement of the actin filaments that interdigitate with
myosin filaments. Upon activation of the motors, the actin filaments are drawn inwards, which results in contraction. Also disordered cytoskeletal networks
can generate net contractile stresses~\cite{SzentGyorgyi:1951vx}. The dynamics of this process has been studied in reconstituted systems of filaments and
motors~\cite{Backouche:2006ig,Smith:2007gc,Bendix:2008im,Foster:2015gf,Linsmeier:2016ff,Schuppler:2016gd}. In a disordered network, contraction starts at
the boundaries of a gel slab and then propagates into the gel's interior.

Reconstituted networks have also been shown to self-organize into asters and vortices~\cite{Nedelec:1997uh} and spindle-like
assemblies~\cite{Surrey:2001vb}. Minimal networks of two microtubules, molecular motors, and passive crosslinkers, showed that such molecules provide a
minimal module for generating stable overlaps between antiparallel filaments as observed in spindle midzones~\cite{Bieling:2010im,Lansky:2015fe}.
Mircotubule organizing centers have been found to position themselves in microfluidic chambers either through
polymerization-depolymerization~\cite{Holy:1997uq,FaivreMoskalenko:2002ii} or through molecular motors at the chamber walls pulling on the
filaments~\cite{Laan:2012bn}.

A somewhat different aim is followed in gliding assays, where filaments at high density move on substrates covered with
motors~\cite{Butt:2010kx,Schaller:2010cq} and which are more akin to self-propelled particles.

For investigating motor-filament systems, different theoretical approaches have been developed. Stochastic simulations aim at accounting for the various
constituents individually~\cite{Nedelec:2007bn,Dasanayake:2013io}. Kinetic descriptions take a mean-field approach and describe the system state in terms
of densities of the various components~\cite{Kruse:2000wla,Liverpool:2003gb,Kruse:2003vr}. Finally, phenomenological descriptions mostly neglect molecular
properties and focus on symmetries and conservations laws~\cite{Kruse:2004il,Kruse:2005fy}.

\paragraph{Kinetic descriptions of filament networks} In the kinetic approach and in the limit of a purely viscous system, filaments are typically assumed to be rigid rods. The
state of the filament network can be captured by the density $c$ of filament plus-ends. This density depends on the position $\mathbf{r}$ of the plus-end,
the orientation $\mathbf{\hat u}$ of the filament with $\mathbf{\hat u}$ pointing form the plus- to the minus end and $\mathbf{\hat u}^2=1$, and the
filament length $\ell$. For the motors it is often appropriate to distinguish between the densities $m_b$ of motors bound to filaments and $m_u$ of unbound
motors. The time evolution of the density is then given by a continuity equation
\begin{align}
\label{eq:kineticGeneral}
\partial_tc + \nabla\cdot\mathbf{j}_\text{trans}+\mathbf{\hat u}\times\nabla_\mathbf{\hat u}\cdot\mathbf{j}_\text{rot}+\partial_\ell j_\ell &= S\\
\partial_tm_b+ \nabla\cdot\mathbf{j}_\text{mot} & = R\\
\partial_tm_u - D\Delta m_u & = -R.
\end{align}
Here, $\mathbf{j}_\text{trans}$ is a translational current that is due to the action of molecular motors and filament assembly at the plus end,
$\mathbf{j}_\text{rot}$ a rotational current that accounts for changes in filament orientation, and $j_\ell$ a current describing the net effect of
filament assembly on the filament length. The source term $S$ accounts for filament degradation, whereas filament nucleation is captured by a boundary
condition on $j_\ell$ at $\ell=0$. The current $j_\text{mot}$ describes the flux of bound motors, $D$ is the diffusion constant of unbound motors, and the
source term $R$ describes the binding and unbinding dynamics.

It can be useful to distinguish between different filament populations, for example, to account for microtubules with shrinking and growing plus ends or
for kinetic differences between filaments with capping proteins bound or not. In the simplest case filaments form a bundle and have a fixed length. If the
bundle is aligned with the $x$-axis, Eq.~(\ref{eq:kineticGeneral}) can be written as
\begin{align}
\partial_t c(x,\pm\textbf{e}_x) & = -\partial_x j^\pm.
\end{align}

In the viscous limit, attention is typically restricted to motor-mediated interactions between filament pairs. In the case of a bundle as above, the
current takes the form~\cite{Kruse:2003vr}
\begin{align}
j^\pm(x)&=\!\!\int_{-\ell}^\ell\!\!\! d\xi\left[v^{\pm\pm}(\xi)c^\pm(x+\xi)+v^{\pm\mp}(\xi)c^\mp(x+\xi)\right]c^\pm(x),
\end{align}
where the motor-induced sliding velocities obey $v^{\pm\pm}(\xi)=-v^{\pm\pm}(-\xi)$ and $v^{+-}(\xi)=-v^{-+}(-\xi)$ to respect momentum conservation. The
motor density is in this case assumed to be homogenous. A similar form of the interaction kernel can be obtained from analogies with collision terms used
for granular materials~\cite{Aranson:2005ky}. -- Kinetic approaches have also been used to describe (viso)elastic motor-filament
systems~\cite{Guenther:2007cy,Peter:2008ec,Lenz:2012ki}

The kinetic approach has been used to study motor-induced contraction of filament bundles~\cite{Kruse:2000wla} and the stability of isotropic filament
solutions~\cite{Liverpool:2003gb}. Instabilities rely in both cases on interactions between filaments with orientations $\mathbf{\hat u}_1$ and
$\mathbf{\hat u}_2$ such that
\begin{align}
\mathbf{\hat u}_1\cdot \mathbf{\hat u}_2>0.
\end{align}
On a molecular level, such interactions rely on end effects, for example, on motors getting stuck at a filament end~\cite{Nedelec:1997uh} and also lead to
the generation of net mechanical stresses in isotropic networks.

\begin{figure*}
\includegraphics[width=0.9\textwidth]{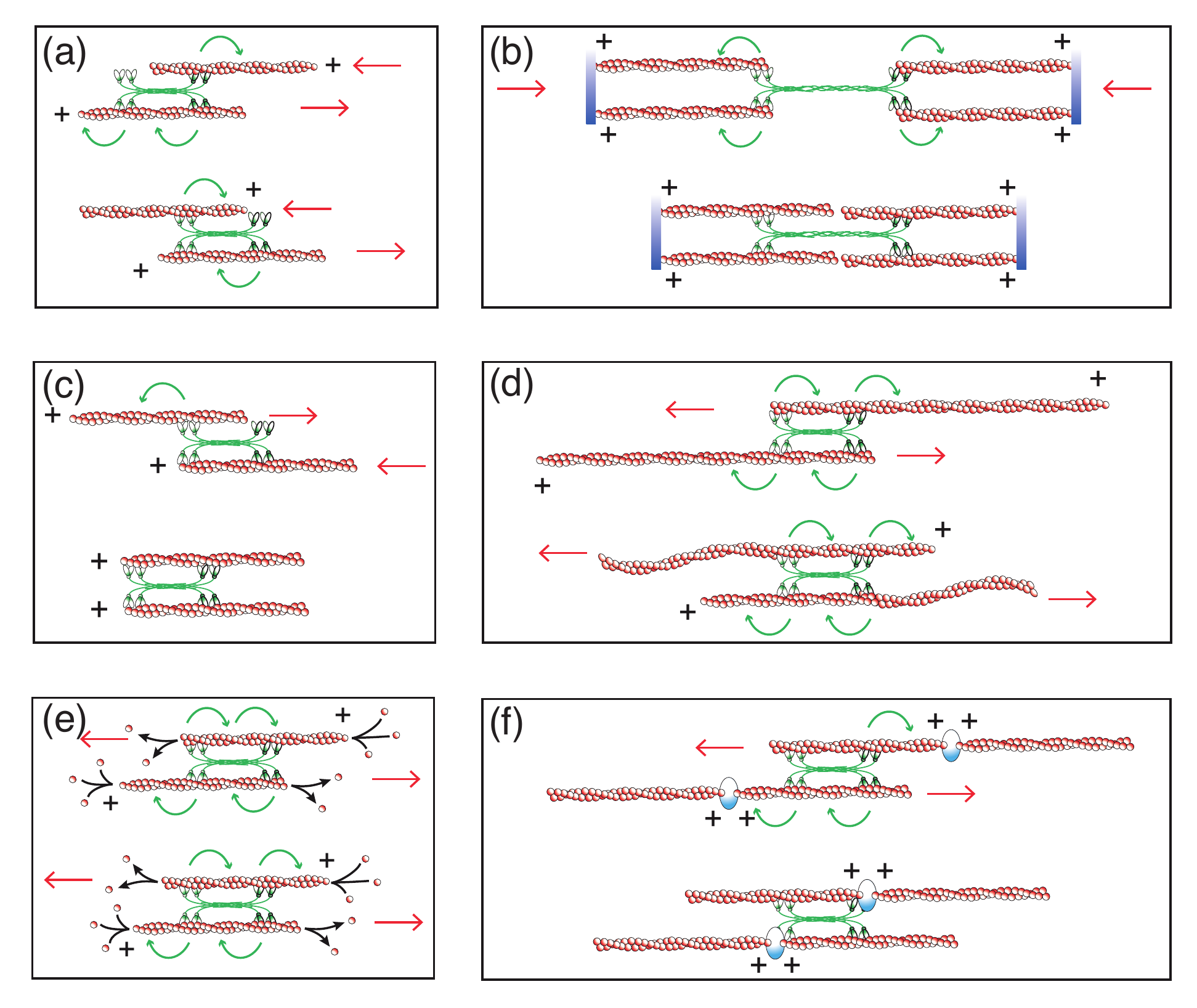}
\caption{\label{fig:contraction}Possible mechanism of net stress generation for rigid filament pairs. (a) Antiparallel filaments first contract then expand
implying no net stress. (b) Arrangements of actin filaments and motors in a muscle sarcomere leading to contractions only. (c) If motors stall at the
filament ends, net stresses are generated in parallel filament pairs. (d) The buckling of filaments can break the symmetry between contraction and extension
of antiparallel filaments. (e) Filament treadmilling leads to to extended times of contractile overlaps. If these times exceed the time a motor is bound,
net stresses are generated. (f) Biopolar structures generate a structure that is reminiscent of the filament arrangement in a sarcomere.}
\end{figure*}

\paragraph{Mechanisms of stress generation} In the kinematic framework, mechanical stresses can be calculated by analyzing the momentum
flux~\cite{Kruse:2003vr}. In a homogenous filament bundle, antiparallel filaments do not generate a net contractile stress as there are as many contraction
as extension events, Fig.~\ref{fig:contraction}(a). In skeletal muscle this problem is solved by the sarcomeric arrangement of the actin filaments,
Fig.~\ref{fig:contraction}(b). If motors stall at the filament ends, then interactions between parallel filaments generate a net stress,
Fig.~\ref{fig:contraction}(c). Another mechanism depends on nonlinear filament elasticity: as filaments buckle more easily than they are stretched,
contraction is favoured over extension generating a net contractile stress~\cite{Lenz:2012df}, Fig.~\ref{fig:contraction}(d), and filament buckling has
indeed been observed in reconstituted actomyosin bundles~\cite{Thoresen:2011bv}. Other mechanisms can be envisioned~\cite{Lenz:2014ic}: myosin motors are
not point-like but form themselves minifilaments. If they prefer to be aligned with the filaments they will generate net contractile stresses as has been
found in stochastic simulations~\cite{Dasanayake:2011iq,Dasanayake:2013io}. In similar simulations, it was found that motors can also generate net
extensile stresses, because motors are more likely to link two filaments that have a finite overlap and are so persistent as to stay bound until they fully
extend a filament pair~\cite{Gao:2015gn}. Finally, filament treadmilling can contribute to the generation of filament overlaps that favour
contraction~\cite{Oelz:2015bq}, Fig.~\ref{fig:contraction}(e), even in absence of active crosslinkers~\cite{Zumdieck:2007id}.

The above mechanisms rely essentially on accounting for local differences in the motor distributions. In addition, the local organization of the filament
network has a significant impact on contractility~\cite{Ennomani:2016jn}. In bundles, filaments could form bipolar structures, Fig.~\ref{fig:contraction}(f),
for example, through motors linking filament plus ends, which would lead to contraction as was proposed for contractile rings~\cite{Wollrab:2016hr}. In
higher dimensions these structures correspond to asters.

\subsubsection{Hydrodynamics of motor-filament networks}

\paragraph{Hydrodynamics of active polar gels} Many interesting features of motor-filament networks can be discussed without referring to a specific
molecular mechanism of stress generation by using the formalism of generalized hydrodynamics outlined in Sec.~\ref{sec:generalizedHydrodynamics}~\cite{deGroot:1985ue}. Within this approach, the
cytoskeleton is one instance of an active polar gel~\cite{Kruse:2004il,Kruse:2005fy}. In this context, "active" refers to the coupling of mechanical
stresses to a chemical reaction (ATP hydrolysis), although the class of active matter is somewhat broader~\cite{Marchetti:2013bp}. Active polar gels are
defined on one hand by conserved quantities, namely, the gel components, momentum, and angular momentum. Note, that one typically assumes the system to be
at constant temperature, such that energy is not conserved. On the other hand, polar filaments can locally align to generate macroscopic polar order, such
that polarity provides an order parameter of a broken continuous symmetry.

The fluxes appearing in the conservation laws could be obtained form microscopic descriptions as the kinetic theories presented above. In the framework of
nonequilibrium thermodynamics, however, one uses phenomenological expressions. To this end, one first identifies the pairs of conjugated generalized
(thermodynamic) currents and generalized (thermodynamic) forces. An expression for the currents is then obtained by expanding them in terms of the forces
up to linear order.

In the simplest version, an active gel can be described as an effective one-component fluid that is coupled to the hydrolysis of ATP. If the densities
$n_T$, $n_D$, and $n_P$ of ATP and its hydrolysis products ADP and P$_i$, respectively, are spatially homogenous, then the conservation laws
are~\cite{Kruse:2004il,Kruse:2005fy}
\begin{align}
\partial_t\rho +\partial_\alpha \rho v_\alpha &=0\\
\partial_t g_\alpha + \partial_\beta\sigma^\mathrm{tot}_{\alpha\beta} &= f^\mathrm{ext}_\alpha\\
\dot n_\text{T} = - \dot n_\text{D} = - \dot n_\text{P} &= r.
\end{align}
Here, $\mathbf{g}=\rho\mathbf{v}$ denotes the momentum density and $\mathsf{\sigma}^\mathrm{tot}$ is the corresponding momentum flux density. It is equal
to the mechanical stress tensor. Externally applied forces $\mathbf{f}^\mathrm{ext}$ provide possible momentum sources. Angular momentum has not been
listed explicitly, which is appropriate in absence of external torques that do not result from the external forces. In addition, there is a dynamic
equation for the evolution of the polarization field $\mathbf{p}$. The part $F_d$ of the free energy associated with distortions in the polar field is
commonly taken to be
\begin{align}
F_d = &\frac{1}{2}\int d^3\mathbf{r}\left\{K_1\left(\nabla\cdot\mathbf{p}\right)^2 + K_2\left[\mathbf{p}\cdot\left(\nabla\times\mathbf{p}\right)
\right]^2+\right.\\
&\left.K_3\left[\mathbf{p}\times\left(\nabla\times\mathbf{p}\right)\right]^2\right\}.
\end{align}
Here, $K_1$, $K_2$, and $K_3$ are the Frank elastic constants for splay, twist, and bend, respectively.

The conjugated pairs of forces and fluxes are then $v_{\alpha\beta}\leftrightarrow\sigma^d_{\alpha\beta}$, $p_\alpha\leftrightarrow h_\alpha$, and
$r\leftrightarrow\Delta\mu$, where $v_{\alpha\beta}=\left(\partial_\alpha v_\beta+\partial_\beta v_\alpha\right)/2$ are the components of the rate of
strain tensor, $\mathsf{\sigma}^d$ is the deviatory stress, $\mathbf{h}=-\delta F_d/\delta\mathbf{p}$ the molecular field conjugate to the polarization,
and $\Delta\mu$ the the difference between the chemical potentials of ATP and its hydrolysis products.

The total stress has a hydrostatic part $\mathsf{\sigma}^e$ -- the Erickson stress, which is the generalization of the hydrostatic pressure to the case of
a polar fluid~\cite{Ericksen:1962de} -- and a deviatory part, such that $\sigma^\text{tot}_{\alpha\beta}= \sigma^d_{\alpha\beta}-\frac{1}{2}\left(p_\alpha
h_\beta-p_\beta h_\alpha\right)+\sigma^e_{\alpha\beta}$, where $\mathsf{\sigma}^d$ is the symmetric part of the deviatory stress and
$-\frac{1}{2}\left(p_\alpha h_\beta-p_\beta h_\alpha\right)$ its antisymmetric part. The constitutive equation for the symmetric part of the deviatory
stress can be divided in three components:
\begin{align}
\mathsf{\sigma}^d_{\alpha\beta} &= \mathsf{\sigma}^\text{visc}+\mathsf{\sigma}^\text{dist}+\mathsf{\sigma}^\text{act}.
\end{align}
Here, the viscous stress $\mathsf{\sigma}^\text{visc}$ is that of a Stokesian fluid $\sigma^\text{visc}_{\alpha\beta}=2\eta v_{\alpha\beta}$ and
$\mathsf{\sigma}^\text{dist}$ the stress resulting from distortions in the polar field,
\begin{align}
\sigma^\text{dist}_{\alpha\beta} &= \frac{\nu}{2}\left(p_\alpha h_\beta+p_\beta h_\alpha\right) + \bar\nu_1p_\gamma p_\gamma\delta_{\alpha\beta}.
\end{align}
The expression for $\mathsf{\sigma}^\text{dist}$ is the same as for a nematic liquid crystal, but with the polarization replaced by the director field.
Finally, the stress component $\mathsf{\sigma}^\text{act}$ resulting from activity is
\begin{align}
-\sigma^\text{act}_{\alpha\beta} &= p_\alpha p_\beta\zeta\Delta\mu +\bar\zeta\Delta\mu \delta_{\alpha\beta}+ p_\gamma
p_\gamma\tilde\zeta\Delta\mu\delta_{\alpha\beta}.
\end{align}
The equation for the evolution of the polarization vector is
\begin{align}
\frac{D}{Dt}p_\alpha &= \frac{1}{\gamma}h_\alpha + \lambda_1 p_\alpha\Delta\mu-\nu_1p_\beta v_{\alpha\beta}-\bar\nu_1 v_{\beta\beta},
\end{align}
which contains the active term $\lambda_1p_\alpha\Delta\mu$. Here, the coefficients $\nu_1$ and $\bar\nu_1$ are imposed by the Onsager reciprocity
relations. This hydrodynamic approach to active gels has been generalized to elastic~\cite{Banerjee:2011ks} and to multicomponent viscoelastic active polar
gels~\cite{Joanny:2007ci,Guenther:2007cy,CallanJones:2011vx}.

\begin{figure}
\includegraphics[width=0.5\textwidth]{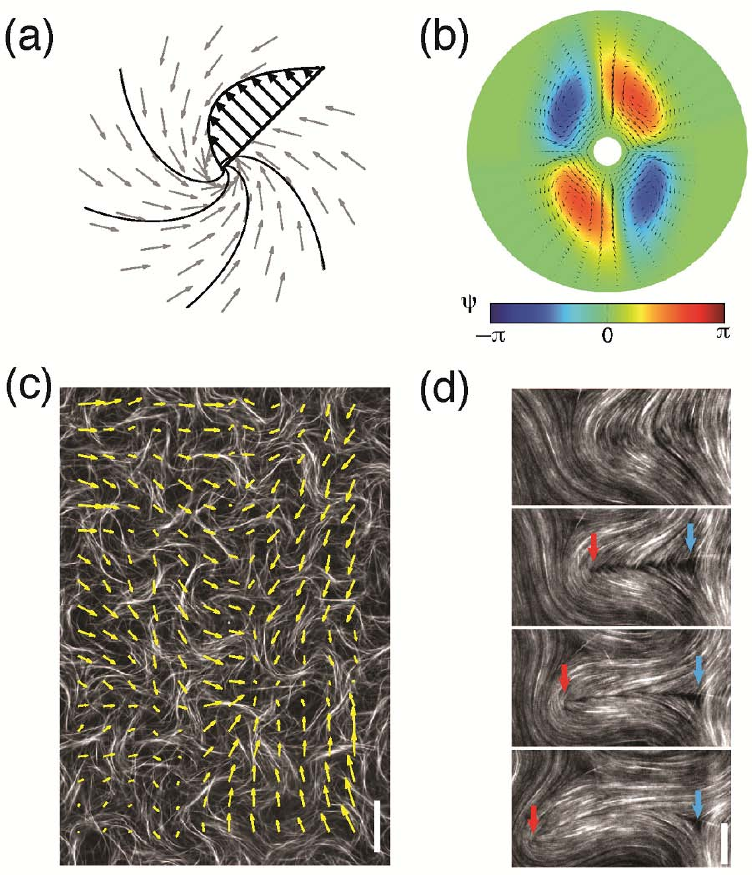}
\caption{\label{fig:flows}Spontaneous flows in active polar gels. (a) Illustration of the spontaneous circular flow around a spiral point defect.
From~\cite{Kruse:2004il}. (b) Flows around vortices in a ring. Arrows indicate the flow, colours the orientation angle $\psi$ of the polarization field with
respect to the radial direction. From~\cite{Neef:2014ez}. (c) Spontaneous flow field (arrows) in a reconstituted system of microtubules and motor complexes.
Scale bar: 80~$\mu$m. (d) Generation and separation of two topological point defects of charges $\pm1/2$ in the system (c). Scale bar: 20~$\mu$m. (c) and
(d) from~\cite{Sanchez:2012gt}. }
\end{figure}

\paragraph{Spontaneous flows} One of the most spectacular properties of polar active gels is their ability to spontaneously generate flows. Beyond a
critical activity, a stationary state with homogenous polar order will be unstable against small perturbations~\cite{Voituriez:2007jy}. In this case,
gradients in the polar order parameter will develop, which in turn lead to a spontaneous flow. In a Taylor-Couette geometry, where the active gel is
confined in the interstitial space between two coaxial cylinders, this flow can set the two cylinders into relative rotational
motion~\cite{Furthauer:2012iu}. The resulting torque-rotational velocity relation can display regions with multiple unstable branches and coexistence of
states with rotations in opposite direction. For higher activities, secondary instabilities have been reported that lead to the emergence of topological
point defects and possibly chaotic behavior~\cite{Neef:2014ez}, Fig.~\ref{fig:flows}(b), that has been observed in extensile active
nematics~\cite{Sanchez:2012gt}, Fig.~\ref{fig:flows}(c). These phenomena are reminiscent of the dynamics of bacterial suspensions confined in circular
domains~\cite{Wioland:2013jm,Lushi:2014fn}. Gradients in the polar order parameter also generate flows around spiral defects~\cite{Kruse:2004il},
Fig.~\ref{fig:flows}(a).

\paragraph{Topological defects} Topological defects in the polarization field readily develop in reconstituted filament-motor
systems~\cite{Sanchez:2012gt,Keber:2014fh}. Asters and vortices that are stable defects in equilibrium, give way to spirals beyond a certain activity
threshold and spontaneously start to rotate~\cite{Kruse:2004il,Kruse:2005fy}. Defects of topological charge $\pm1/2$ in active nematics form spontaneously,
Fig.~\ref{fig:flows}(d), and annihilate~\cite{Sanchez:2012gt} and can show oscillatory behavior when confined to a spherical surface~\cite{Keber:2014fh}. The
dynamics of toplogical defects in active nematics has been studied numerically~\cite{Giomi:2013ky,Thampi:2013cu} and analytically~\cite{Pismen:2013ie}.

\paragraph{Spontaneous actin waves} In a number of diverse adhering cells, the actin cytoskeleton has been found to spontaneously represent waves. On one
hand there are actin polymerization waves. Beginning with \textit{Dictyostelium discoideum}~\cite{Vicker:2000wj}, such waves have now been observed in
various cell types~\cite{Gerisch:2004im,Weiner:2007cl} and are often linked to cell motility~\cite{Allard:2013hu}. Also circular dorsal ruffles, which are
protrusions on the upper side of an adhering cell  and for which no function is currently known, result from polymerization waves~\cite{Bernitt:2015iy}.
The basic underlying mechanism seems to be a negative feedback between actin filaments and the activity of a nucleation promoting
factor~\cite{Weiner:2007cl,Doubrovinski:2008fa,Carlsson:2010uj}, which leads to a dynamics that is reminiscent of excitable
systems~\cite{Whitelam:2009dt,Ryan:2012bq,Bernitt:2017hs}.

On the other hand, there are spontaneous actin waves that depend in an essential way on stresses generated by molecular
motors~\cite{Giannone:2004vu,Doebereiner:2006cr,Barnhart:2011if}. A generic mechanism for such waves that result from the coupling of a regulator to an
active gel has been studied in~\cite{Bois:2011kx,Kumar:2014fx}. Similarly, they are generic properties of active elastic materials with stress-dependent
motor regulation~\cite{Guenther:2007cy,Radszuweit:2013ku} or in presence of turnover~\cite{Dierkes:2014kl}. In combination with filament treadmilling,
motors can also generate waves in actomyosin bundles~\cite{Torres:2010dp,Oelz:2016gt,Wollrab:2016hr} as were observed in cytokinetic rings in fission
yeast~\cite{Wollrab:2016hr}. Lateral waves observed during the spreading of fibroblasts~\cite{Giannone:2004vu,Doebereiner:2006cr} have been proposed to
result from cytoskeleton-membrane interactions~\cite{Shlomovitz:2007kc,Zimmermann:2010iy,Gholami:2012ky}.

\paragraph{Cell migration} Roughly, the crawling of cells on solid substrates comprises extension of a flat veillike protrusion called lamella and its
anchoring to the substrate, release of surface attachments in the cell's rear, and forward locomotion of the cell body containing the nucleus. Physical
analysis focussed on one hand on specific aspects of this integrated process. Notably, the extension of lamellae by actin polymerization was described in
terms of a "Brownian ratchet"~\cite{Mogilner:1996ip} or through blebbing, a process that leads to the detachment of the cell membrane from the actin
network and its bulging out due cytosolic pressure~\cite{Charras:2008gf}.  On this basis, the force-velocity relation of cell-crawling has been
analyzed~\cite{Carlsson:2003jo,Weichsel:2010dl,Schreiber:2010fy}. In the frame of the substrate the actin flows from the leading edge towards the cell
body, a process called retrograde flow. Hydrodynamic analysis of this phenomenon suggests that it results form contractile activity of myosin
motors~\cite{Kruse:2006gq}. The combined effect of polymerization and contraction on the force-velocity relation was studied in~\cite{Recho:2013fb}.

On the other hand, the experimental observation of crawling cell fragments~\cite{Euteneuer:1984ut,Malawista:1987wx} suggested that the actin cytoskeleton
can autonomously generate cell migration. As a consequence, droplets of active gels were analyzed. In this context, one has to deal with a dynamic
boundary. Studies with a sharp boundary therefore focused on cytoskeletal dynamics in regions of fixed shape~\cite{AWhitfield:2014in} or on the stability
of circular shapes~\cite{CallanJones:2008ue}, although also full crawling was considered~\cite{Doubrovinski:2011wg}. Alternatively, phase-field models have
been used~\cite{Shao:2010uk,Ziebert:2012ij}.

A phase field is an auxiliary field that equals one in the cell interior and zero outside. The dynamics of the phase field $\psi$ is given
by~\cite{Shao:2010uk,Ziebert:2012ij}
\begin{align}
\partial_t\psi &= D_\psi\Delta\psi + \frac{\delta F}{\delta \psi} + \text{coupling~terms}
\end{align}
The diffusion terms notably sets the surface tension associated with the phase-field boundary. The free energy $F$ is commonly taken to be quartic in the
phase field such that $\delta F/\delta \psi=\kappa\psi\left(1-\psi\right)\left(\psi-\delta\right)$. Here $\delta$ can be dynamically adjusted to maintain a
constant cell volume $\delta=\frac{1}{2}+\epsilon\left(d^3\mathbf{r}\psi\left(\mathbf{r}\right)-V_0\right)$, where $v_0$ is the target volume
size~\cite{Ziebert:2012ij}. The coupling terms describe the interaction of the phase field with the actin cytoskeleton. A common choice is
$-\beta\mathbf{p}\cdot\nabla\psi$ that confines the interaction to the ``cell'' boundary.

The phase-field approach has been used to study spontaneous cell polarization~\cite{Shao:2010uk,Ziebert:2012ij}, the effects of
adhesion~\cite{Shao:2012jb,Ziebert:2013eg}, the migration of cells on micropatterns~\cite{Camley:2013un}, and the effects of substrate stiffness on
migration~\cite{Ziebert:2013eg,Lober:2013ti}. In these studies, the contractile stresses generated by the actin network play an essential role.
Alternatively, spontaneous polymerization waves can orchestrate the cytoskeleton to generate cell crawling~\cite{Weiner:2007cl,Doubrovinski:2011wg}. These
waves can generate erratic motion through a deterministic mechanism~\cite{Dreher:2014gr}.

As an alternative to adhesion-based motility, cells can also move by ``flowing and squeezing''~\cite{Laemmermann:2008ib}. In this case, the cell does not
establish adhesion sites with a substrate. The necessary coupling to the environment can be obtained from pushing on the environment~\cite{Hawkins:2009kz}.
Furthermore, spontaneous actin flows either in the bulk~\cite{Tjhung:2012ej,Recho:2013hq,AWhitfield:2014in} or below the cell
membrane~\cite{Hawkins:2011kl} can generate migration. Similarly, asymmetric contraction of the cytoskeletal network can push the cytosol to extend the
leading edge~\cite{CallanJones:2013dm}, which is similar to blebbing.

\paragraph{Cortex instabilities} In animal cells, the actin cytoskeleton forms a thin layer below the cell membrane, which is known as the actin cortex. It
determines the cellular shape and mechanical properties~\cite{Chalut:2016el}, hosts vital structures like contractile rings that, for example, cleave the
cell during division, and plays an important part in endocytosis as well as in the formation of cellular protrusions. Its physical properties have been
probed by atomic force microscopy~\cite{Matzke:2001jp,Pesen:2005hq} and laser ablation~\cite{Saha:2016ie} The cortex thickness has been measured to be of
about 200nm~\cite{Clark:2013ef}, which is much smaller than the length of many cortical actin filaments~\cite{Fritzsche:2016fx}. By making an analogy with
pre-wetting, it was suggested that myosin motors are eventually at the origin of the well-defined actin cortex~\cite{Joanny:2013fm}.

Gradients and anisotropies in tension generate flows of cortical actin~\cite{Mayer:2010kt}. These flows can lead to formation of contractile
rings~\cite{Salbreux:2009fp,Turlier:2014hq} as well as alignment of the filaments in the ring~\cite{Reymann:2016kp}. Asymmetries in cortical tension make
the position of the contractile ring unstable and can lead to ring oscillations~\cite{Sedzinski:2011ef}. Oscillations have also been observed for
actomyosin rings~\cite{Paluch:2005da} and of cellular shapes~\cite{Salbreux:2007fc} in nondividing cells. A framework for analyzing cortex-driven shape
changes has been formulated for active elastic thin shells~\cite{Berthoumieux:2014eo}. In general, however, motion of actomyosin rings depends on ring
contractility but also cortical flows~\cite{Behrndt:2012gy}. Furthermore, cortical flows help to establish cell polarity by transporting certain proteins
from one cell end to the other~\cite{Goehring:2011bi}. Remarkably, the actomyosin cortex in \textit{Caenorhabditis elegans} embryos generates chiral
torques, which generates counterrotating cortical flows that are used to establish the left-right symmetry of the developing
organism~\cite{Naganathan:2014fc}. Finally, let us note that the contractility of thin active gel layers, be it the actin cortex or a cell monolayer, can generate instabilities that lead to three dimensional shape changes~\cite{Hannezo:2011ko,Shyer:2017kt,Ideses:2018dn}.


\section{Neural networks and brain function}

Understanding the brain function is a grand goal for biology. The brain consists of a tightly connected network of billions of nerve cells (called neurons)~\cite{Abbott2001Book}. From a physical point of view, a nerve cell or an individual neuron is an electrically excitable unit. For a strong enough excitation above a threshold, the cell sends out an electrochemical pulse referred to as an action potential. This signal travels along the axon, which is a linear extension of the nerve cell. The transport of an action potential along an axon is quantitatively described by the Hodgkin-Huxley model~\cite{Hodgkin}, the essence of which is captured by the FitzHugh-Nagumo model~\cite{FitzHugh,Nagumo}. At the end of the axon, the action potential can be transmitted to other nerve cells that receive the signal in extensions called dendrites. The coupling between an axon and dendrites occurs through synapses that chemically excite or deexcite the postsynaptic nerve cell by releasing biomolecules called neurotransmitters.

The activity of the brain is represented by the electrical activity of the whole neural network rather than that of individual neurons. The brain function closely associated with the activity is then determined by the underlying neural networks \cite{Abbott2008-489,Hopfield1982-2554}. Therefore, one needs to explore the underlying neural network dynamics for specific functions. The physical and quantitative under- standing of global brain functions such as learning and memory, decision making, and attention, as well as associated nonequilibrium neural network dynamics, are still challenging at present \cite{Abbott2001Book}. We will quantify the nonequilibrium landscape and flux and associated nonequili- brium thermodynamics to explore these brain functions. For reviews of other aspects of dynamics of neural networks, see~\cite{Abbott2001Book, Amit}.

\subsection{Learning and memory}

To theoretically study cerebral processes, certain neural networks have been introduced where in neuron dynamics can be simplified. The state of neuron $i$ is given by the continuous variable $u_i$ representing its electrical potential and connects to another neuron $j$ with strength $T_{ij}$. The state of neuron $i$ changes due to the input from other neurons, a leak of ions that will bring an unstimulated neuron back to its resting potential $u_i=0$ and a possible external input current $I_i$, such that
\begin{align}
\label{eq:NeuralNetworkEq}
C_i\frac{du_i}{dt} &= \sum_{j\neq i}T_{ij}f_j(u_j)-\frac{1}{R_i}+I_i,
\end{align}
where $C_i$ and $R_i$, respectively, denote the capacity and resistance of neuron $i$ and where $f_j$ is a monotonically increasing sigmoidal function such
that neuron $i$ receives input from neuron $j$ only if the ``activity'' of the latter is above a certain threshold. An important subclass of such networks
are Hopfield networks for which the weights are symmetric, $T_{ij}=T_{ji}$ \cite{Hopfield1982-2554}.

The attractors of such a network are often interpreted as patterns stored in the network. Given a certain distribution of input currents $I_i$ the network
should settle into an attractor that associates this input with a previously learned pattern, where ``learning'' refers to adjusting the connection
strengths $T_{ij}$ such that certain inputs yield distinct activation patterns of the network. There are several different learning algorithms, that is,
dynamics for the $T_{ij}$ that yields the desired network properties. In the case of a symmetric network with $T_{ij}=T_{ji}$,
the underlying dynamics is determined by the gradient of a
potential energy \cite{Hopfield1982-2554} defined as
\begin{align}
\label{eq:energyHopfieldNetwork}
E&=-\frac{1}{2}\sum_{i,j} T_{ij}f_i(u_i)f_j(u_j)+\sum_i\frac{1}{R_i}\int_0^{u_i}\xi f'_i(\xi)d\xi\nonumber\\
&\quad\quad+\sum_i I_if_i(u_i)
\end{align}
with $dE/dt\le0$. Consequently, $E$ is a Lyapunov function of the system and it will always settle in one of the steady state attractors \cite{Hopfield1982-2554}. Symmetric connections imply an underlying neural network with behavior determined by purely potential energy. Memory is stored in the neural connection patterns forming the basins of attractions on the landscape. Learning can then be understood as a way of retrieving information from the initial queue near specific memory basins.

In the realistic neural networks of the brain, neural con- nections are not symmetric, that is, $T_{ij}\neq T_{ji}$ so the energy $E$, Eq.~(\ref{eq:energyHopfieldNetwork}), is no longer a Lyapunov function. Still, Eq.~(\ref{eq:NeuralNetworkEq}) describes the neural network dynamics, which now also depends on the nonequilibrium potential landscape related to the steady-state probability distribution and steady-state rotational curl probability flux breaking the detailed balance. The rotational curl flux of neural networks emerges when neural connections are asymmetric \cite{Yan2013PNAS}. As a consequence, continuous line attractors can emerge \cite{Yan2013PNAS}. These attractors could provide a physical origin of associations between memories by flux.

\subsection{Cycling of sleep phases}

The associations can be shown in an example connecting different
phases during sleep, where periods of rapid eye movement (REM) alternate with non-REM periods~\cite{McCarleyMassaquoi1986AmJPhysiol}. The two phases are regulated by the interaction of two neural populations. The main contribution for the underlying circuit of REM sleep is an activation-repression loop inferred from the experimental studies ~\cite{McCarleyMassaquoi1986AmJPhysiol}. The dynamics follows the equation of motion:
$\frac{dx}{dt}= a A(x) x S_1(x) - b B(x) x y$ and $\frac{dy}{dt}=-c y + d x y S_2(y) $. Here x and y represent the activities of REM-on and REM-off neural population, respectively. The constants $a$, $b$, $c$, $d$ and the sigmoidal functions $A$, $B$, $S_1$, and $S_2$, respectively, specify the interaction strengths between the populations~\cite{Yan2013PNAS}.
From a 
stochastic version of the equation of motion for x and y
, one can obtain the
Mexican hat-shaped landscape as a continuous close line attractor 
from the steady state solution of the corresponding Fokker-Planck equation. The 
rotational curl flux driving the REM sleep flow can be directly derived from the force decomposition, Sect.~\ref{sec:noneqpotentials}. 
In addition, it allows one to assess the stability of the limit cycle in terms of the landscape topography, that is, the height of the Mexican hat potential's center, and the frequency 
of the REM/non-REM cycle as a function of the neurotransmitter (acetylcholine and norepinephrine)
release~\cite{Yan2013PNAS}. It is shown that the nonequilibrium rotational curl flux
from asymmetrical neural connections is crucial for generating the REM sleep rhythm while for symmetrical neural connections oscillations can not appear~\cite{Yan2013PNAS}.

\subsection{Brain decision making}
One particular aspect of cognition is decision making. Studies in monkeys have linked the process of decision making  to the neural activity on a certain part of the cerebral cortex~\cite{Shadlen1996-628,Shadlen2001-1916,Roitman2002-9475,Huk2005-10420}.
In these experiments, trained monkeys were presented, for a few seconds, a
pattern of randomly arranged dots that moved coherently in one direction against a background of randomly appearing and disappearing stationary dots.  Some time after the patterns were switched off, the monkeys were
asked to make a decision about the direction of motion of the dots. 

\paragraph{The physics of decision making}
The neural network underlying decision making in this motion discrimination task consists of two populations of nerve cells competing with each other for
selecting the leftward or rightward direction~\cite{Shadlen2001-1916,Roitman2002-9475, Wong2006-1314,Wong2007-Neural}. The two populations self-activate and at the same time mutually inhibit each other. The sensory input current $I_{\textrm{motion},i}$ into population $i=1,2$ depends on the fraction of coherently moving dots or `coherence' $c$ as $I_{\textrm{motion},i} = \hat{I}(1 \pm c)$, where $\hat{I}$ is the input in absence of coherent motion and the plus and minus signs are respectively used, when the motion is in the preferred direction of the population or opposite to it. The total input currents are then given by
\begin{align}
I_{\textrm{tot},1} &= J_{11}S_1 - J_{12}S_2 + I_0 + I_{\textrm{motion},1}\\
I_{\textrm{tot},2} &= -J_{21}S_1 + J_{22}S_2 + I_0 + I_{\textrm{motion},2}.
\end{align}
Here, $I_0$ is the average background synaptic input, $J_{ij}$ the synaptic coupling constants, and $S_i$ the average gating variables. Their value is determined dynamically through
\begin{align}
\dot{S}_i &= -\frac{1}{\tau_S}S_i + \gamma(1-S_i)r_i
\end{align}
with characteristic times $\tau_S$ and $\gamma$ and where $r_i$ is the firing rate of the neural population $i$. It is essentially zero below a threshold value and then increases almost linearly as a function of the total input current~\cite{Shadlen2001-1916,Roitman2002-9475, Wong2006-1314,Wong2007-Neural}.

\begin{figure*}[!ht]
\includegraphics[width=1.0\textwidth]{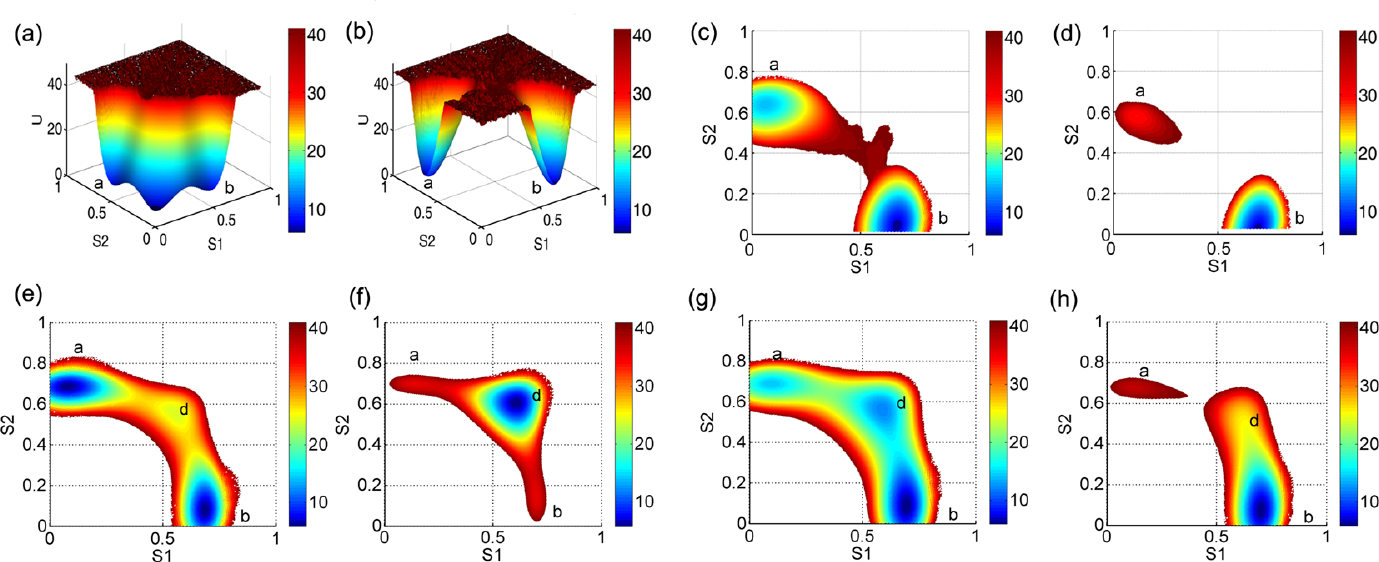}
\caption{ Nonequilibrium potential landscapes for brain decision making and mind changes (a)(b) Potential landscapes at the zero coherence level for different stimulus inputs for decision making, $\mu_0  =$ 0 and 30 Hz,
respectively.
c denotes undecided state, and a and b denote decided states. (c)(d)
Potential landscapes at the non-zero coherence level, where $\mu=
 = $ 30 Hz and the coherence $c= $ 0.2 and 0.65, respectively.
 (e)(f) Potential landscapes with different large inputs for mind changes.
at zero coherence level. The strength of stimulus $\mu_0
 = 50$ and $65$ Hz respectively. (g)(h) The potential
 landscapes with different large inputs for mind changes.
at different coherence levels.  ($\mu_0= 55$) when coherence $c'= 0.02$ and $0.12$, respectively. In all subgraphs, parameters $a = 269.5, b = 108$, and
$D = 3.6 \times 10^{-7}$) (from Ref. \cite{Yan2016CPB}).} \label{Yan2016CPB_fig2}
\end{figure*}

Stochastic dynamics in the nonequilibrium landscape unveils the physical mechanism behind decision making~\cite{Yan2016CPB}, Fig.~\ref{Yan2016CPB_fig2}{a}. In the absence of a stimulus, $\hat{I}=0$, there are three stable attractors corresponding to the undecided state and the decided states. In the latter one of the populations has a high activity, otherwise all activities are low. With increasing stimulus, the stability of the undecided state decreases until it eventually becomes unstable, such that only the two decided states remain and the animal has to make a decision, Fig.~\ref{Yan2016CPB_fig2}{b}.  For non-coherent patterns, $c=0$, the decision is random, whereas for $c\neq0$ it is biased towards the correct decision, because the basin of attraction of the correct decided state grows while the that of the incorrect decided state shrinks. Furthermore, as the stimulus is reduced, the barriers around the decided states still remain for some time, which endows the systems with some memory for the decision made. Once an incorrect decision is made, it takes much longer time to go over the barrier to reach the incorrect basin.
This explains why it takes longer decision time for error trails than correct trails. These findings are in agreement with the results from monkey experiments~\cite{Roitman2002-9475, Mazurek2003-1257,Shadlen2001-1916, Wong2006-1314,Wong2007-Neural}.

\paragraph{Speed, accuracy and dissipation tradeoff in decision making}

The process of decision making can be optimized for several quantities, notably, accuracy, speed, and dissipation. One can quantify the decision time by the corresponding mean first passage time from one undecided state to the decided state. One can also quantify the performance or
accuracy of the neural network through a path integral
method~\cite{Wang2010JCP}, by defining the accuracy of
decision-making task as the ratio in probabilities between the
optimal correct path and the error path. The
dissipation in terms of entropy production is directly related to the rotational curl flux. As stated previously, this measures the system's distance from equilibrium and can therefore be quantified for decision making \cite{Yan2016CPB}.

Let us now focus on speed-accuracy-dissipation tradeoff mechanism by varying input threshold.
If speed is the main concern of decision-making, there is an optimal
decision speed with almost minimum dissipation cost and reasonable accuracy. Higher accuracy requires longer decision time (slower decision speed). If accuracy
is the major concern of the decision-making, the dissipation cost and decision time are higher than optimum at the best accuracy, while there is a suboptimal
accuracy with optimal speed and dissipation cost. When dissipation cost is the main concern for decision-making, the decision accuracy may not be the best at the
least dissipation cost, a faster speed can still be reached. One sees that reasonable accuracy performance can be reached with minimum dissipation cost and fast decision
time \cite{Yan2016CPB}.


\paragraph{Mechanisms of mind changes} Changes of mind occur often in making decisions. In this scenario, the initial choice can be altered. Decisions can be changed in two cases. For an incoherent stimulus $c=0$, the system will settle in one of the two decided states as the strength of the stimulus is increased, Fig.~\ref{Yan2016CPB_fig2}(e). Upon further increase of the stimulus, a new attractor appears with high activity in both populations and the system will eventually settle in this state, Fig.~\ref{Yan2016CPB_fig2}(f). As the stimulus is again reduced, the high activity state disappears and the system will eventually settle in one of the decided states. However, this state will not necessarily match the original one; the decision can be changed. The emergence of a new attractor with increasing input strength has been observed experimentally~\cite{Resulaj-2009}. Studies suggested that the changes of mind might be due to the unprocessed information before the first decision \cite{Resulaj-2009}.

For coherent inputs, $c>0$, the correct choice state is more attractive and changes from an initial correct choice are unlikely to occur. However, when
the network makes a wrong decision at the beginning, changes can occur relatively easily. This is because although there is a second chance (new
center basin state) to make a decision due to large stimulus input, it is still more likely to be attracted to the
stronger basin of attraction for the correct choice, as shown in Fig.~\ref{Yan2016CPB_fig2}(g,h). This is supported by experimental data showing that the probability of changes to the wrong choice from the
correct one decreases monotonically with increasing coherence~\cite{Resulaj-2009, Albantakis2011-Changes, albantakis2012multiple}. Therefore the process of changes of mind
can be described in three steps: making the initial decision, then attracted to the new basin state and at last
making a different decision. As seen, a change of mind can be
understood with the landscape topography.



\section{The genetic basis of organismal progression}
\label{sec:differentiation}

Two major processes transform the appearance and capabilities of organisms during their lifetime: development and ageing. Development is often accompanied by morphogenesis and D'Arcy Thompson's seminal book On Growth and Form \cite{ThompsonBook} early on emphasized the need of physics for understanding morphogenetic processes. We refer the reader to recent reviews on the physics of morphogenesis for a description of the current state of this field ~\cite{Lecuit:2007cw,Mirabet:2011cw,Sun:2011cf}. We instead will focus on genetic aspects of development and ageing that so far receive less attention among physicists and present examples of how nonequilibrium landscapes can help to deepen our understanding of these processes.

\subsection{Stem cell differentiation}

Stem cells are undifferentiated cells
capable of differentiating into specialized cells.
Although a fertilized egg has the potential to develop into all the cell types of a body, differentiation typically occurs in a sequence of several steps, such that cells emerging at various stages of this process become more and more specialized. In 1957, Waddington suggested a pictorial way to visualize the developmental process in terms of a ball rolling down an increasingly fragmenting valley~\cite{Waddington}. Although intuitive, this picture lacks a physical foundation and has no quantification.  The Waddington landscape  has received recent attention among physicists and other scientists for global quantification of the development and reprogramming~\cite{Jiang2008NCB, Chickarmane2008PlosOne, Huang2009SCDB, Wang2010BJ,Wang2011PNAS, Feng2012SR, Li2013PlosCB, Sasai2013PlosCB, Xu2014PlosOne2, Li2014JRSI, Li2015CR, Sasai2015SR}. 

\begin{figure*}[!ht]
\includegraphics[width=1.0\textwidth]{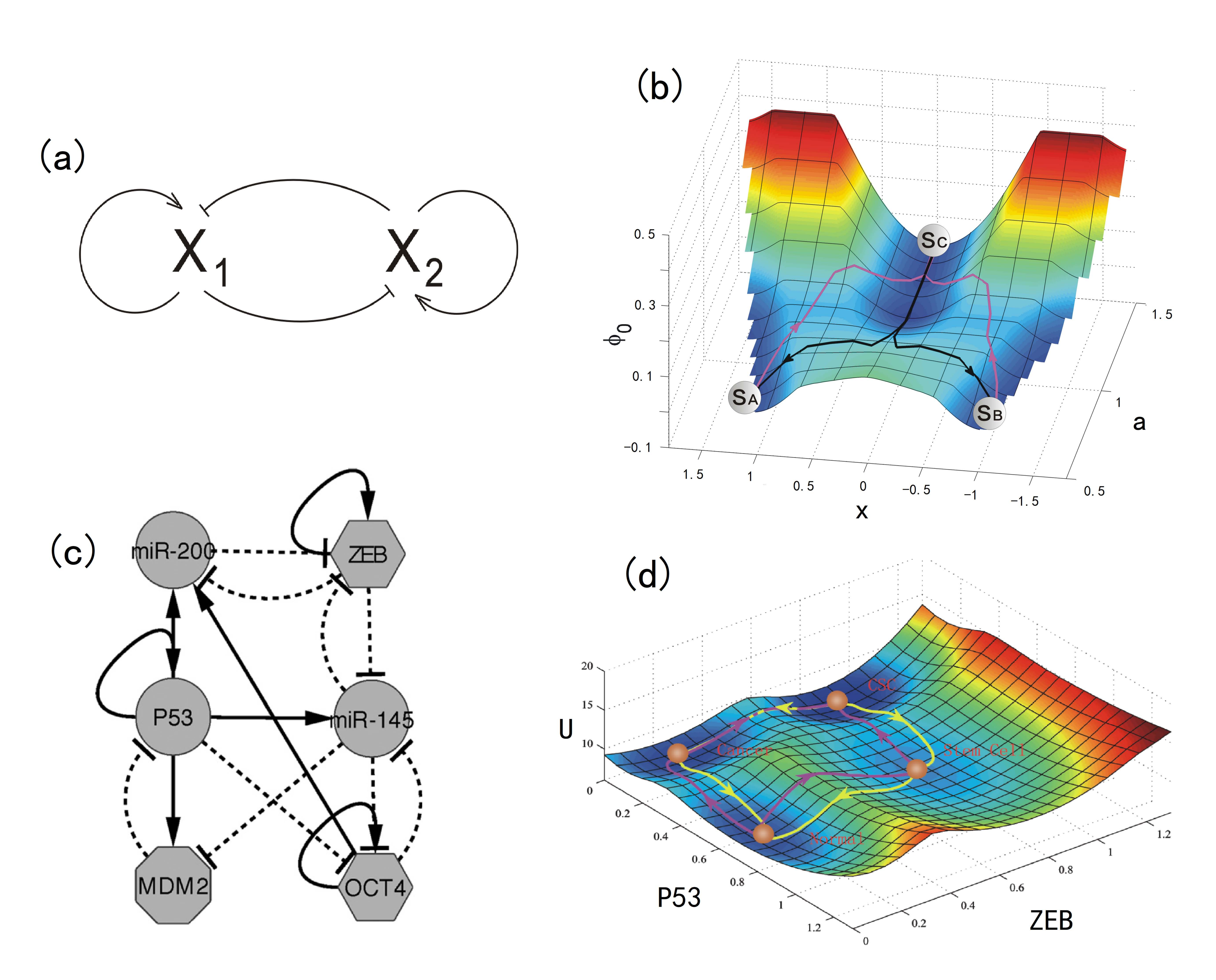}
\caption{ Stem cell differentiation and cancer-development. (a) The wiring diagram of core gene regulatory motif of differentiation and development. (b) The quantified Waddington developmental landscape and dominant transition paths for  differentiation, reprogramming and trans-differentiation paths. $S_c$ denotes stem cell state while $S_a$ and $S_b$ denote differentiated cell states. (c) The wiring diagram of the core gene regulatory motif of cancer and development. (d) The landscape and the dominant transition
paths between different cell states for cancer and development.
(from Ref. \cite{Xu2014PlosOne, Li2015CR}).} \label{Xu2014PlosOne_fig5}
\end{figure*}

Biologists have thought for long that differentiation was irrevocable.
 This view was shattered after the identification of the core genetic network underlying differentiation. A typical core motif involves two self activating genes mutually repressing each other ~\cite{Jiang2008NCB, Chickarmane2008PlosOne, Huang2009SCDB, Wang2010BJ,Wang2011PNAS, Feng2012SR, Li2013PlosCB, Sasai2013PlosCB, Xu2014PlosOne2, Li2014JRSI, Li2015CR, Sasai2015SR}, such as PU.1-GATA1 gene pair for  common myeloid
progenitor (CMP) differentiating to either myeloid cell or
erythroid cell in blood cell formation, Oct4 and Cdx2 gene pair for the inner cell mass/trophectoderm
lineage decision, and Nanog and Gata6 gene pair for
segregation of primitive endoderm and epiblast within the inner
cell mass,   Fig.~\ref{Xu2014PlosOne_fig5}a. Another example of the gene motif involves the mutual regulations of the transcription factors Oct-4 and NANOG~\cite{chambers2007nanog,takahashi2006induction,kalmar2009regulated}. They both activate themselves. NANOG activates Oct-4, whereas Oct-4 activates NANOG at low and suppresses it at high levels, Fig.~\ref{Xu2014PlosOne_fig5}a. This network topology generates a bimodal distribution of NANOG expression~\cite{kalmar2009regulated}. Although the possibility of noise-induced transitions between the two principle states cannot be excluded, the existence of two subpopulations is rather thought to result from excitable dynamics of the regulatory network~\cite{kalmar2009regulated}.
 Remarkably, controlled upregulation of NANOG leads to the reversal of a differentiated cell into a pluripotent stem cell (e.g. embryonic stem cells are pluripotent)~\cite{takahashi2006induction}, thus revoking 
 the previous paradigm that differentiated cells cannot return to the pluripotent state.  
 Waddington's picture is unsuitable for describing this process. Instead, nonequilibrium landscapes and fluxes provide an appropriate framework for an in depth analysis.

The dynamics of gene regulatory networks can be captured by
\begin{align}
 \frac{dX_i}{dt} &=-K_i X_i + \sum_{j} \frac{a_{ij}X_j^n}{S^n+X_j^n}+ \sum_{j} \frac{b_{ij} S^n}{S^n+X_j^n}. \label{equ:ECS}
\end{align}
Here, $X_i$ denotes the expression level of gene $i$. It changes by degradation at rate $K_i$, activation, and repression. The activation strength of gene $j$ on gene $i$ is given by $a_{ij}$, whereas $b_{ij}$ quantifies the repression strength; $a_{ij}=b_{ij}=0$ if gene $j$ does not regulate the expression of gene $i$. The parameter $S$ represents the ``activation threshold" and $n$ quantifies the cooperativity of the regulations~\cite{Li2013PlosCB}.

The dynamics underlying differentiation was studied by analyzing Eq.~(\ref{equ:ECS}) for a gene regulatory network motif of two self-activating and mutually repressing genes ~\cite{Jiang2008NCB, Chickarmane2008PlosOne, Huang2009SCDB, Wang2010BJ,Wang2011PNAS, Feng2012SR, Li2013PlosCB, Sasai2013PlosCB, Xu2014PlosOne2, Li2014JRSI, Li2015CR, Sasai2015SR}. As discussed previously, this gene motif appears often in stem-cell differentiation and reprogramming~\cite{Xu2014PlosOne}.  The cell starts from the stem-cell state basin and eventually transforms to differentiated state basins during the developmental process. Here the developmental direction is dictated by the change of the effective self-regulations, which was unspecified in the original Waddington picture \cite{Waddington, Wang2011PNAS}. The differentiation process can be viewed as the evolution of the landscapes along development \cite{Wang2011PNAS}. 

The effects of development and of external interventions for dedifferentiation or reprogramming were captured by changing the interaction strengths $a_{ij}$ and $b_{ij}$. In this way bifurcations were generated that correspond to the (de)differentiation process. The analysis showed that the differentiation and reprogramming pathways are typically irreversible, which is in contrast to Waddington's picture, Fig.~\ref{Xu2014PlosOne_fig5}. A detailed understanding of these pathways can guide the design of reprogramming pathways. 

To this end, a more complete gene regulatory network underlying a human stem-cell differentiation has been explored resulting in optimal reprogramming paths that are consistent with experiments~\cite{Li2013PlosCB}. Furthermore, two attractors corresponding to two different cell types can coexist and the rate of noise or input-induced switches between them can be quantified~\cite{Xu2014PlosOne}. The paths between two differentiated states might, but need not pass through a stem-cell like state~\cite{Wang2011PNAS,Xu2014PlosOne2}, making the direct trans-differentiation possible. This is quite important since reprogramming often encounters cancer state ~\cite{takahashi2006induction}. These findings are particularly relevant in the context of the heterogeneity of stem-cell differentiation due to environmental and epigenetic influences~\cite{Feng2012SR, Sasai2013PlosCB, Li2014JRSI, Sasai2015SR}.

\subsection{Aging}

Aging  has been thought of as an inevitable process of continuous decay
of physiological functions, which eventually leads to death, but experiments on model organisms show that ageing can be significantly delayed by suitable
genetic manipulations and in appropriate environments~\cite{Kirkwood2005Cell, Gems2013ARP, Kenyon2010Nature}. This suggests that the aging process is regulated and
programmed. Therefore, finding out the underlying genetic regulations and environmental influences is vital although challenging at the moment. One can
study aging using \textit{C elegans} as an example.

\begin{figure*}[!ht]
\includegraphics[width=1.0\textwidth]{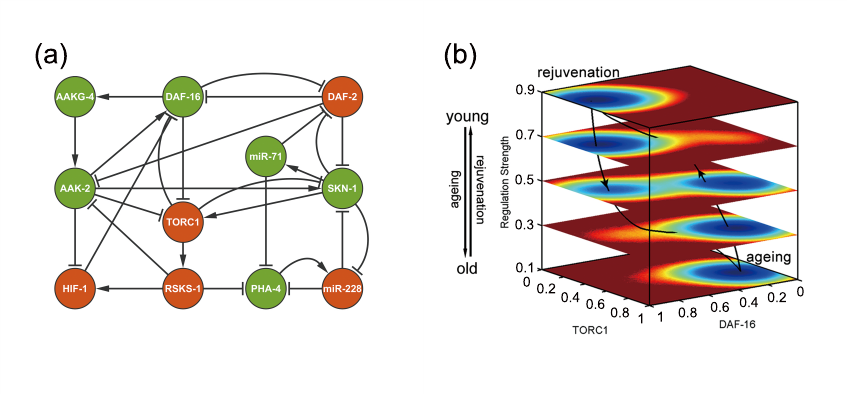}
\caption{ Gene network and nonequilibrium landscape of ageing and rejuvenation. (a) The wiring diagram of the core gene regulatory network of ageing of {\it C. elegans}.  (b) Dynamical landscape of {\it C. elegans} aging process. The horizontal coordinates denote the gene expression levels of DAF-16 and TORC1, the vertical
coordinate denotes the
regulation strength at which DAF-16 inhibits TORC1 
Rejuvenation and aging attractors are labeled as Rejuvenation and Ageing, and black lines denote the
optimal  paths between the rejuvenation and aging attractors upon changes in regulation
strengths.
(from Ref. \cite{Zhao2016JRSI}).} \label{Zhao2016JRSI_fig5}
\end{figure*}

Based on experimental studies of pathways with an impact on ageing a network of 11 genes and miRNAs involving DAF-2, DAF-16, SKN-1,  AAK-2, AAKG-4 genes, TORC1, RSKS-1, PHA-4,  HIF-1, miR-71 and miR-228 was established~\cite{Zhao2016JRSI}, Fig.~\ref{Zhao2016JRSI_fig5}a. Its dynamics is given by Eq.~(\ref{equ:ECS}) with appropriate connection strengths $a_{ij}$ and $b_{ij}$. The nonequilibrium landscape displays two attractors. In the ``ageing'' state, dominantly genes with lifespan-limiting effects are expressed, whereas in the ``rejuvenating'' state genes that enhance the lifespan prevail, Fig.~\ref{Zhao2016JRSI_fig5}b. Since the ageing state is more stable than the rejuvenating state, most worms are expected to age ``normally''; only a small fraction has an extended lifespan. The network, though, can switch between ageing and rejuvenation following
genetic or environmental interventions. On a more molecular level, further analysis suggested that self-degradation of lifespan-limiting and longevity-promoting genes leads to an increased stability of the ageing and rejuvenation states, respectively. This finding is consistent with experiments~\cite{Kenyon1993Nature,Lee2003Science,Apfeld2004GenesDev,Samuelson2007GenesDev,Kimura1997Science}. The study also suggests, why the ageing state becomes more probable with increasing lifetime. The negative regulation of the target of rapamycin complex 1 (TORC1), which plays an important function in monitoring the metabolic state of a cell and in regulating protein synthesis, by DAF-16, which has a dramatic effect on the lifespan of \textit{C.~elegans}~\cite{Kenyon1993Nature, wolff2006trifecta}, might become weaker due to the accumulation of damages. Along with a weakening of this connection, the dominant state can switch from rejuvenation to ageing. When increasing this negative regulation strength, the process reverts. Moreover, ageing and
rejuvenation switching paths can be quantified although they do not overlap much due to the presence of the nonequilibrium rotational curl flux. This indicates at
least in principle there is a possibility for reverting the ageing process through interventions (such as increasing the DAF-16 negative regulation to TORC1) \cite{Zhao2016JRSI}. The first hint of such success is
from the stem cell reprogramming discussed earlier where the differentiated cells can be turned back to the iPS progenitor cells \cite{Yamanaka2006Cell, Wang2011PNAS}.

Further analysis of the ageing and rejuvenation attractors suggests that self-degradation of lifespan-limiting and longevity-promoting genes leads to an
increased stability of the ageing and rejuvenation states, respectively, which is consistent with
experiments~\cite{Kenyon1993Nature,Lee2003Science,Apfeld2004GenesDev,Samuelson2007GenesDev,Kimura1997Science}.

Whereas the ``rejuvenation'' state in \textit{C. elegans} rather slows down or arrests ageing instead of reverting it, there are a few organisms that can revert to earlier developmental stages. Notably, the jelly fish {\it Turritopsis dohrnii} has normal ageing process (from young to old), but can change from the sexually mature medusa stage, in which they live as individuals, back to the sexually immature and colonial polyp stage (from old to young)~\cite{bavestrello1992bi, piraino1996reversing}. This process can be repeated as the old to young and young to old oscillation continues. In this way this jelly fish can live forever, unless an accident, a predator, or a disease interrupts these cycles. The regulatory network  in Fig.~\ref{Zhao2016JRSI_fig5}a is capable of producing sustained limit-cycle oscillations, where it switches periodically between ageing and rejuvenating phases~\cite{Zhao2016JRSI}. Stability of the limit cycle path is guaranteed by the Mexican hat-shaped landscape while the nonequilibrium rotational curl flux guarantees the stability of the oscillation flow and therefore the possibility of immortality through the forever oscillations. If the oscillation dynamics can emerge in more complex biological systems such as animals or human, this will provide new perspectives to understand and control the aging process for ourselves.

\section{Cancer}


Cancer is a major death causing disease for human beings. In spite of decades of efforts to understand the mechanisms leading to cancer, many open questions remain~ \cite{CancerBiologyBook}. Several  hallmarks of
cancer have been identified~\cite{Weinberg2000Cell, Weinberg2011Cell} and are the aim of anti-cancer strategies. The processes of tumor growth, their vasculation, or their spreading during metastasis depend strongly on physical properties. It is thus of no surprise that physicists are heavily involved in understanding cancer~\cite{Welter:2009ee,RamisConde:2009kp,Wirtz:2011cc}. Physical cancer treatment is still routinely employed. Beyond these nowadays obvious connections between cancer and physics, nonequilibrium concepts can be used to unravel the genetic and epigenetic conditions for the development of tumors and to explore new strategies of curing the disease~\cite{Welter:2013gs}. Although cancer is still mostly viewed as a disease caused by mutations, there is growing evidence from a physical point of view that the focus on genetics is too restricted and that environmental aspects have to be taken into account~\cite{Kauffman,Gatenby2003,Huang2009SCDB,Ao2008MH,Creixell2012NB,Bivort2009,Wang2007CMLS,Basan2009HFSPJ,Lu2013PNAS, Xing2013BJ,Lu2014CR,Li2014PNAS, Li2014JRSI, Li2015CR, Chen2016SR, Yu2016PlosOne}. As a consequence, cancer should be thought of as a disease state of the whole gene network.
Environmental changes can lead to changes or imbalances in regulations among genes in the network which  favor the cancer state, implying that cancer
treatment needs to target a collection of key genes and regulations or environmental conditions. Several questions related to this strategy remain
unanswered: How do we quantify the cancer state? Can one go from cancer back to a healthy state? How can we identify key genes and regulators? In this section, it is not our aim to give a comprehensive review of the physics of cancer. Rather we highlight how nonequilibrium concepts, notably, nonequilibrium landscapes and rotational curl fluxes as well as the homeostatic pressure, have advanced our understanding of this major death causing disease.

\subsection{Quantifying the landscape of cancer}

\begin{figure*}[!ht]
\includegraphics[width=1.0\textwidth]{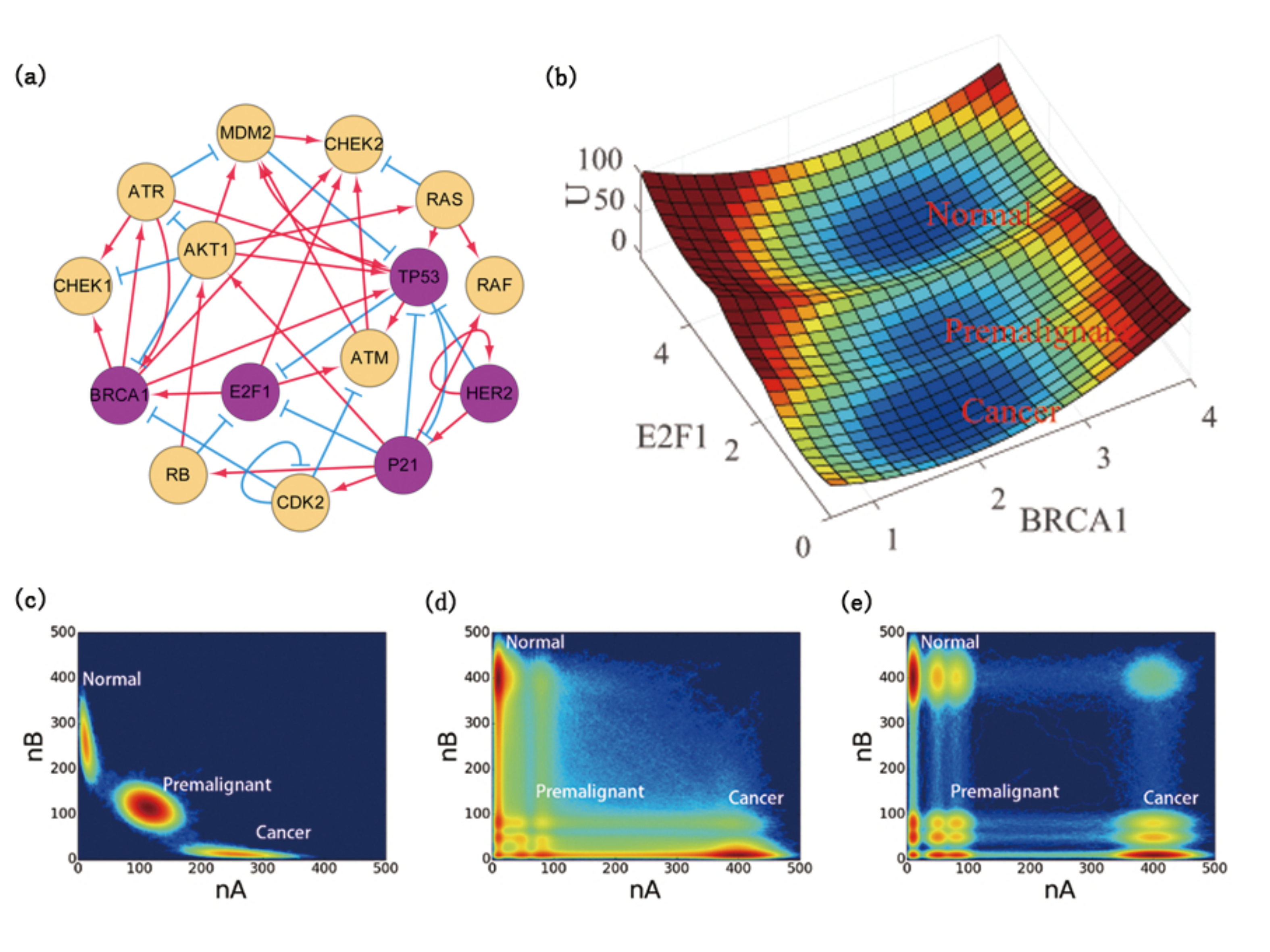}
\caption{ Gene network and nonequilibrium landscapes of cancer (a): Underlying gene regulatory network for breast cancer. $->$ represents activation regulations while $-|$ represents repression regulations. (b) The tristable landscape of the breast cancer gene regulatory network. (c)(d)(e): Landscapes from fast to slow epigenetic regulations.
(from Ref. \cite{Yu2016PlosOne, Chen2016SR}).} \label{Yu2016PlosOne_fig2}
\end{figure*}

To illustrate the application of nonequilibrium landscapes to cancer, consider breast cancer for which a core gene regulatory network consisting of 15
genes was constructed~\cite{Yu2016PlosOne}. This core network consists of oncogenes
BRCA1, MDM2, RAS, HER2; tumor suppressor genes
as TP53, P21, RB; kinases as CHEK1, CHEK2, AKT1,
CDK2, RAF, for cell cycle regulations; the transcription
factor E2F1; and ATM, ATR, important for early signal
transduction through cell-cycle checkpoints. The wiring diagram of the network is shown in \cite{Yu2016PlosOne}, shown in Fig. \ref{Yu2016PlosOne_fig2} (a).

The dynamics of the gene
regulatory network is captured by similar type of equation as Eq.~(\ref{equ:ECS}).

The landscape projected on the expression levels of the oncogene \textit{BRCA1} and the transcription factor E2F1, which is a marker for breast cancer, exhibits three attractors, Fig.~\ref{Yu2016PlosOne_fig2}b. They correspond to the normal, the cancer, and a premalignant state and the respective gene expression levels associated with the attractors are consistent with experimental findings~\cite{Lu2013PNAS, Xing2013BJ, Yu2016PlosOne}. In comparison to the premalignant state, the attractors of the normal and the cancer state are much more stable indicating the importance of detecting the disease in early stages: whereas appropriate treatment might be able to switch back from the premalignant to the normal state, the transition to the cancer state is practically irreversible. The dominant pathways of switching can be identified and used to
quantify the process of how the normal state changes to the cancer state and vice versa.

Beyond these general statements, the landscape analysis can provide more specific information. For example, changes in the expression of the central tumor suppressor gene \textit{TP53} change the depths of and the barriers between the attractors. Notably, an increased repression of \textit{TP53} facilitates the transition first to the premalignant and then to the cancer state, which eventually has the dominant basin of attraction~\cite{Yu2016PlosOne}. Global sensitivity analysis based on the barrier heights allows one to identify key genes and regulations for breast cancer formation and dysfunction. Four key
regulations (HER2$\dashv$TP53, TP53$\rightarrow$ATM, ATM$\rightarrow$MDM2, CDK2$\dashv$BRCA1) and six key genes (HER2, TP53, ATM, MDM2, BRCA1 and CDK2) are
identified. These could serve as the targets for the network based drug discovery.

\subsection{Cancer and Development}

A hallmark of cancer is the abnormal growth of cells \cite{Weinberg2000Cell, Weinberg2011Cell}. Development at the cellular level often refers to the differentiation process from primary stem cells \cite{Waddington, Wang2011PNAS}.A hint of the possible connection between cancer and differentiation lies in the fact that cancer often regrows after the radiation and chemo treatments  \cite{Marotta2009COGD}. The possibility of the existence of the seeds for cancer in the form of cancer stem cells was explored recently \cite{Lobo2007ARCDB}. To understand the underlying mechanism of the cancer stem cell and the relationship between cancer and develop- ment, one needs to explore the underlying regulatory inter- actions among genes.

There are intimate connections
between gene regulatory networks for development and for cancer, for example, involving the tumor suppressor TP53  and its suppressor MDM2 as well as ZEB
and OCT4, which play a role in differentiation. ZEB is known to
be a major player in the epithelial-to-mesenchymal transition, often linked to cancer metastasis and formation of CSCs. The miRNA regulates both
cancer and development and therefore mediate the interactions between the cancer and developmental genes. The core gene regulatory motif for cancer and development is illustrated in \cite{Li2015CR} Fig.~\ref{Xu2014PlosOne_fig5}(c).

The dynamics of this core regulatory motif can be described by Eq.~(\ref{equ:ECS}). An analysis of the corresponding nonequlibrium landscape reveals four attractors
corresponding to the stem-cell, cancer stem-cell,
healthy differentiated, and differentiated cancer states~\cite{Li2015CR}, Fig.~\ref{Xu2014PlosOne_fig5}(d).
Normal differentiated state emerges with high TP53 and low ZEB expressions while the differentiated cancer state emerges with low TP53 and low ZEB expressions.
Stem cell state emerges with high TP53 and high ZEB expressions while cancer stem cell state emerges with low TP53 and high ZEB expressions.
Based on the landscape topography, the stem cell is most likely to transit into the normal differentiated state
from which it can move to the cancer cell state. However, the stem cell can also change into a cancer stem cell, which then provides another route into the cancer
state. From the landscape analysis it was also found that, consistent with experiment \cite{Li2015CR, Marotta2009COGD, Lobo2007ARCDB}, the cancer cell state re-emerges after eliminating all cancer cells. Furthermore, it helped to identify in this network motif the key elements responsible for generating new cancer seeds. These findings can help to design a new way for regulating the cancer seeds and a promising strategy for anti-cancer
therapy targeting at cancer stem cells

\subsection{Cancer heterogeneity}

Cancers are often heterogeneous. This is a critical issue for radiation and chemotherapy, because a radiation dose or single drug might not be able to kill all cancer related cells in a heterogeneous population~\cite{Marusyk2012NRC}. Heterogeneity might be due to genetic differences between the cells that result from mutations accumulating during cancer progression. Alternatively, and maybe more importantly, intra-cellular heterogeneity might be primarily caused by epigenetic modifications~\cite{Shackleton2009Cell}. These are modifications of the DNA through DNA methylation and accessory proteins like histones that do not change the genome, but affect, for example, DNA organization and transcription. Notably, they can lead to regulatory delays. As a consequence, a larger variety of states can emerge~\cite{Chen2016SR}

The idea of heterogeneity was illustrated by a core gene motif of cancer with mutual repressions and self activations. The genes produce proteins and proteins regulate genes and determine whether the genes are turned on or switch off. When the regulations of the proteins to the genes are fast (slow) compared to the production/degradation of the proteins, then the proteins and genes are inseparable (separated) and can be treated as the same (distinctly different) identity. It was found \cite{Chen2016SR} in Fig. \ref{Yu2016PlosOne_fig2} (c-e) that when
regulatory binding/unbinding is fast compared to  synthesis/degrdation (adiabatic limit), three states quantified by the basins of attractions emerge
including the normal state, cancer state and an intermediate premalignant state. When regulatory binding/unbinding is comparable or even slower than the
synthesis/degrdation corresponding to the epigenetic case (nonadiabatic limit) with extra time scales involving histone remodification and DNA methylation,
the heterogeneity emerges. One observes both premalignant and cancer state basins are surrounded by significant number of shallower and less stable state
basins. \cite{Chen2016SR}. The effective weakening of the interactions among genes through epigenetics illustrated here in terms of longer regulation time compared to the synthesis/degradation can lead to less constraints and more freedom for each gene and therefore the emergence of more states and substates in terms of metastable basins around major basins on the landscape.
This may provide a way to understand and control cancer heterogeneity by targeting the epigenetics and environments.

\subsection{Homeostatic pressure}

Primary tumors are rarely lethal, but cells can leave a tumor and invade other parts of the organism, in which they subsequently metastasize, that is,  produce secondary tumors. These cells are mainly transported by the blood stream. However, the distribution of metastases is not fully determined by the blood flow pattern as the receiving tissue must in some sense be compatible with the metastatic cell. This phenomenon has been captured by the seed-and-soil hypothesis~\cite{CancerBiologyBook}.

This hypothesis can be conceptualized by the homeostatic pressure. It is a tissue-inherent quantity that describes the pressure exerted by an expanding tissue of proliferating, growing, and dying cells \cite{Basan2009HFSPJ}. A planar interface between two tissues of different homeostatic pressures will move into the direction of the tissue with the lower homeostatic pressure. For a spherical clump of cells, also interfacial stresses enter the picture.

If cell growth is independent of the size of the tissue, then cell spheroids will expand in a surrounding tissue only if they exceed a certain size \cite{Basan2009HFSPJ}. Taking into account the stochastic nature of cell growth, division, and death, the homeostatic pressure provides a quantitative conceptualization of the seed-and-soil hypothesis. To be useful for therapy, biochemical or immunological means of affecting the homeostatic pressure need to be uncovered.


\subsection{Cancer and immunity}

Tumor cells express antigens and are thus prone to be eliminated by the immune system~\cite{Weinberg2011Cell}. This avenue is exploited by cancer immunotherapy, which has achieved spectacular results for specific types of cancer. ~\cite{Sabado2015CancerImmunTherapy}
However, cancer can dysfunction the immune system. This leads to two hallmarks of cancer immunity, avoiding immune destruction and tumor promotion inflammation \cite{Weinberg2011Cell}.
A profound understanding of the relation between cancer and the immune system remains elusive. 

Theoretical studies of the complex interaction between cancer and the immune system that take spatial aspects into account are commonly based on active particles~\cite{Bellomo:2008eh}. Such descriptions can be fairly comprehensive, but the large number of details that are accounted for, make a through analysis rather difficult and often prohibit an understanding of the fundamental principles. In contrast, non-spatial models that are typically formulated in terms of ordinary differential equations for the interaction between tumor and immune or healthy cells are simple enough to be comprehensively analyzed~\cite{Earn2011BMB, Bellone2014Bioinf,Hahnfeldt2013IF}.

Cancer and immune cells can communicate and influence each other either through direct contact or via cytokines, which are a small signaling molecules secreted by cells, notably, during an immune response. The dynamics of the respective cell and cytokine concentrations can be written in a similar way to Eq.~(\ref{equ:ECS})~\cite{Li2016ArXiv}, with several essential modifications: the degradation rate of a type of cells depends on the concentration of other cell types and of cytokines; while the net regulations for cells from others take the additive form, the net regulations for cytokine concentrations from others take the multiplicative form. Similarly to gene regulation networks the proliferation of cells and the secretion of cytokines can be either enhanced or diminished by the presence of other cell types and cytokines. The concentration dependencies of the various rates is given in terms of Hill functions and the coupling coefficients encode for the network structure.

The nonequilibrium landscape for a network consisting of one cancer cell type, 12 immune cell types, and 13 types of cytokines
has three attractors corresponding to a healthy state and states of, respectively, low and high tumor cell concentration~\cite{Li2016ArXiv}. In the healthy state, the innate immune response leads to an increased presence of natural killer cells, a type of white blood cells with the task to destroy infected cells. In the two other states,
the adaptive immune response leads to an increase in the concentration of a kind of white blood cells, namely CD8+ cells. However, their concentration is higher (lower) for the state with the lower (higher) concentration of tumor cells~\cite{Onuchic2014PNAS,Li2016ArXiv} 
suggesting that the adaptive immune system is suppressed by cancer cells.

\begin{figure*}[!ht]
\includegraphics[width=1.0\textwidth]{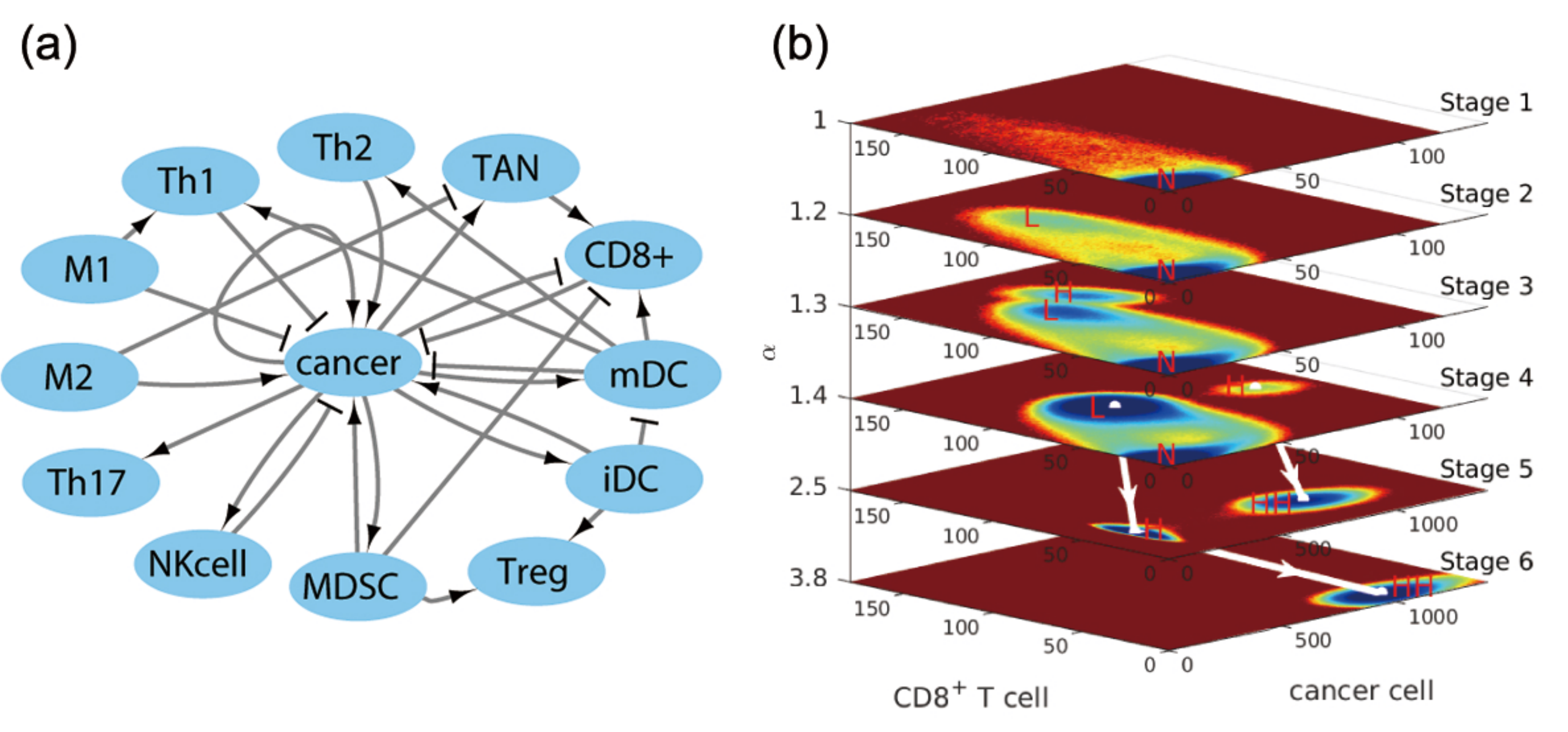}
\caption{Gene network and nonequilibrium landscape of cancer-immune system (a) A core cancer-immune cell-cell interaction network. (b)Cancer-Immune landscapes at various stages of tumor development
(from Ref. \cite{Li2016ArXiv}).} \label{Li2016ArXiv_fig4}
\end{figure*}

The interaction between cancer and the immune system depends on the state of 
progression as the interactions between different cells or
cells and cytokines are modified in the environment of a developing tumor. The various effects of the immune system in different stages of developing tumors is known as cancer immunoediting~\cite{Schreiber2002NI}. 
Correspondingly, the nonequilibirum landscape attractors of the immune system-cancer network change with changing interaction strengths and several stages can be distinguished \cite{Li2016ArXiv},
Fig.~\ref{Li2016ArXiv_fig4}b. At stage 1, only the healthy 
state state attractor exists. It is controlled by the immune system and nascent tumors are repressed, corresponding to the
elimination phase of cancer immunoediting.
At stage 2 a low cancer expression state emerges, and in stages 3 and 4 a low and a high cancer expression state
emerge in
addition to the normal state. These three stages correspond to the equilibrium phase of cancer immunoediting, which is the phase persisting the longest.
In
stages 5 and 6 only the low and high cancer expression states remain, corresponding to the escape phase of cancer immunoediting, when the cancer has escaped the organism's immune response. 
Important immunotherapy
targets are predicted from the landscape approach from the topography through global sensitivity analysis including three types of immune cells
(mature dendritic, natural killer, and CD8+ T-cells) and two types
of cytokines (IL-10 and IL-12)~\cite{Li2016ArXiv}. The oscillation behavior of immune-cancer network dynamics was also expected in some cases. \cite{Li2016ArXiv}

\section{Population dynamics and ecology}

Living organisms are highly social by nature and often coordinate with each other to generate collective behaviors in space and time. Studying the dynamics of population and ecology provided some early examples of nonequilibrium dynamics. In the following, we highlight some recent developments on microbial population and ecology.

\subsection{Populations of microorganisms} 

Population dynamics is typically studied through field research; however, it often requires significant effort and suffers from a lack of control over environmental conditions. By contrast, microbial populations are much more amenable to manipulation and quantification while still possessing intricate dynamics. Thus, synthetic microbial populations have been recently exploited as model systems to study population dynamics~\cite{brenner2008engineering,xavier2011social,grosskopf2014synthetic,de2014synthetic,chuang2009simpson,Gore2009Nature,kong2018designing,ozgen2018spatial,shou2007synthetic}. Microorganisms can establish coordination and create collective behaviors among populations by secreting and detecting chemicals. One common way to generate population behaviors is using quorum sensing, a mechanism that enables cells to sense the density of their peers and respond accord- ingly~\cite{miller2001quorum,kong2014programming}.

\begin{figure*}[t]
\includegraphics[width=1.0\textwidth]{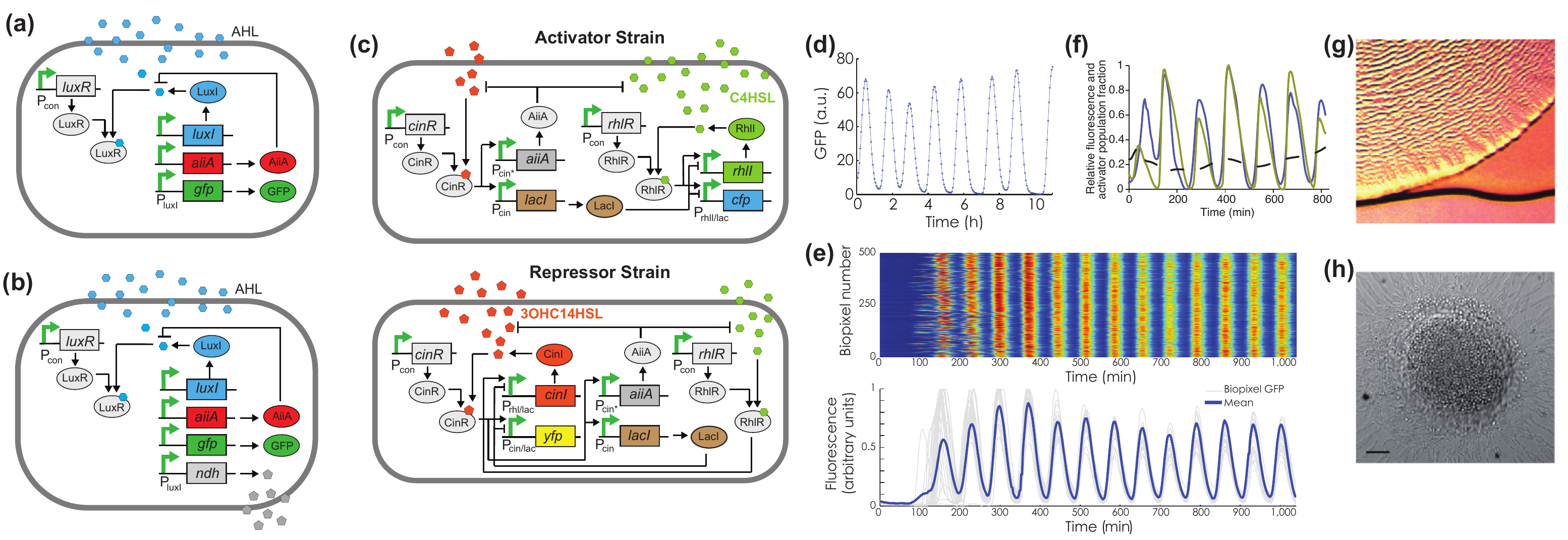}
\caption{Microbial oscillations at the population and multi-strain levels.
(a) A gene circuit that enables synchronized oscillation of individual
bacterial colonies~\cite{danino2010synchronized}. (b) A gene circuit that synchronizes the oscillation of thousands of bacterial
colonies~\cite{prindle2012sensing}  (c) A gene circuit that allows stable oscillations of two bacterial strains~\cite{chen2015emergent}. }
\label{fig-exp-pop-com}
\end{figure*}

A gene network based on quorum sensing was designed and implemented in \textit{E.~coli} to generate synchronous population oscillations ~\cite{danino2010synchronized}. In this network (Fig.~\ref{fig-exp-pop-com}a), expression of the genes $luxI$ and $aiiA$ is controlled by the P$_{lux}$ promoter. The enzyme LuxI synthesizes the quorum sensing molecule N-Acyl homoserine lactone (AHL), which activates the P$_{lux}$ promoter after binding to LuxR and, hence, promotes its own expression. In contrast, AiiA degrades AHL, which provides a negative feedback on LuxI production. In each cell, such a topology leads to the oscillation of the activity of the P$_{lux}$ promoter. As AHL can diffuse from cells to neighboring cells, it couples individual cells and generates synchronous oscillation of thousands of cells in a square region of 10$^4~\mu$m$^2$ in size~\cite{danino2010synchronized}.
The distance over which the population synchronizes depends on the diffusion constant of the quorum sensing molecule. Using a similar approach where H$_2$O$_2$ was used as a signaling mechanism to overcome the slow diffusion obstacle for long-range coupling, synchronized oscillations were observed for several millions of bacteria across a distance of 5~mm could be synchronized (Fig.~\ref{fig-exp-pop-com}b)~\cite{prindle2012sensing}. The same oscillation mechanism that involves activation and repression can also be realized through multiple strains~\cite{chen2015emergent} (Fig.~\ref{fig-exp-pop-com}c).

In addition to synthetic populations, natural organisms exhibit remarkable collective behaviors that are far from equilibrium. The formation of multicellular life forms from unicellular microorganisms is a representative class of such processes~\cite{claessen2014bacterial,lyons2015evolution}. Multicellularity can arise from simple cellular aggregation as a consequence of incomplete separation after cell division. Another way collective behavior arises is through dynamic aggregation of previously individual cells, which involves cellular communication and differentiation, partitioning of tasks, and spatial organization. Well-studied examples of the latter route are fruiting body formation of myxobacteria and of the slime mold  \textit{Dictyostelium disoideum}~\cite{munoz2016myxobacteria,zusman2007chemosensory}. In nutrient-limited conditions, cells communicate through multiple modes of interactions and self-organize into complex, three-dimensional structures. Within the fruiting bodies, a subset of the cells differentiates into nonreproductive cells, while the remaining cells become reproductive spores.

\subsection{Ecology}

\begin{figure*}[t]
\includegraphics[width=1.0\textwidth]{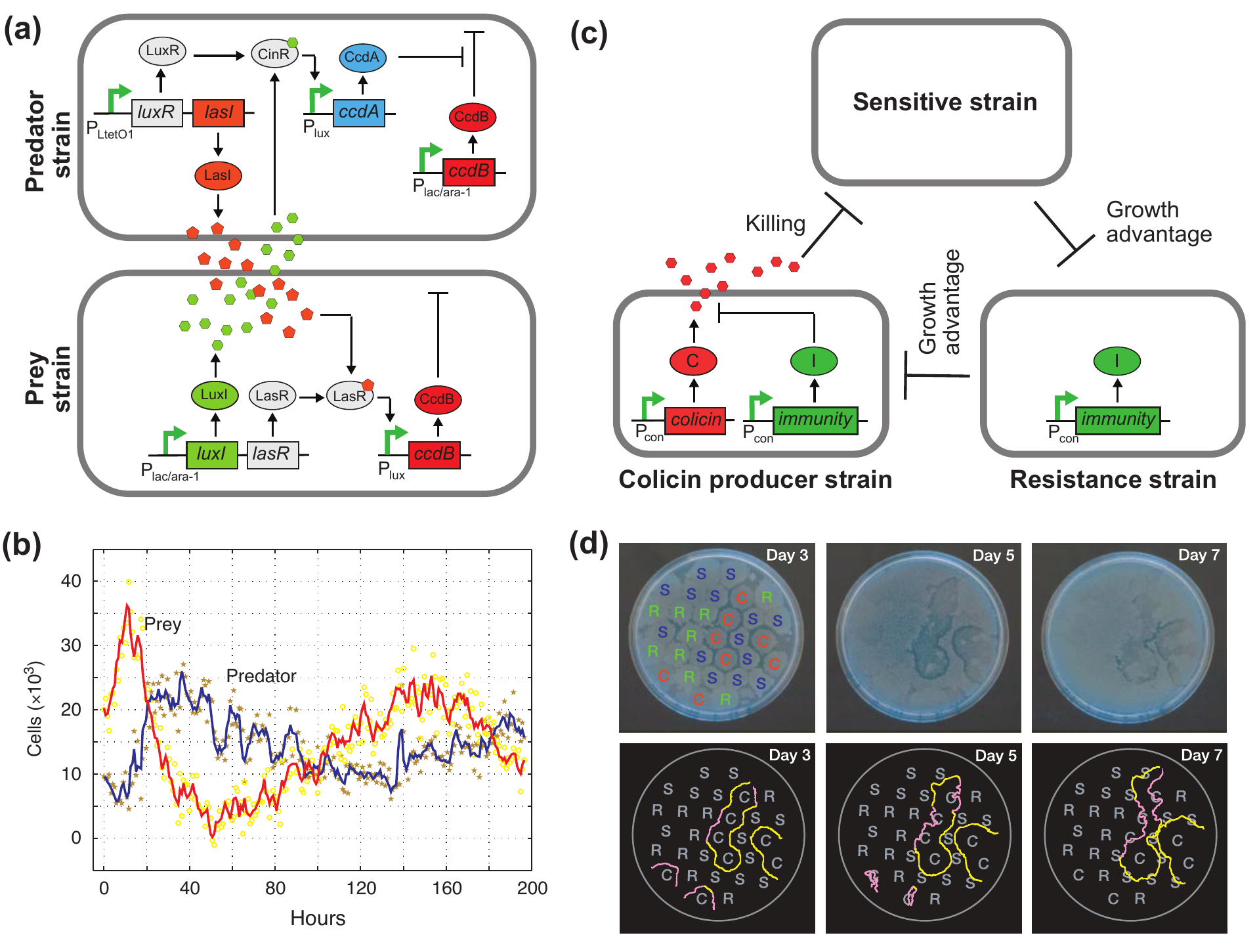}
\caption{Microbial ecosystems. (a) A synthetic predator-prey system~\cite{balagadde2008synthetic}. (b) A three-strain ecosystem
resembling a rock-paper-scissor game~\cite{kerr2002local}. }
\label{fig-exp-eco-evo}
\end{figure*}

Populations are typically not isolated; instead, they often compete and cooperate with populations of other species in nature. This is a topic of ecology~\cite{Murray,Vandermeer}. Predator-prey systems have been of particular interest in this field since the work of Lotka and Volterra~\cite{Lotka,volterra1927variazioni}. Their model showed that the systems can generate various dynamics including sus- tained oscillations. Observed in animal populations in nature~\cite{Murray}, such dynamics have been recently observed in populations of engineered bacteria that utilize quorum sensing machineries (Fig.~\ref{fig-exp-eco-evo}a)~\cite{balagadde2008synthetic}. Experiments in microchemostats confirmed distinct types of population dynamics, namely, coexistence, extinction, and oscillation. The underlying landscape for a global description of the dynamics was also quantified~\cite{Xu2014PlosOne,li2011potential}. 

Beyond well-mixed cultures, microbial populations have also been used to study spatial ecology. An example is provided by a system of three \textit{E.~coli} strains that populate a petri dish~\cite{kerr2002local}. The ecosystem consists of a strain producing the toxin colicin (C), a strain sensitive to colicin (S), and a strain resistant to it (R) (Fig.~\ref{fig-exp-eco-evo}b). In this setting, strain C kills strain S by releasing colicin, strain S in turn outgrows strain R because it does not synthesize resistance proteins, and strain R has a higher fitness than strain C as it does not produce toxins. Together these strains form a fitness advantage loop, resembling an ecological version of the rock-paper-scissors game. The corresponding experiments showed that a single species rapidly dominates in the well-mixed case while, on plates, the ecosystem exhibited coexistence~\cite{kerr2002local}.

\subsection{Landscape and flux analysis of ecosystems}

 One of the central questions of ecology concerns the coexistence of species. Under which conditions is this possible? How many species can coexist in a given environment? This translates into the question of whether or not an ecosystem is stable. The analysis of stability against small perturbations is standard and in some cases Lyapunov functions have been found~\cite{Lotka,Volterra,Harrison,Goh1,Goh2,Hsu,Holling,Murdoch,Hastings}. A general way of assessing the global stability of states, however, is still missing. The theory of nonequilibrium landscapes and fluxes turns out to be useful in this context~\cite{Wang2006PLOSCB,Wang2007BJ,Wang2008PNAS,Lapidus2008PNAS,Zhang2012JCP,Xu2013Non,Xu2014PlosOne}. The intrinsic nonequilibrium landscapes provide
 Lyapunov functions for quantifying global stability of the ecosystems \cite{Xu2014PlosOne}. We will illustrate with few examples. In the following, $C_1$ and $C_2$ denote the sizes of two populations.

\paragraph{Predator and prey}

\begin{figure*}
\includegraphics[width=1.0\textwidth]{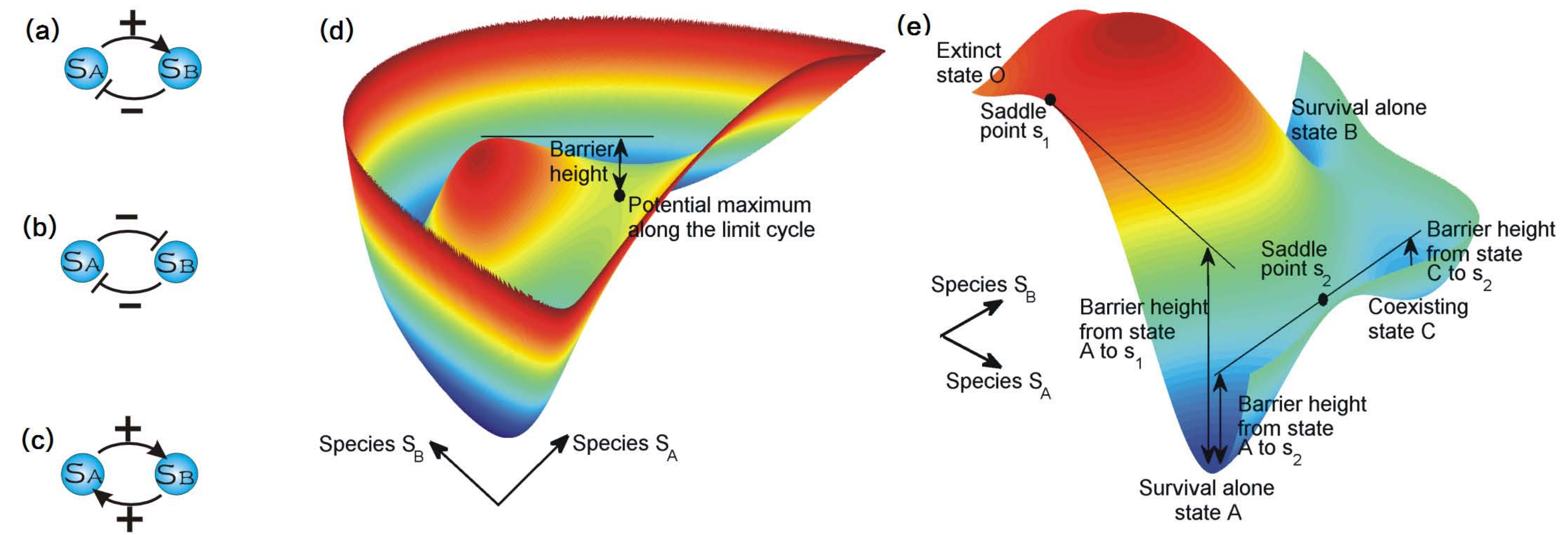}
\caption{
The schematic diagram for the ecological models and associated landscapes. (a)Predation model. (b)Competition model. (c)Mutualism model. (d) Limit cycle attractor landscape for predation model. (e) Multiple attractors for competition and cooperation models (from Ref. \cite{Xu2014PlosOne}).
}\label{Xu2014PLOSONE_fig2}
\end{figure*}

Consider the following model for a population $C_1$ of predators and $C_2$ of prey~\cite{Murray}:  
\begin{align}
\frac{dC_{1}}{dt}&=C_{1}(1-C_{1})-\frac{aC_{1}C_{2}}{C_{1}+d} \\%
\frac{dC_{2}}{dt}&=bC_{2}(1-\frac{C_{2}}{C_{1}}).
\end{align}
In absence of predators, the prey population evolves according to the logistic growth model. The predators feed on the prey as quantified by the parameter $a$ and $d$ denotes the prey population size at which it is consumed by the predators with  their maximum rate. The population of the predators grow also according tho the logistic growth model. The ratio of the birth rates of both populations is $b$ and the capacity of the system for prey is $C_1$.
When the number of predators increases, more preys will be eaten up. The shortage of food will lead to the reduction
on the number of the predators. This leads to the
growth of the preys, which will promote the growth on
the number of predators from the rich sources of preys.
This is the origin of the limit cycle predator-prey dynamics. The nonequilibrium landscape has the shape of a Mexican hat, Fig.~\ref{Xu2014PLOSONE_fig2}(a), revealing an oscillatory state that is globally stable, while the rotational curl flux enables the stability of the oscillation flow \cite{Xu2014PlosOne}.

\paragraph{Cooperation and competition}

Cooperation and competition between two species can be described by~\cite{Russian}:
\begin{align}
\frac{dC_{1}}{dt}&=C_{1}(C_1-L_1)(1-C_{1})+a_1 C_{1}C_{2}
 \nonumber\\
\frac{1}{\alpha}\frac{dC_{2}}{dt}&= C_{2}(C_{2}-L_2)(1-C_{2})+a_2
C_1C_2
\end{align}
The factors $(C_i-L_i)$, $i=1,2$, modifying the logistic growth model assure that the population size does not drop below $L_i$ in absence of the other species. The terms proportional to $C_1C_2$ describe the interaction between the two species, which is cooperative if $a_1,a_2>0$ and competitive if $a_1,a_2<0$. The parameter $\alpha$ quantifies differences in the growth rates of two species. Four different steady states are possible: extinction ($C_1=C_2=0$), mutual exclusion (either $C_1=0$ or $C_2=0$), or coexistence. Through the corresponding basins of attraction, the potential landscapes determine, which of these states is stable for a given set of parameters, Fig.~\ref{Xu2014PLOSONE_fig2}(b) \cite{Xu2014PlosOne}.

 \section{Evolution}

Evolution is the essence of biology. After Darwin had laid out the principles of evolution by variation and natural selection~\cite{Darwin}, a significant fraction of subsequent research has focused on quantifying evolutionary processes. Concerned with the process of adaptation, Sewall Wright introduced the concept of an adaptive landscape for evolution~\cite{Wright, Fisher, Rice, Ewens, Yuri}.
Adaptation or ``shifting-balance'' in Wright's terms, then refers to reaching summits in this landscape, by random genetic drift due to mutations and selection. Therefore, the evolutionary dynamics follows its gradient until an optimum is reached. The parallels to energy minimization in physics are obvious. The virtues and shortcomings of the adaptive landscape metaphor are still debated~\cite{Rice,Pigliucci:2008iy}. %

The central results of quantitative genetics is ``Fisher's fundamental theorem'', which states that the speed of evolution in terms of change in time of the average fitness keeps increasing and equals to the genetic variance of the fitness in the population. Still, some critical issues remain. The most prominent ones include: How general is Wright
and Fisher's evolution theory and how can it explain that the evolution can continue without stop?

A key assumption in Wright and Fisher's theory is that selection force is independent of the relative proportion that a gene variant, ``allele'', appears in
a population at a specific site or ``locus'' on the chromosome. In other words, the selection driving force is independent of the interactions among gene species. This is so called the allele frequency independent selection, a condition that is also known as linkage equilibrium (LE) \cite{Shraiman}. However, in general, the selection can have allele frequency dependence. This is apparent for example in coevolution where two or more species mutually affect their evolution. Wright and Fisher's theory breaks down for scenarios such as coevolution. It can not explain open-ended evolution that is clearly observed in bacterial colonies that have been evolving under constant physical and chemical conditions for tens of thousands of generations \cite{Elena:2003fr} and to some extent in a molecular evolution experiment, where a molecular species constitutes the building blocks for a new species, which in turn can form another molecular species and so on \cite{Worst:2016ip}.

The key to resolve above issues lies in the fact that the
adaptive landscape is not the only driving force for the
general evolution dynamics as Wright and Fisher theory
stated.
In the following we discuss how a generalization of the adaptive landscape to include nonequilibrium fluxes as an additional driving force for the general evolution dynamics overcomes the restrictions of the selection being independent of the interactions among gene species assumed by Wright and Fisher~\cite{Zhang2012JCP, Xu2017JTB}.

\subsection{Single-locus multi-allele evolution}

Consider a single gene at a fixed position or ``locus'' on a chromosome of a ``diploid'' organism, such that each chromosome is present in two (non-identical) copies. For each locus there are several different DNA sequences or ``alleles'' present in a population. The dynamics of the fractions $x_i$ of allele $A_i$, $i=1,\ldots,n$ in the population is determined by its fitness $w_i$. In an individual, the genotype $A_iA_j$ has the fitness $w_{ij}$, which depends on both alleles, such that $w_i = \sum_{j=1}^{n}  w_{ij}x_j$. The population's mean fitness is then $\bar w \equiv \sum_{i,j = 1}^{n} w_{ij}x_i x_j $. The probability $P$ of having the relative allele frequencies $\{x_i\}_{i=1,\ldots,n}$ evolves according to
\begin{align}
\label{FPE4E}
\partial_t P & = - \nabla \cdot \left[ \left(\mathbf{F}^{S} + \mathbf{F}^{M}\right)P - D \nabla \cdot (\mathsf{G} P) \right],\\
\intertext{where}
F_i^S &= \frac{x_i\left(x_i-\bar w\right)}{\bar w}\\
\intertext{describes the effects of natural selection and}
F_i^M &= \sum\nolimits_{j=1}^n x_j m_{ji} - x_i \sum\nolimits_{j=1}^n m_{ij}
\label{eq:mutationForce}
\end{align}
captures the effects of mutations with $m_{ij}$ being the rate of mutation from allele $A_j$ to $A_i$. The diffusion term accounts for genetic drift that results from stochasticity of the reproduction process~\cite{Kimura1997Science, Rice,Yuri}. Here, $G_{ij} = x_i (\delta_{ij} - x_j)$
and $D = 1/(4N_e)$ with $N_e$ being the effective population size.

In the limit of small fluctuations, i.e., a large population size, $D\ll1$ or $N_e\gg1$, the landscape satisfies to lowest order
%
\begin{eqnarray}\label{HJE4E}
(\mathbf{F}^{S} + \mathbf{F}^{M}) \cdot \nabla \phi_0 + \nabla \phi_0 \cdot \mathsf{G} \cdot \nabla \phi_0 = 0
\end{eqnarray}
where $\phi_0$ is the leading order term in an expansion of $U$ in terms of $D$. This equation is the Hamilton-Jacobi equation~(\ref{HJE}) and $\phi_0$ thus a Lyapunov function for the dynamic system Eqs.~(\ref{FPE4E})-(\ref{eq:mutationForce}).

In case the fitness of each genotype is independent of the allele frequencies and in absence of mutations, the steady state flux vanishes, $\mathbf{J}_{ss} = 0$, and the intrinsic landscape $\phi_0$ is
\begin{align}
\phi_0 &= -\frac{1}{2} \mathrm{ln} \bar{w}
\end{align}
\cite{Zhang2012JCP}. Furthermore, one can show that $d\bar{w}/dt = \mathbf{F}\cdot\nabla\bar{w} = - 2 \bar{w} \mathbf{F}\cdot\nabla\phi_0 = 2 \bar{w} \nabla \phi_0 \cdot \mathsf{G} \cdot \nabla \phi_0 \geq 0$ \cite{Zhang2012JCP}. Consequently, the mean fitness is a Lyapunov function of the dynamics and hence puts Wright's metaphor onto a solid ground.

In the general case, where the fitness can depend on the allele frequencies, one finds $\mathbf{F}^{S} + \mathbf{F}^{M}= - D \nabla \cdot (\mathsf{G} U) + \mathbf{J}_{ss}/P_{ss} $, where $U=-\ln P_{ss}$ is the nonequilibrium landscape~\cite{Zhang2012JCP} while $G_{ij}=C_i(\delta_{ij} - C_j)$ is from the sampling feature of the genetic drift and D gives the scale of the fluctuations. The driving force of evolution can thus be decomposed into the
gradient of the landscape associated with the steady state probability 
and the steady state probability flux, which is typically different from zero due to interactions between individuals and hence allele-frequency dependent selection. Note
that for $J_{ss}\neq0$ the nonequilibrium landscape is no longer directly related to 
the fitness landscape. Consequently, states with lower mean fitness may have a higher probability.

We now turn to Fisher's theorem. Consider the adaptive rate for general evolution dynamics:
\begin{eqnarray}\label{dphi0dt}
d{\phi _0}/dt
&=& - \nabla \phi_0 \cdot {\bf D } \cdot \nabla \phi_0   \\ \nonumber
&=& - \mathbf{F} \cdot {\bf D }^{-1} \cdot \mathbf{F} + \mathbf{V} \cdot {\bf D }^{-1} \cdot \mathbf{V} .
\end{eqnarray}
where $\mathbf{V}$ is the steady state probability flux velocity defined as $\mathbf{V}=\mathbf{J}_{ss}/P_{ss}$. The diffusion matrix ${\bf D }$  describes the sampling nature of the random mating. In fact, the $d \phi_0({\bf G }) / dt$ is related to the genetic variance:
$
V_A(w^{(i)}) /({\bar w}^{(i)})^2 = \ 2 \ \mathbf{F}^{(i)} \cdot ({\bf G }^{-1})^{(i)} \cdot \mathbf{F}^{(i)} ,
$
where $V_A(w^{(i)}) = 2 \sum\nolimits_{k=1}^{n_i} C_k^{(i)} (w_k^{(i)} - \bar{w}^{(i)})^2$ is the genetic variance \cite{Zhang2012JCP}. One can see
\begin{eqnarray}\label{eFTNS}
\frac{d{\phi_0({\bf G })}}{dt} = - \frac{1}{2} \frac{V_A(w)}{{\bar w}^2} + \mathbf{V}({\bf G }) \cdot {\bf G }^{-1} \cdot \mathbf{V}({\bf G }) .
\end{eqnarray}
Under frequency-independent selection, the intrinsic flux velocity is zero $\mathbf{V}(\mathbf{D}) = 0$, the detailed balance is preserved (equilibrium) and the intrinsic potential $\phi_0 = -(1/2) \mathrm{ln} \bar{w}$. This reduces to the Fisher's fundamental theorem of natural selection that adaptation rate is monotonic and only depends on genetic variance, ${d\bar w}/{dt} = V_A(w)/{\bar w}$. As seen, the Eq.(\ref{eFTNS}) works for the general evolution beyond equilibrium case with nonzero flux breaking the detailed balance. Thus, Eq.(\ref{eFTNS}) generalizes the Fisher's fundamental theorem of natural selection. The adaptive rate for general evolution depends on both the genetic variance proposed by Fisher and the intrinsic flux velocity $\mathbf{V}$ resulting from the complex biotic interactions which breaks the detailed balance, missing in the Fisher's theorem. \cite{Zhang2012JCP}.

The new landscape and
flux theory for evolution provides a nature explanation of coevolution with direct implications to Red Queen hypothesis with forever evolution even when reaching the optimum of adaptive landscape (such as limit cycle). \cite{Zhang2012JCP, VanValen} While reaching the evolution optima once attracted to the oscillation path, the evolution still proceeds due to
the rotational curl flux driving force originated from the biotic interactions. This gives the origin of non-zero genetic variance $\frac{1}{2} \frac{V_A(w)}{{\bar w}^2} = \mathbf{V}({\bf G }) \cdot {\bf G }^{-1} \cdot \mathbf{V}({\bf G })$ even at the evolution optima $\frac{d{\phi_0({\bf G })}}{dt}=0$. Therefore, natural selection can influence certain species to change their allele frequencies leading to genetic variance
even if the overall population reaches its optima \cite{Zhang2012JCP}.


\subsection{Multi-locus multi-allele evolution }

The interactions among genes are critical in understanding the evolution. The different loci representing the locations of the genes are not independent.
In particular, recombination
provides an additional way of changing alleles on a chromosome
and the fitness of an allele might depend on the
genetic background, a phenomenon called epistasis.
These interactions can lead to the evolution linkage disequilibrium.  They need to be considered in addition to the well known selection, mutation, migration and random mating for multi-locus multi-allele evolution.

Some models on multi-locus evolution were suggested with certain limitations \cite{Rice, Ewens,Yuri}. For example, the adaptive landscape is not
quantified in the most general evolution scenarios \cite{Rice, Ewens,Yuri, Shraiman}; only time dependent adaptation was considered \cite{Lassig1,
Lassig2}. Certain adaptive landscape approach has not been directly applied to multi-locus evolution. \cite{Ao2005Life} In fact,  Wright, Fisher and quasi-linkage
equilibrium
(QLE) theories can only be applied to special evolutionary scenarios.

For multi-locus multi-allele evolution, allele frequencies alone do not have enough information for quantifying
genotype frequencies. Gamete frequencies can be used instead  at different loci
in multi-locus-multi-allele evolution \cite{Yuri, Rice, Ewens}, where ``gamete'' refers to the set of alleles at the $L$ loci under consideration, such that gamete $\mathbf{i}\equiv(i_1, i_2,\ldots,i_L)$ has allele $A^j_{i_j}$ at locus $j$. One can derive the driving forces of the evolution.

The gamete frequency $x_\mathbf{i}$ evolves according to
\begin{align}
\frac{dC_\mathrm{i}}{dt}&=\mathbf{F}^S_\mathbf{i}+\mathbf{F}^M_\mathbf{i}+\mathbf{F}^R_\mathbf{i}.
\end{align}
Here, the first two terms describe the effects of natural selection and of mutations similar to the single-locus case. The effect of recombination is captured by the last term, which reads
\begin{align}
\mathbf{F}^R_\mathbf{i}=-{\sum_{Q}}^\prime r_{Q}D_{\mathbf{i},Q},
\end{align}
where the sum extends over all subsets of loci other than the empty set or the full set of all $L$ loci. Furthermore, $r_Q$ is the rate of recombination for set $Q$ and $D_{\mathbf{i},Q}=C_{\mathbf{i}}-C_{\mathbf{i}_Q}C_{\mathbf{i}_{\overline{Q}}}$ the linkage disequilibrium coefficient for locus
group
$Q$. Here, $C_{\mathbf{i}_Q}$ denotes the total frequency of all gametes that are identical to $\mathbf{i}$ at the loci in $Q$ and $C_{\mathbf{i}_{\overline{Q}}}$ that of all gametes that are identical to $\mathbf{i}$ at all loci not in $Q$. ~\cite{Rice, Ewens,Yuri}

The evolution of gamete frequency under selection was given as(\cite{Rice, Ewens,Yuri}):
$
F_\mathbf{i}^{S}=C_\mathbf{i}(w_\mathbf{i}-\overline{w})
$
where $w_\mathbf{i}$ denotes the marginal fitness of the gamete $\mathbf{i}$ and the $\overline{w}$ denotes the total fitness of all gametes in the $L$
loci system. Evolution of gamete frequencies under mutation  was also studied ~\cite{Shraiman}. One can take all these driving forces together to study the evolution of gamete frequencies.

The genetic variance of gamete fitness under selection and recombination can be shown as(\cite{Rice, Ewens,Yuri}):
\begin{align}
\label{eq:FisherGeneralized}
\frac{d \overline{w}}{\partial t}
=V_{A}+V_{R}
\end{align}
 $V_{A}=2\sum_{\mathbf{i}}C_{\mathbf{i}}(w_{\mathbf{i}}-\overline{w})^2$ represents the total gametic variance from natural selection and
 $V_{R}=-2\sum_{\mathbf{i}}\sum_{Q,Q\neq \emptyset,L}w_{\mathbf{i}}r_{Q}D_{\mathbf{i},Q}$ represents the epistatic gametic variance from the linkage
 disequilibrium of the loci(\cite{Rice, Ewens,Yuri,Shraiman}). One can see that the mean fitness increases as the recombination decreases. The generalized
 form of Fisher's fundamental theorem presented here considers the additional contribution from linkage disequilibrium due to recombination. \cite{Rice,
 Ewens,Yuri,Shraiman,Xu2017JTB}


\subsection{Evolution adaptive landscape and flux under different evolution scenarios}

The nonequilibrium landscape and flux theory can be applied to the general case of multi-locus multi-allele evolution after a suitable generalization of Eq.~(\ref{FPE4E}), which also includes recombination force $\mathbf{F}^R_\mathbf{i}$ and others such as epistasis. We discuss the results in various evolutionary scenarios listed below.

\paragraph{Absence of recombination ($r_Q=0$) and mutations ($m=0$)}

Under non-epistatic selection and genetic drift from random mating in the population of a multi-locus-multi-allele system, the Hardy-Weinberg principle and linkage equilibrium is achieved \cite{Rice, Ewens,Yuri, Xu2017JTB}: all gamete frequencies are the products of the frequencies of the constituting alleles. This reduces effectively to the one locus multi-allele evolution dynamics \cite{Zhang2012JCP}. For allele frequency independent selection, the Wright and Fisher theory works. For allele frequency dependent selection, the Wright and Fisher theory breaks down. The evolution dynamics is no longer determined by the adaptive landscape alone, but also by the curl flux due to the biotic interactions. The red queen hypothesis can be explained by the presence of curl flux driving the evolution and giving the genetic variations at optimal adaptation \cite{Zhang2012JCP}.

\paragraph{Presence of recombination ($r_Q>0$), absence of epistasis ($\epsilon_{ij}=0$) and mutations ($m=0$)}

When the fitness matrix is additive without epistasis, Wright's fitness landscape concept and the generalized Fisher's fundamental theorem can still be applied \cite{Rice, Ewens,Yuri,Shraiman}. The mean fitness never decreases because the additive form of fitness leads to $V_R=0$. However, this does not necessarily mean that the rotational curl flux vanishes.  In fact, the curl of the recombination force from gamete frequency evolution $\nabla \times (\mathbf{G}^{-1} \cdot \mathbf{F}_R )$ can be none-zero. Therefore, for recombination $r_Q>0$ and the epistatic selection $\epsilon_{ij}=0$, the evolution adaptive dynamics is determined by the gradient potential or the mean fitness, and the non-zero flux \cite{Xu2017JTB}.

\paragraph{Presence of recombination  ($r_Q>0$) and epistatic selection ($\epsilon_{ij} \neq0$), absence of mutations ($m=0$)}

When both recombination and non-zero epistasis effects (non-additive fitness matrix) are present for evolution, the dynamics becomes intrinsically nonequilibrium \cite{Akin,Hastings1981}.  Even under gamete/allele frequency independent selection, with  recombination $r_Q>0$ and epistasis $\epsilon_{ij}\neq0$, the dynamics is determined by the gradient of the evolution landscape and non-zero flux which breaks the detailed balance \cite{Xu2017JTB}. The recombination and epistasis can contribute to the breakdown of the detailed balance which leads to the breakdown of the Wright and Fisher theory for evolution.

\paragraph{Presence of mutations ($m\neq 0$)}

The mutations for the multi-locus-multi-allele evolution are often frequency-dependent. Therefore, the mutation can also lead to nonequilibrium behavior, giving another source for breaking detailed balance \cite{Xu2017JTB}.

\paragraph{Quasi-linkage equilibrium ($r_Q\gg0$, $m=0$, $\epsilon_{ij}=0$)}
If selection is weak, such that adaption is much smaller than recombination, 
the linkage disequilibrium exponentially decreases due to recombination. A state called
quasi-linkage equilibrium (QLE) emerges~\cite{Rice}. 
QLE is a good approximation for multi-locus evolution at high recombination rates and in the absence of epistasis. When the system relaxes to QLE \cite{Rice,Shraiman}, a generalized Fisher's law holds approximately.
The dynamics can then be simplified as depending on allele frequency rather than gamete frequency \cite{Nagylaki1993,Nagylaki1999}.
The landscape and flux theory works beyond these restrictions of weak selections and the QLE \cite{Xu2017JTB}. It is important to note that in  general,  the mean fitness and the optimal probability of the state do not coincide. As a result, the adaptive fitness should be quantified by the potential landscape rather than mean fitness since the landscape reflects directly the probability of the state. Furthermore,  the adaptation evolution dynamics is determined by both the landscape gradient and rotational curl flux breaking the detailed balance originated from the recombination, mutation, epistasis, or gamete/allele frequency dependent selection \cite{Xu2017JTB}.

\paragraph{Red Queen Hypothesis}

Fisher's and Wright's analysis imply that evolution will eventually come to a halt, because the maximum fitness is reached. This does not need to be the case as can be illustrated by the coevolution of a predator and a prey species~\cite{vermeij1994evolutionary,dieckmann1995evolutionary}. Assume that the predator captures prey by running faster and by being able to spot them against the background. Improving the speed and the ability to spot the prey increases the predator's fitness. As a result the prey species will evolve its speed and its camouflage to survive. Inversely, if the prey improves these traits, the predator will evolve in response to run faster and to spot the prey better. Such a competition causes `arms race' between the two species, which can lead to sustained oscillations of the species' genotype frequencies.

This evolutionary process represents a case of the Red Queen hypothesis~\cite{VanValen}, which explains the persistence of sexual reproduction and recombination: It provides an accelerated rate of evolution of a species and hence allows it to outcompete its predators and parasites. In fact, the hypothesis has been experimentally supported by a number of coevolution examples such as plant-pathogen systems \cite{clay1996red} and parasite-fish ecosystems \cite{lively1990red}.

 While the Red Queen hypothesis challenges the adaptive landscape theory of evolution of Wright \cite{Wright} and Fisher \cite{Fisher}, the associated co-evolution scenario fits naturally into the picture of the nonequilibrium landscape and flux theory. It allows for evolution to continue even if the physical environment is invariant or the landscape reaches the optimum~\cite{Zhang2012JCP, Xu2017JTB}. The origin of such forever evolution was suggested to be the rotational curl flux breaking the detailed balance as a result of gamete/allele frequency dependent selection, mutation, recombination, or epistasis ~\cite{Zhang2012JCP, Xu2017JTB}. Another possible cause of open-ended evolution was suggested by molecular evolution experiments~\cite{Worst:2016ip}, which can be interpreted as that the evolution of a species suddenly opens the possibility of evolving new traits and thus occupation of a new ecological niche.


\subsection{Evolutionary game theory}

Evolutionary game theory provides a framework for exploring the origin of the large variety of human and animal behaviors~\cite{Maynard1973Nature,Nowak2004Book, Sandholm2009Encyclopedia}. Originally, game theory focuses on the study of cooperation and competition strategies adopted from rational decision-makers. Evolution game theory was born by applying the
game theory to evolutionary biology and population dynamics for exploring the strategic interactions among large populations of agents ~\cite{Hofbauer2011PSAM,Sandholm2009Encyclopedia}. The aim for the use of game theory is
to understand the variety of human and animal behaviors~\cite{Maynard1973Nature,Nowak2004Book, Sandholm2009Encyclopedia}. In evolution biology, one of the important questions is how can cooperators survive when they can be taken advantage of by ``cheaters''? Experimental efforts in yeast \cite{Gore2009Nature} showed  that the cooperators can survive even in the presence of cheaters, the interactions between the two are through a feedback loop. The evolutionary consequences of the cooperative inactivation of antibiotics by bacteria has recently been explored experimentally \cite{Gore2013MSB, Gore2015MSB}.
In a game
, each player receives
a specific payoff following an encounter with another player with the
amount of the payoff depending on the actions chosen by the players
~\cite{Hofbauer2011PSAM,Sandholm2009Encyclopedia,Cason2003JET}. For example, in the classic Prisoner's Dilemma the two players can cooperate or not. They both receive a reward $\mathit{R}$ as payoff if they cooperate, and a punishment $\mathit{P}$ if both do not cooperate. In case, one player chooses to cooperate, whereas the other does not, the latter receives a temptation payoff $\mathit{T}$ and the former a ``sucker'' payoff $\mathit{S}$. The various payoffs satisfy $\mathit{T}>\mathit{R}>\mathit{P}>\mathit{S}$ to reflect the intuitive notions associated with the game.

In repeated games, where players have several subsequent encounters, different strategies can be defined. In the example of the Prisoner's Dilemma, one could always cooperate, always defect cooperation, or play `tit-for-tat', where the player cooperates on the first encounter and cooperates or defects on any other encounter based on the action of the other player in the previous encounter. This strategy requires a memory and thus has some cost. The payoff matrix conveniently summarizes the results of encounters of players with various strategies. Let $C_i$ denote the fraction of players with strategy $i$ and $A_{ij}$ the payoff for a player employing strategy $i$ upon an encounter with a player using strategy $j$. Then $\mathsf{A}\mathbf{C}$ gives the average payoff for each strategy. In the example
\begin{align}
\label{A_Co}
\mathsf{A} &= \left(  \begin{array}{ccc}
\mathit{P} & \mathit{P} & \mathit{T} \\
\mathit{P}-\mathit{c} & \mathit{R}-\mathit{c}  & \mathit{R}-\mathit{c} \\
\mathit{S} & \mathit{R}  & \mathit{R}
\end{array}\right).
\end{align}

Stationary distributions of strategies for non-cooperative games like the Prisoner's Dilemma, in which nobody can can gain by changing only her strategy are called Nash equilibria~\cite{Nowak2004Book} and characterizing such ``optimal'' solutions is an important task of game theory. The local stability of Nash equilibria has been studied and for very simple models Lyapunov functions have been found to characterize global stability~\cite{Sandholm2009Encyclopedia}. In general, though, it remains at present a challenging task.

Evolutionary games are repeated games where the (average) payoffs determine the fitness of a strategy. Explicitly, the time evolution of the fractions $C_i$ follows the rule that population $C_i$ grows if their average payoff is above the mean and shrinks in the opposite case. Mutations are implemented through rates of switching from one to an alternative strategy. The dynamics can be written as
\begin{align}
\frac{dC_i} {dt}&= \sum_{j} C_i f_i(C) Q_{ij} - C_i \bar{f}.
\end{align}
where $f_i = \sum_j A_{ij}C_j$ is the average payoff (or fitness) of population $i$ and $\bar f=\mathbf{C}\cdot\mathsf{A}\mathbf{C}$ the mean population payoff (or fitness)~\cite{Nowak2004Book, Bladon2010PRE, Allen2012BMB}. For a uniform mutation rate $\mu$, $Q_{ii}=1-2\mu$ and $Q_{ij}=\mu$ if $i\neq j$.

For the Prisoner's Dilemma and starting with a random initial distribution of strategies, at first the defectors will typically win. Then, however, a small population of
tit-for-tat players will invade the game and replace the defectors. In certain regions of parameter space, subsequently, cooperators will take over, which in turn will be replaced by
defectors and so on. This cycle 
has been interpreted to mimic oscillations between war and
peace in animal or even human species~\cite{Nowak2004Book}.

\begin{figure*}[!ht]
\includegraphics[width=1.0\textwidth]{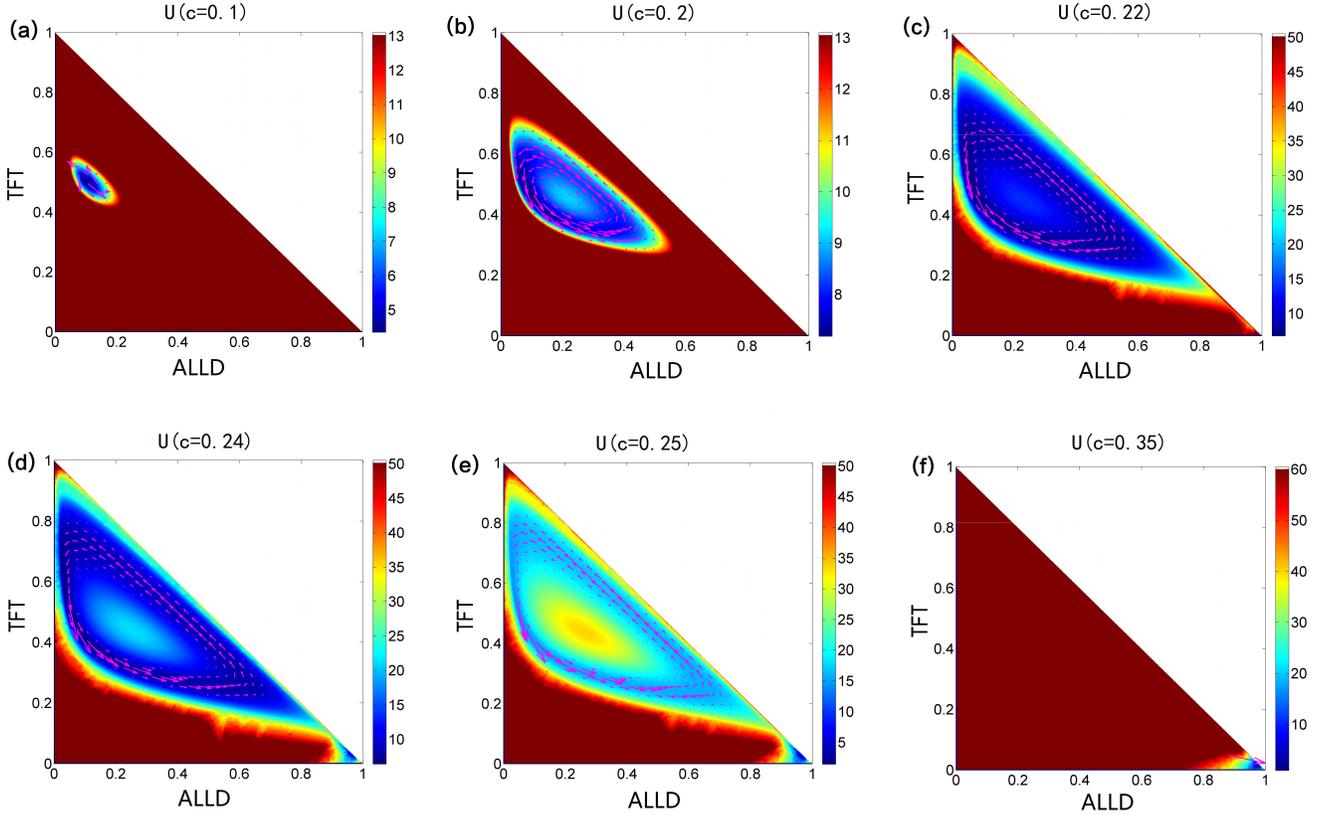}
\caption{ The nonequilibrium landscapes for game theory of repeated prisoner's dilemma for different cost $\mathit{c}$. (a) Monostable peace state where Tit-for-tat strategy dominates  at small cost $\mathit{c}=0.1$. (b)(c)(d)(e) Limit cycle oscillations between peace state with tit-for-tat strategy and war  state with all-defect strategy at $\mathit{c}=0.2$,  $\mathit{c}=0.22$, $\mathit{c}=0.24$ and $\mathit{c}=0.25$. (f). As cost increases and monostable war state with all-defect strategy at large cost ($\mathit{c}=0.35$) emerges.
(from Ref. \cite{Xu2016ArXiv2}).} \label{Xu2016ArXiv2}
\end{figure*}

The nonequilibrium landscape for the above game reveals that, for a small cost $\mathit{c}$ of the tit-for-tat strategy, the cooperator strategy is the most stable one and has the largest basin of attraction~\cite{Xu2016ArXiv2}, Fig. \ref{Xu2016ArXiv2}.
As the cost $\mathit{c}$ increases, the basin of the non-cooperator state
increases and a mixed strategy state appears. In case, the payoff for defectors increases, 
the defector state gains stability and, eventually, becomes the most stable state. 
Similarly,
increasing the reward $\mathit{R}$ or the punishment $\mathit{P}$ favors the cooperative state. A Lyapunov function can be found quantifying the global stability of the
 system dynamics~\cite{Xu2016ArXiv2}. In this way, cooperative behavior can emerge in populations of selfish individuals. Similarly competing traits can coevolve in species.

\section{nonequilibrium economy}

Although typically not considered to be part of Biology, economy reflects biological activities at human scales. An important aspect of economy is the
balance between supply and demand~\cite{Walras,Marshall}. Traditionally,
the focus has been on economic equilibrium, when supply and demand are balanced, but
this approach cannot explain economic cycles of growth and stagnation or decline, let alone economic crises~\cite{Marx,Goodwin,
Schumpeter,Minsky,Fisher1,Keynes,Keen1,Keen2}.
Thus, nonequilibrium economy is necessary. Furthermore,
the driving forces of economy need to be identified and quantified \cite{Zhang2016ArXiv}.

In conventional economy, the supply and demand are often assumed to be monotonic with respect to price under complete competition market or close to
equilibrium condition economy \cite{Marshall}. As a result, only one equilibrium emerges. The economic dynamics is then described by the shift of
this equilibrium point. However, nonequilibrium economic behavior such as inflation and under-employment, and overproduction etc are present ~\cite{Keynes,Keen1,Keen2}. This
will often lead to non-monotonic relationship of demand and supply with respect to the price ~\cite{Marshall,Mas,beckmann1969}.
As a result, multiple stable states and even limit cycle
can emerge. The competition and monopoly/oligopoly model provides a good example to illustrate this \cite{Zhang2016ArXiv}.

Experience tells us that the price of a good increases if the demand exceeds the available supply and that the production of a good increases with the
price of the good, but decreases with the available supply~\cite{Marshall,Mas,beckmann1969}. This leads to the following dynamic system for the price $P$
and the quantity $Q$ of a good
\begin{align}
\frac{d P}{d t}&=(F(P)-Q) \\
\frac{d Q}{d t}&=(P - C(Q)).
\end{align}
Here, 
$F(P)$ describes the demand of the good as a function of its current price and
$C(Q)$ is the marginal cost of producing the good, that is, the cost of producing one more unit. For
convenience, $P$ and $Q$ can take on any real value. Positive values are obtained after an appropriate shift of the origin.

Specific choices of
the demand and cost functions are $F= (-1+a)P+c$ and $C(Q) =  (d+bQ^2)Q$, where $a<1$ and $b\ge0$~\cite{Zhang2016ArXiv}. The
demand is monotonously decreasing with increasing price, but the cost function presents a nonlinear dependence on the quantity of the good, which can be
nonmonotonous. Typically, the cost for producing another unit decreases with the amount units produced. The case that the production cost increases with the supply is encountered, for example, when income effect becomes significant ( storage cost often encountered in industries and agricultural productions).
Due to the nonlinearity of the cost function, two stable steady states can appear if $a<0$. One of them corresponds to the case, when 
customers buy the commodity at a low price due to a rich supply of products. The other corresponds to a monopoly/oligopoly, where customers will still purchase it even at a higher price.
When $0<a<1$,
 limit cycles can emerge with coherent oscillations between competition and monopoly. As the demand slope changes, the market can switch from monopoly to competition or vice versa.  The resulting underlying bistable landscape topography through barrier height between monopoly and competition basins can be used to quantify the stability and switching of these states. For limit cycle dynamics, the resulting Mexican hat shaped landscapes guarantees the stability of the oscillation path, while the rotational curl flux drives the oscillation flow between the monopoly and competition ~\cite{Zhang2016ArXiv}.

\begin{figure*}[t]
\includegraphics[width=0.8\textwidth]{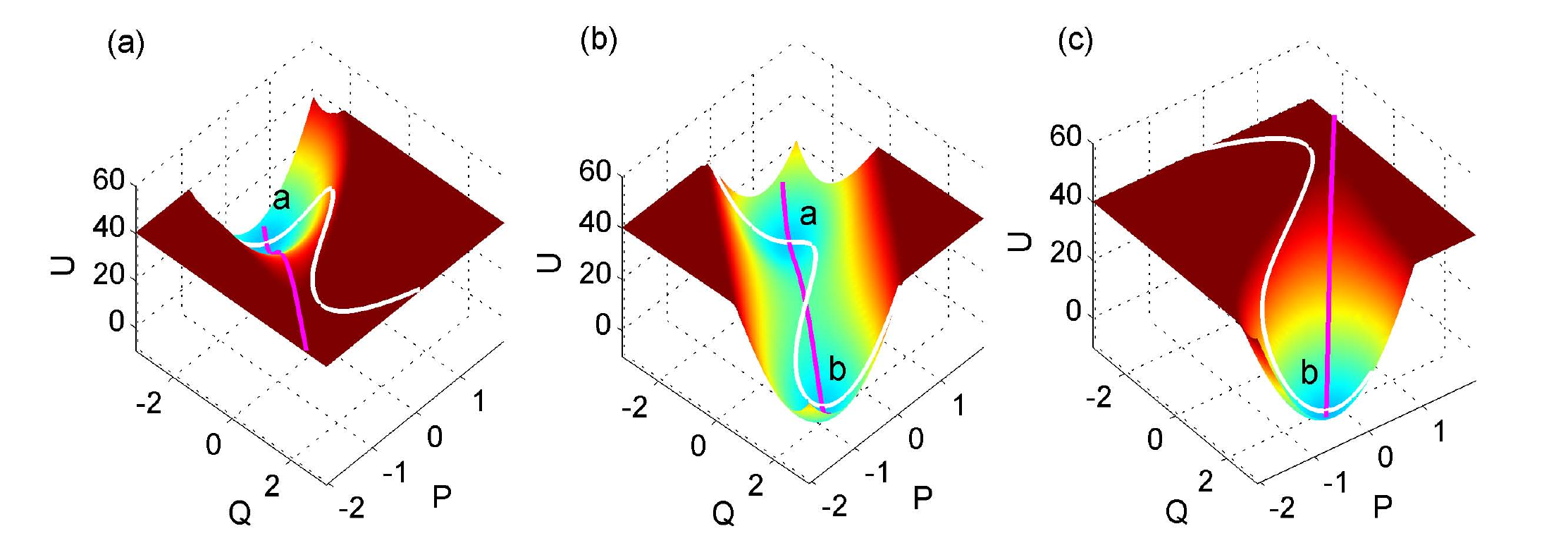}
\caption{ Nonequilibrium landscapes and shifted demand curves (purple lines) in a bistable economic model for the quantity $Q$ and the price $P$ of a commodity, from~\cite{Zhang2016ArXiv}. (a) The left shifted (decreased) demand curve and monopoly state dominant landscape.
(b) The middle located demand curve and monopoly/competition bi-stable state coexisting landscape.
(c) The right shifted (increased) demand curve and competition state dominant landscape.
}
\label{Zhangkun2016ArXiv_fig17}
\end{figure*}

 The stability of these states can be determined by the nonequilibrium landscape ~\cite{Zhang2016ArXiv}, as shown in Fig. \ref{Zhangkun2016ArXiv_fig17}. It changes as the demand curve is shifted. when the demand curve shifts to the right over
certain value, the basin of attraction of the competition state becomes deeper and more prominent relative to the monopoly state. Eventually only
competition state survives shown in Fig. \ref{Zhangkun2016ArXiv_fig17} (a). The model returns to the conventional supply and demand model where only one equilibrium state typically appears. When demand
increases, more sellers join the production for goods, more competitors form a more competitive environment leading to bistability shown in Fig. \ref{Zhangkun2016ArXiv_fig17} (b). When demand decreases, more sellers leave the
commodity market. This leads to a less competitive environment with monopoly/oligopoly shown in Fig. \ref{Zhangkun2016ArXiv_fig17} (c).

 The driving force of nonequilibrium economy is determined by both landscape and curl flux. ~\cite{Zhang2016ArXiv}. While the landscape topography provides the
quantifications and stability of economic states, the flux representing the nonequilibriumness can help to shape the dynamics. Furthermore, the flux leads to certain states unstable but help to maintain the stable flow among states. The global sensitivity analysis based on landscape and flux can be used to key element
finding for economic stability. \cite{Zhang2016ArXiv}. Furthermore, due to the presence of the flux, new states can emerge in the nonequilibrium  economy beyond the single equilibrium state assumed in the conventional equilibrium economic theory (such as the emergence of the monopoly/oligopoly state or limit cycle from the competition state or vice versa).
This discussion is quite general and can be applied to other nonequilibrium economical studies.

\section{Outlook}

As we have seen, the landscape and flux theory as well as generalized hydrodynamics provide frameworks for studying large classes of systems that are out of thermodynamic equilibrium. Notably, these concepts provide insight into the physical foundations of biological processes ranging form efficient electron transport in biomolecules to cellular dynamics and tissue development. Beyond these scales, they apply to the dynamics of populations and whole ecosystems including the behavior of human societies as well as biological evolution. Although not limited to such cases, the landscape and flux theory is particularly suited for describing systems with a finite number of degree of freedom, whereas generalized hydrodynamics provides a particularly powerful approach to collective phenomena in spatially extended systems. Our review could only give a glimpse of successful applications to biological processes and we expect many more fundamental insights into the phenomenon of Life.

In particular, further physical analysis should shed light on the question what separates living systems from other physical (or chemical) systems out of thermodynamic equilibrium. Beyond its physical aspects, living beings and their assemblies are often associated with qualities such as function, information processing, or consciousness. How can -- existing as well as future -- physical concepts be linked to them? For example, how do organizational principles of nonequilibrium systems constrain and guide the evolution of functions that provide ever better fits to the environment of a species? How does the unfolding of genetic information during the development of organisms depend on general laws governing the dynamics out of thermodynamic equilibrium? There is an already fruitful connection between the theory of information and that of statistical physics. It seems safe to speculate that the links between the two will further tighten and lead to new insights into the efficiency, speed and energy cost of information processing. We expect that the study of cell signal transduction, neural networks in the brain, and organisms will play a leading role in this endeavor~\cite{Dill, Bialek, Levenchenko, Yan2016CPB, Zeng2016ArXiv} that has only begun.

Molecular biology, which owes a large deal to physics and physicist, provides us with a rather detailed picture of the molecular inventory. It is a daunting set of a large variety of highly complex molecular machines, some of which have been characterized individually in awesome detail. On the other hand, although there is already much known about biological processes on larger scales, much of it remains to be described. Recent years have seen striking progress in experimental techniques, which allow us now to follow the embryonic development of organisms or the dynamics of bird flocks in great detail and further advances can be expected. The situation is thus somewhat opposite to that in the 19th century, when the molecular properties of materials were essentially unknown, whereas their macroscopic properties could be measured. In spite of this difference, the question of how to bridge the gap between the microscopic dynamics and the macroscopic thermodynamics/function across spatial and temporal scales attracts researchers of biological systems now as much as it did attract condensed matter physicists in the 20th century. Some progress has been made in understanding the macromolecular organization such as genome folding, transcription and translation machines, and molecular motors ~\cite{Sasai2012BJ,Wolynes2015PNAS}. Also, efficient methods for large-scale simulations of cellular networks and the whole cell at various temporal and spatial scales are being developed~\cite{Schulten2011PlosCB}. Time will tell, whether computer simulations will be the golden path towards an understanding of the relation between the microscopic and the macroscopic behavior of a biological system. In any case, new ideas and concepts are probably needed to reach this goal.

On the other hand, new experimental methods and tech- niques need to be developed to investigate the mechanism and function of biological objects. For example, to study the dynamics of cells and cellular networks, in vivo measurements of the kinetic rates are crucial and necessary. Experimental explorations are not only important to quantify the deviation from equilibrium and the role of nonequilibriumness for the function of a biological system. They are also crucial for probing and verifying fundamental laws of nonequilibrium physics such as landscape and flux, as well as thermodynamic cost and dissipation in the context of single molecule enzyme dynamics, single molecular motors, the regulation dynamics of gene motifs, cell cycle, and spatial organization of the cells, and of brain function. Furthermore, low throughput and high throughput data from experiments can help us to pin down the underlying mechanisms and nonequilibrium physics for the subjects of interest, for example, single-cell data for understanding function and diseases and connectome (a comprehensive map of neural connections in the brain) from understanding the brain function. From these studies, new biological functional phases or new forms of active matter as a result of the nonequilibriumness and environmental changes can be uncovered. This may provide opportunities to design functions even beyond the living world.

Nonequilibrium physics will be important for biological applications such as enzyme dynamics, metabolism, gene regulations, structure, function and dynamics of cells, physiology, cancer, differentiation and development, immune, aging and other human diseases, evolution or ecology, sociology, human networks, economics, even perhaps psychology and politics, to name but a few.

With the ever increasing possibilities to manipulate and interrogate biological systems, a vast playground lies at our feet. It has the potential to produce gargantuan amounts of data that will dwarf the already enormous sets currently produced every day. Without a conceptual framework guiding experiments, the sheer quantity of the data risks to severely obstruct our advances in understanding life. The physics of nonequilibrium systems will play a crucial role in this quest, aiding design of future experiments and providing a guide for data analysis. The concepts presented in this review are just the beginning.

\section*{Acknowledgement}

The authors wish to thank our mentors as well as past and current collaborators for sharing their insights and enthusiasm in our exploration of the nonequilibrium behavior of biological systems. They are too numerous to be listed here.
XF  was supported in part by National Natural Science Foundation of
China (NSFC-91430217) and MOST, China. (2016YFA0203200 and 2013YQ170585).
TL was supported in part by National Science Foundation (1227034, 1553649), Office of Naval Research (N000141622525) and Department of Energy (DE-SC0018420, DE-SC0019185). KK was supported in part by Deutsche Forschungsgemeinschaft through SFB 1027. JW was supported in part by National Science Foundation (USA: NSF-76066, NSF-1808474) and National Institutes of Health (NIH-1R01GM124177-01A1). JW would like to thank Dr. Li Xu for help on editing.


\end{document}